# Periodic Review of the Committee on Radio Astronomy Frequencies hosted by ESF

*Public Version*

March 2022

*(With amendments made in September 2023)*

# *Authors*

*– CRAF Chair –*

*Benjamin Winkel (MPIfR)*

*– CRAF Stakeholders Chair –*

*Simon Garrington (JBO, STFC)*

*– JIVE Liaison for frequency management –*

*Francesco Colomer, Waleed Madkour, Agnieszka Slowikowska*

*– Management team –*

*Pietro Bolli (INAF), Michael Lindqvist (OSO),*
*José Antonio López-Pérez (Observatorio de Yebes, IGN),*
*Leif Morten Tangen (Kartverket Geodetisk Observatorie Norway),*
*Ivan Thomas (CNRS-INSU, Nançay Radioastronomy Facility & Paris Observatory),*
*Peter Thomasson (JBO, STFC), Roel Witvers (ASTRON)*

*– CRAF Secretary –*

*Joe McCauley (Trinity College, Uni Dublin)*

*– CRAF Members (including observers and designated members) –*

*Marta Bautista (Observatorio de Yebes, IGN), Miguel Bergano (Instituto de Telecomunicações, Polo de Aveiro), Vladislavs Bezrukovs (VIRAC), Fabio Giovanardi (INAF), Hayo Hase (IVS, BKG Germany), Karel Jiricka (Astronomical Institute of the Czech Academy of Sciences), Gyula I. G. Józsa (MPIfR), Juha Kallunki (Aalto Uni, Academy of Finland), Christophe Marqué (STCE, Royal Observatory of Belgium), Derek McKay (Aalto Uni, Academy of Finland), Axel Murk (Institute of Applied Physics , Uni Bern), Vincent Pietu (IRAM), Vincenza Tornatore (Politecnico di Milano), Busang Sethole (SARAO), Marian Soida (Astronomical Observatory, Jagiellonian Uni Kraków), Boris Sorokin (SKAO), Gie Han Tan (ESO), Adrian Tiplady (SARAO), L. Viktor Tóth (Hungarian Academy of Science), Federico Di Vruno (SKAO), Susanne Wampfler (CSH, Uni Bern), Andrew Williams (ESO), Serge Yerin (Institute of Radio Astronomy of National Academy of Sciences of Ukraine)*

*– Collaborators –*

*Justin Bray (JBO, STFC), Axel Jessner (MPIfR), Nectaria Gizani (THERMOpYlae-HOU), Christian Monstein (IRSOL, Uni Locarno), Mike Peel (Imperial/IAC/ULL), João Salmim Ferreira (Associação RAEGE Açores), Harry Smith (Consultant), Giorgios P. Veldes (THERMOpYlae-UTH), Pawel Wolak (Institute of Astronomy, NCU, Torun)*

# Executive Summary

The Committee on Radio Astronomy Frequencies (CRAF) is an Expert Committee of the European Science Foundation. It aims to provide a cost-effective single voice on frequency protection issues for European radio astronomy observatories and research institutes, achieving a significantly greater impact than that achievable by individual national institutions. By working together, European observatories and institutes can profit from synergy effects, cover many more topics, and learn from each other. CRAF was founded in 1988 and has since then been engaged with the International Telecommunication Union (ITU), in particular its Radiocommunication Sector (ITU-R), and the European Conference of Postal and Telecommunications Administrations (CEPT) and its European Communications Committee (ECC). Almost all work in spectrum management is organised around the 3–4-year cycle of the world radiocommunication conferences (WRCs) hosted by the ITU-R. To take account of new developments, the Radio Regulations (RR) are revised at the WRCs, which usually results in many 'downstream' activities at regional and/or national levels. The four-week long WRCs are the culmination of hundreds of meetings and thousands of preparatory documents. The ITU-R Radio Regulations contain rules for (multi-lateral) spectrum management and the important frequency allocation table, which defines the radio services that are permitted to operate in a given frequency band.

The ITU-R recognised radio astronomy as a new service (the radio astronomy service; RAS) in 1959 and allocated several bands to it, which are eligible for protection by national administrations. Allocations to radio services (including RAS) in a certain frequency interval can either be primary or secondary, with the secondary service in a band having only limited rights to claim protection. All RAS observations are passive, i.e., receive-only. The so-called "passive bands", in which all active emissions are prohibited (RR footnote No. 5.340), are the most strongly protected. Some additional bands have a footnote in the RR, which urges administrations to protect certain frequencies, in which RAS operates, but where it has no allocation ("5.149-only" bands). This is the weakest form of protection, and national administrations determine how seriously this footnote is considered for the protection of radio astronomy.

There are different modes of radio astronomy observations, which are mainly categorised by the type of the received emission. Cosmic sources can emit not only across a broadband continuum of frequencies, but also at very specific ones – spectral lines. While for continuum spectra, astronomers would have some freedom to change the observed frequencies – as long as the gaps between the measured bands are not too large – spectral line frequencies are dictated by nature. Even worse, if the sources of interest are significantly redshifted owing to the expansion of the universe, the received frequencies can fall outside the RAS bands. One consequence of this is that radio astronomers regularly observe at frequencies, which are not allocated or assigned to the RAS. This is called 'opportunistic observing' and obviously works only on a best-effort basis, as one cannot claim protection for measurements outside the RAS bands.

The international spectrum management organisations, ITU-R and CEPT, are input-document driven. This means that any activities, discussions, and outcomes are solely based on text that was provided before a meeting. Thus, for effective work, input documents of other organisations, companies and administrations need to be analysed in advance, proper responses must be worked out and fed back into the groups. This often involves the preparation of spectrum compatibility calculations, which analyse the conditions required for two radio services to co-exist. Radio astronomy, as a passive service, is usually the victim service in such calculations. Based on such studies, contributions for use in regulatory texts (e.g., the ITU-R Radio Regulations, Recommendations, and Resolutions, or CEPT/ECC Decisions) have to be prepared by CRAF delegates. Last but not least, the CRAF input documents need not only to be presented in the meetings but also require substantial lobbying for the RAS positions, in order to maximise the support from national administrations. It is important to note that while CRAF has a seat at the ITU, only the national administrations have a vote and hence the lobbying element is an important one. The preparation, presentation, and advocacy for these input documents to safeguard radio astronomy in Europe is a primary role for CRAF, and (at the ITU and CEPT level) is largely delegated to CRAF by radio observatories, through their CRAF members.

In the period 2011–2021, which is under review, CRAF produced a large number of input documents and other publications and participated in many spectrum management meetings, both at ITU-R and CEPT levels. Between 2011–2019, typically one to two dozen of documents were submitted per year. Since 2020 this figure has more than doubled. The same is true for the number of meetings attended, which has exceeded 50 per year since 2020. This could only be realised by a major re-structuring of CRAF's management and procedures in 2018, introducing a dedicated management team and so-called work-item teams, each of which focuses on a particular topic, such as satellite systems or mobile phone networks. The re-structuring resulted in CRAF encouraging more of its members to become actively engaged, which was necessary to manage the steadily increasing work-load that was experienced over the past ten years.

Work on most spectrum management topics is a long-term activity. Many of the issues have kept generations of CRAF delegates busy, some even from the formation of the Committee in 1988. Among the most labour-intensive tasks are those related to

- Satellite systems, in particular Iridium and the more recent mega-constellations,
- International Mobile Telecommunication (IMT, aka cell phone networks),
- Protecting the passive bands (RR No. 5.340), in which no emissions are permitted,
- Short range Radars and ultra-wide band devices, and
- Future protection of the Geo-VLBI system VGOS.

Often, CRAF successfully defended RAS bands and observatories. However, given the significant commercial interests at stake, and the legal and lobbying resources which large national and international corporations can support, not all battles can be won. CRAF has been very successful so far in achieving protection from air- and spaceborne transmitters with the exception of the Iridium satellite system, whose company has largely ignored regulation and licence conditions. Examples of successes include:

- Helicopter Radar at 77 GHz,
- Planned implementation of no-fly zones to protect RAS stations from cell-phone equipment on aerial vehicles, even for 2$^{nd}$ harmonics emissions,
- Additional mitigation measures were implemented to stop interference from GLONASS,
- Protection of European stations from the Starlink and OneWeb mega-constellations was established – in some countries CRAF even successfully managed to achieve additional protection,
- Protection at 150 and 408 MHz from S-PCS systems was established.

CRAF's track record with respect to the IMT is very mixed:

- Protection was established for the 5.149-only bands at 3.35 GHz,

- The secondary RAS band 2655–2690 MHz and the 5.149-only band 1718.8–1722.2 MHz was lost in CEPT countries to IMT,
- Protection of the passive band 2690–2700 MHz was ensured from IMT in the adjacent band,
- WRC-19 allocated the bands 24.25–27.5 GHz and 40.5–43.5 GHz to IMT, two very important frequency ranges featuring a plethora of molecular spectral lines that were so far well observable in an opportunistic manner (e.g., without a RAS allocation); at least within Europe local coordination is required for the two affected RAS bands at 23.6–24 GHz and 42.5–43.5 GHz thanks to CRAF's work,
- At WRC-23 the fundamentally important 5.149-only band at 6.65 GHz, which features a Methanol spectral line, will likely be lost to IMT and thus no longer available for the RAS[1].

The RAS also is under heavy pressure with respect to the passive (RR No. 5.340) bands, where all emissions by active services are prohibited. More and more spectrum management stakeholders try to undermine this very strict regulation in order to enable large-bandwidth applications. CRAF opposes this wholeheartedly. At present approximately half of the CEPT administrations are in favour of relaxing the status of 5.340 bands, while the other half is (still) willing to protect RAS[2]. At the ITU-R, the atmosphere is even less favourable to RAS.

CRAF has strong links to all spectrum management organisation and fora, including the ITU-R, the CEPT, but also other regional RAS organisations and the international RAS group, IUCAF (Scientific Committee on Frequency Allocations for Radio Astronomy and Space Science). CRAF collaborates very closely with the Square Kilometre Array Observatory (SKAO), which is a CRAF Observer institution. Many joint input documents have been produced and there is cooperation on the development of software for spectrum compatibility calculations. SKAO is also one of two main sponsors of the IAU Centre for the Protection of the Dark and Quiet Sky from Satellite Constellation Interference. CRAF is a top-tier supporter of the Centre.

Another success story is the expertise in developing increasingly sophisticated compatibility calculations. While lobbying and arguments of economic value are important, the 'bread and butter' of successful spectrum sharing is to work out the technical conditions under which effective coexistence of two or more services is possible. CRAF has undertaken considerable work to produce a Python-based software package, named pycraf, to ease and streamline the creation of compatibility calculations. Moreover, this software is open-source and widely available, representing a major step forward in the transparency and reproducibility of these important and complex calculations.

Despite all the successes, there are also some weak spots, which the Committee should address in the next years. One major issue for the Committee is the huge volume of work. Other stakeholders in spectrum management, most often large corporations operating or selling IMT, satellite and airborne services, have a significant advantage regarding technical, legal and lobbying resources. Despite being the victim service, it usually falls on RAS representatives to provide compatibility studies. There are only two fundamental solutions to this: (1) further increase the budget spent on spectrum management, such that CRAF and/or its member institutes can hire additional people, and/or (2) set priorities on the topics to be covered. While the second option seems reasonable and cost-effective, it should be considered with care. Once RAS has lost access to a band, it will likely never be able to get it back again, because once commercial services begin to use a band there are financial consequences of moving frequency. Deciding the priorities for protecting bands is difficult and these are bound to evolve in time: the scientific requirements for future experiments will depend on discoveries yet to be made.

---

[1] As of September 2023, it is still unclear whether CEPT will endorse the use of this band for an IMT identification. Some administrations consider to enable this frequency range for WiFi applications, which would likely be somewhat easier to coordinate with in the vicinity of RAS stations. This is currently under study in CEPT/ECC SE45 and PT1 groups.

[2] It is noted that in 2023, the use of some 5.340 frequencies above 100 GHz was made possible by ECC Decision (22)03 for a very limited range of applications. Other future uses were highly discouraged in a statement sent to ETSI, which they should consider in new industry standards.

It is CRAF's view that the scientific community as a whole must be more involved in spectrum management issues. This means increased lobbying for spectrum (not only for funding of new instruments) on the one hand, but also public outreach on the other. The case of the Starlink mega-constellation has made it very clear that there are many people who do not want to sacrifice their cultural heritage of a dark night sky. Radio astronomy could build on this and try to shape the public opinion to increase support for scientific use of the spectrum. Based on the high level of expertise of our committee members, it is clear that we must take on a key role in this endeavour. The CRAF management team is fully aware that the Committee must increase its efforts in this "community management", not only with public outreach activities, but also by putting more resources into informing the radio astronomical community and convincing our colleagues to become more involved.

*Disclaimer: This is the public version of the self-evaluation report. A few paragraphs, which touch financial or personnel topics, were removed. Furthermore, as the original report was published in early 2022, more than a year ago, we added several footnotes in places throughout the report, where newer information was available and deemed relevant. If you have questions or want to get involved, please contact the CRAF management team via email: contact(at)craf.eu.*

# Glossary

| | |
|---|---|
| 5G | 5th generation mobile communications standard |
| AI | (WRC) Agenda Item |
| AAS | Active Antenna Systems (phased arrays, used for beam-forming antennas) |
| ALMA | Atacama Large Millimeter Array (Observatory) |
| AV | Aerial vehicles |
| BEM | Block Edge Mask (i.e., how strong the out-of-band signal must be suppressed) |
| BS | Base Station (of the IMT) |
| C&C | Command & Control |
| CEPT | European Conference of Postal and Telecommunications Administrations |
| CG | Correspondence Group |
| CORF | Committee on Radio Frequencies of the US National Research Council |
| CPM | (World Radiocommunication) Conference Preparatory Meeting (of the ITU-R) |
| CRAF | Committee on Radio Astronomy Frequencies |
| D&QS | Dark & Quiet Skies for Science and Society (Workshops) |
| EAS | European Astronomical Society |
| EC | European Commission |
| ECC | Electronic Communications Committee (CEPT group for spectrum management) |
| ECC/CPG | ECC (World Radiocommunication) Conference Preparatory Group |
| ECC/FG-WT | ECC/Forum Group on Wind Turbines |
| ECC/FM44 | ECC/Frequency Management Group 44 (Satellite Communications) |
| ECC/PT1 | ECC/Project Team 1 (IMT Matters) |
| ECC/SE7 | ECC/Spectrum Engineering Group 7 (Compatibility and sharing issues of mobile systems) |
| ECC/SE21 | ECC/Spectrum Engineering Group 21 (Unwanted emissions and receiver characterisation) |
| ECC/SE24 | Spectrum Engineering Group 24 (Short Range Devices) |
| ECC/SE40 | Spectrum Engineering Group 40 (Space Service compatibility issues) |
| EESS | Earth-Exploration Satellite Service |
| EPFD | Equivalent Power Flux Density (Studies) |
| ESA | European Space Agency |
| ESF | European Science Foundation |
| ESIM | (Land, maritime or aerial) Earth Station in Motion (these are mobile terminals but usually not hand-helds) |
| ESO | European Southern Observatory |
| EVGA | European VLBI Group for Geodesy and Astrometry |
| FM | CRAF Frequency Manager |
| FSS | Fixed Satellite Service (of the ITU-R) |
| GADSS | Global Aeronautical Distress and Safety System (of the ITU-R) |
| GMDSS | Global Maritime Distress and Safety System (of the ITU-R) |
| GSO | Geostationary Satellite Systems |
| HAPS | High Altitude (stratospheric) Platforms |
| HIBS | (MFCN) Base Stations at high altitudes (e.g., on board HAPS or airplanes) |
| IAU | International Astronomical Union |
| IAU Centre | IAU Centre for the Protection of the Dark and Quiet Sky from Satellite Constellation Interference |
| IMT | International Mobile Telecommunications |
| IRAM | Institut de Radio Astronomie Millimétrique |
| IVS | International VLBI Service for Geodesy and Astrometry |
| ITU-R | International Telecommunication Union (Radiocommunication Sector) |
| ITU-R SG 3 | ITU-R Study Group on Radiowave propagation |
| ITU-R SG 4 | ITU-R Study Group on Satellite services |
| ITU-R SG 5 | ITU-R Study Group on Terrestrial services |
| ITU-R WP 5B | ITU-R Working Party 5B on Maritime mobile service including Global Maritime Distress and Safety System (GMDSS); aeronautical mobile service and radiodetermination service |
| ITU-R WP 5D | ITU-R Working Party 5D on IMT |
| ITU-R SG 7 | ITU-R Study Group on Science services |
| ITU-R WP 7B | ITU-R Working Party 7B on |
| ITU-R WP 7C | ITU-R Working Party 7C on Remote sensing systems and space research sensors, including planetary sensors |
| ITU-R WP 7D | ITU-R Working Party 7D on radio astronomy |
| ITU-R T/G 5-1 | ITU-R Task Group 5-1 (WRC-19 Agenda item 1.13; IMT above 24 GHz) |
| ITU-R T/G 6-1 | ITU-R Task Group 6-1 (WRC-23 Agenda item 1.5; IMT below 1 GHz) |
| IUCAF | The Scientific Committee on Frequency Allocations for Radio Astronomy and Space Science |
| IVS | International VLBI Service for Geodesy and Astrometry |
| JIVE | Joint Institute for VLBI ERIC |
| LEO | Low-Earth Orbit (satellites) |
| LOFAR | Low-Frequency Array (Observatory) |
| LoU | Letter of Understanding |
| LTE (4G) | Long-term evolution (network/technology); cell-phone standard |
| MetAids | Meteorological Aids Service (of the ITU-R) |
| MFCN | Mobile/Fixed Communications Networks |
| MI | (CRAF) Member Institution |
| MoU | Memorandum of Understanding |
| MS | Mobile Service (of the ITU-R) |
| MSS | Mobile Satellite Service (of the ITU-R) |
| MT | (CRAF) Management Team |
| NSF | National Science Foundation (of the United States of America) |
| NGSO | Non-Geostationary Satellite Systems |
| ORP | Opticon-RadioNet Pilot (EU science program for astronomical observatories) |
| PC | Public Consultation |
| PDNR | (ITU-R) Preliminary Draft New Report |
| RAFCAP | Radio Astronomy Frequency Committee in the Asia-Pacific region |
| RAS | Radio Astronomy Service (of the ITU-R) |
| RDS | Radio Determination Service (of the ITU-R) |
| RFI | Radio Frequency Interference |
| RLAN | Radio Local Area Network (aka WiFi) |
| RoP | Rules of Procedure (of the ITU-R Radio Regulations) |
| RQZ | Radio Quiet Zone |
| RR | (ITU-R) Radio Regulations |
| RSPG | Radio Spectrum Policy Group (of the European Commission) |
| S-PCS | Satellite Personal Communication Services (e.g., IoT for remote sensors) |
| SatMoU | CEPT organisation for satellite monitoring |
| SF | (CRAF's) Stakeholder's Forum |
| SKAO | Square Kilometre Array Observatory |
| SRD | Short Range Devices |
| SRS | Space Research Service (of the ITU-R) |
| SWOT | Analysis of Strengths/Weaknesses/Opportunities/Threats |
| SWS | Space Weather Systems |
| ToR | Terms of Reference |
| UAS/UAV | Unmanned Aerial Systems/Vehicles |
| UE | User Equipment (of the IMT); aka smart phones |
| UN | United Nations |
| UWB | Ultra-wide Band (Radar) |
| VGOS | VLBI Global Observing System |
| VLBI | Very Long Baseline Interferometry |
| WI | (CRAF) Working Item (team) |
| WRC | ITU-R World Radiocommunications Conference |
| XO | Extraordinary Meeting |

# 1 Introduction

The Universe in which we live, in particular the night sky, has influenced humanity since the earliest days and astronomy continues to capture people's imaginations all over the world. The ancient astronomers had only visual access to the marvels of the sky, but the rapid technological progress of the past century has opened many new and fascinating windows into the universe for us, mediated by radio waves, x-rays and gamma-rays, ultraviolet and infrared optical radiation, neutrinos, and even gravitational waves. Radio astronomy was the first such window beyond our eyes' optical range. It began in earnest in the years after the second world war and has evolved in step with the ongoing fast progress of radio engineering and computer science. Radio observations have transformed our understanding of the universe in the last seven decades and captured the public imagination. Quasars, radio jets, pulsars, masers, emission of gravitational waves, the Big Bang, images of distant black hole environments to mention just a few of the new discoveries were first revealed by radio astronomy.

To continue this advance with all its potential benefits, it is necessary to operate many observatories with various characteristics at diverse locations and to be able to observe in a large number of frequency bands. Building and operating radio observatories requires sophisticated technology at the forefront of scientific and engineering developments in many fields, and often developing new technologies with applications far beyond astronomy along the way. The main purpose of our instruments is to observe signals from the Universe, but some are also capable of measuring interesting phenomena in the Earth's atmosphere, ionosphere, and magnetosphere, including 'space weather'.

The extra-ordinary sensitivity of radio observatories, however, means that they can suffer significantly from anthropogenic signals, originating from sources on or near the surface of the Earth or in orbit (such as satellites). Many of these sources are transmitting legitimately to provide some service. Prime examples would be communication networks, Radar applications or radio/TV broadcast towers. Naturally, there is a conflict of interest between these active uses of the spectrum and the purely passive radio astronomical observations. In addition, there are also sources of unintended radiation. Every electric or electronic device can emit some electro-magnetic waves, even without a proper antenna. While equipment at observatories can be evaluated, controlled, or modified, other sources at a range of distances from our telescopes, such as power lines or railway lines (with electrification), wind turbines, solar panels and (LED) lighting, are usually beyond our control.

The importance of radio astronomy has long been acknowledged. At worldwide level, the International Telecommunication Union (ITU), in particular its Radiocommunication Sector (ITU-R), is responsible for spectrum allocations and regulation to ensure a fair and well-organised access of all stakeholders to the radio spectrum. The World Radio Administrative Conference in 1959 (WARC-1959) led to new Radio Regulations formally recognising radio astronomy as a radio service (the radio astronomy service; RAS) eligible for protection by member administrations. It also specified the first frequency bands that were exclusively allocated for passive use (no emissions permitted).

Allocations to radio services (including RAS) in a certain frequency interval can either be primary or secondary, with the secondary service in a band having only limited rights to claim protection. All RAS observations are passive, i.e., receive-only. The so-called "passive bands" are the most strongly protected, in which all active emissions are prohibited (RR footnote No. 5.340). Some additional bands have a footnote in the RR, which urges administrations to protect certain frequencies, in which RAS operates, but where it has no allocation ("5.149-only" bands). This is the weakest form of protection, and it is up to national administrations, in terms of how seriously this footnote is taken.

There are different modes of radio astronomy observations, which are mainly categorised by the type of the received emission. Cosmic source can emit broadband continuum, but also spectral lines. While for continuum spectra, astronomers would have some freedom to change the observed frequencies – if the gaps between the measured bands are not too large – spectral line frequencies are dictated by nature. Even worse, if the sources of interest are significantly redshifted, the received frequencies can fall outside of the RAS bands. Most protected bands were chosen based on the most important atomic and molecular spectral lines, such as those originating from neutral atomic hydrogen, the water molecule, methanol, silicon oxide, ammonia, or the hydroxyl radical. One consequence of this is that radio astronomers regularly observe at frequencies, which are not allocated or assigned to the RAS. This is sometimes called 'opportunistic observing'. Obviously, this only works on a best-effort basis, as one cannot claim protection for measurements outside the RAS bands. Modern radio astronomy receivers feature enormous input bandwidths of up to several GHz. While this boosts the possibilities for opportunistic observations it comes with a huge drawback. If just one very strong signal enters the receiver at any frequency, the system can be driven into non-linearities or even into blocking, which effectively stops the observations. Not much can be done to avoid this. Stop-band filters can in principle be used to suppress the strong signals, but such filters usually need to be placed at the very beginning of the receivers, which could severely impact the sensitivity.

Radio astronomy is an important part of humanity's fundamental scientific endeavour, which is immensely significant in its own right, transforming and extending our view of the Universe and our understanding of the processes that led to the formation of stars, planets and life on Earth. In addition, radio astronomy has contributed to many now commonly used technologies (see also Report ITU-R RS.2178-0). The most well-known is the development of the techniques and protocols for WiFi. Other include antenna technology, highly sensitive receivers, and also the very long baseline interferometry (VLBI) technique, which nowadays also powers some of the most important geodetic applications (which in turn are fundamental to a large variety of downstream applications, including the satellite industry). Likewise, radio astronomer's signal and image processing methods entered many other fields, such as medical imaging. All that has been recognised as being in the public interest in many countries around the world and their taxpayers have made major investments in the development of radio astronomy, as it is fully financed by public funds, like most endeavours of non-applied science. It is anticipated that this investment will continue and that even more countries will join major radio astronomical projects.

Radio astronomy requires support at various levels and can only thrive under certain conditions. Governments fund radio astronomical stations, observatories and the specialised scientists and technicians who are needed to develop, operate, and maintain them. It follows that it is in their natural interest that investments are not wasted and that progress in science will continue. One of the fundamental and non-negotiable requirements for that is the adequate access of radio astronomers to the necessary frequency bands in the radio spectrum without harmful interference. Radio astronomers and the communications industry benefit greatly from the inexorable progress in radio, computing, and space technology. However, that creates an ever-increasing number of ways of using the finite radio spectrum and consequently there is an insatiable demand for more frequency bands. The co-existence of scientific services, such as radio astronomy, with a growing industrial use of radio waves needs careful and diligent coordination for which regulatory efforts based on compatibility studies are required. For this aim, the Committee of Radio Astronomy Frequencies (CRAF) was established in 1988 by the European Science Foundation (ESF) Executive Council.

CRAF is one of three regional radio astronomy spectrum management organisations. In the US, there is CORF (Committee on Radio Frequencies of the US National Research Council), in the Asia-Pacific region there is RAFCAP (Radio Astronomy Frequency Committee in the Asia-Pacific region). Furthermore, the International Union of Radio Science (URSI) runs IUCAF (Scientific Committee on Frequency Allocations for Radio Astronomy and Space Science), which is an international committee. The SKAO (Square Kilometre Array Observatory) is also active in spectrum management, but with fewer human resources.

# 2 The Committee of Radio Astronomy Frequencies

## 2.1 Radio Astronomy in Europe

There are many radio-astronomical observatories in European countries, including Russia and Ukraine. The radio telescopes in use can be categorised by type (single dish vs. interferometer), frequency bands supported (from tens of MHz to hundreds of GHz), size (less than a metre per antenna element up to slightly greater than 100-m diameter) and level of professionality (teaching instruments vs. state-of-the-art science facilities). In Table 1, a list of all European radio astronomy stations is presented, plus the one operated by SARAO in South Africa, which is also a CRAF member.

*Table 1: List of European radio astronomy stations.*

| Country | Name | N Latitude | E Longitude |
|---|---|---|---|
| **Belgium** | Humain | 50° 11′ 31″ | 05° 15′ 12″ |
| **Czech Republic** | Ondrejov | 49° 54′ 55″ | 14° 46′ 52″ |
| **Finland** | Metsähovi | 60° 13′ 05″ | 24° 23′ 36″ |
| **France** | Nançay | 47° 22′ 24″ | 02° 11′ 50″ |
| | NOEMA | 44° 38′ 02″ | 05° 54′ 29″ |
| **Germany** | Effelsberg | 50° 31′ 29″ | 06°53′ 03″ |
| | Wettzell | 49° 08′ 42″ | 12° 52′ 38″ |
| **Hungary** | BEST† | 47° 54′ 22″ | 19°32′ 23″ |
| **Italy** | Matera | 40° 38′ 58″ | 16° 42′ 14″ |
| | Medicina | 44° 31′ 15″ | 11° 38′ 49″ |
| | Noto | 36° 52′ 33″ | 14° 59′ 20″ |
| | Sardinia | 39° 29′ 34″ | 09° 14′ 42″ |
| **Latvia** | Ventspils | 57° 33′ 12″ | 21° 51′ 17″ |
| **The Netherlands** | LOFAR* | 52° 55′ | 06° 52′ |
| | Westerbork | 52° 55′ 01″ | 06° 36′ 15″ |
| **Poland** | Torun | 52° 54′ 38″ | 18° 33′ 51″ |
| **Portugal** | Santa Maria | 36° 59′ 07″ | –25° 07′ 34″ |
| **Russia** | Badari | 51° 46′ 10″ | 102° 14′ 00″ |
| | Pushchino | 54° 49′ 20″ | 37° 37′ 53″ |
| | Svetloe | 61° 31′ 56″ | 29° 46′ 54″ |
| | Zelenchukskaya | 43° 47′ 15″ | 41° 34′ 00″ |
| **Spain** | Pico Veleta | 37° 03′ 58″ | –03° 23′ 34″ |
| | Teide Observatory | 28° 18′ 01″ | –16° 30′ 36″ |
| | Yebes | 40° 31′ 29″ | –03° 05′ 13″ |
| **Sweden** | Onsala | 57° 23′ 35″ | 11° 55′ 04 |
| **Switzerland** | Bleien | 47° 20′ 24 | 08° 06′ 43″ |
| **Turkey** | Kayseri | 38° 42′ 37″ | 35° 32′ 43″ |
| **Ukraine** | Yevpatoriya | 45° 11′ 18″ | 33° 11′ 10″ |
| | Kharkiv | 49° 38′ 17″ | 36° 56′ 29″ |
| | Lviv | 51° 28′ 19″ | 23° 49′ 38″ |
| | Odesa | 46° 23′ 44″ | 30° 16′ 19″ |
| | Poltava | 49° 37′ 48″ | 34° 49′ 28″ |
| | Zmiiv | 49° 39′ 57″ | 36° 21′ 13″ |
| **United Kingdom** | Cambridge | 52° 09′ 59″ | 00° 02′ 20″ |
| | Knockin | 52° 47′ 25″ | –02° 59′ 50″ |
| | Darnhall | 53° 09′ 23″ | –02° 32′ 09″ |
| | Defford | 52° 06′ 02″ | –02° 08′ 40″ |
| | Jodrell Bank | 53° 14′ 10″ | –02° 18′ 26″ |
| | Pickmere | 53° 17′ 19″ | –02° 26′ 44″ |
| **South Africa** | SARAO | –30° 43′ 16″ | 21° 24′ 40″ |

†Under construction

*LOFAR core station; there are many more LOFAR sites; see Appendix 7.2

## 2.2 History of the Committee, structure, and membership

From the very beginning radio astronomers had been involved in the consultations on many technical and policy questions relating to the scientific use of radio frequencies and to the coordination with other (active) radio services. In order to coordinate these activities, the Committee on Radio Frequencies of the US National Research Council (CORF) was formed in the early 1960s. A greater need for better communication and coordination among radio astronomers in Europe was seen to be necessary around 1985 and, following an initiative by Dr. Hans Kahlman, Dr. Willem Baan and Dr. Titus Spoelstra, the Committee on Radio Astronomy Frequencies was established by the ESF Executive Council in 1988 as an Expert Committee. For its member institutions – European radio astronomy research facilities and observatories – it shall "provide a cost-effective single European voice on frequency protection issues, achieving a significantly greater impact than that achievable by individual national institutions" (from CRAF Charter). By working together, European observatories and institutes can profit from synergy effects, cover many more topics and learn from each other. As a regional organisation, CRAF does not only work at the ITU-R and national levels, but also engages with the European spectrum organisation, the European Conference of Postal and Telecommunications Administrations (CEPT) and its European Communications Committee (ECC).

The CRAF Charter and Terms of Reference define how CRAF operates:

- Member institutions (MIs) should be involved in radio astronomy or related sciences. New member institutions are accepted by the ESF after consultation with CRAF.
- Member institutions are responsible for financing CRAF via annual contribution payments.
- Each member institution may nominate a committee member for a three-year term, normally renewable.

- Committee members should be experts of standing within one of the communities that CRAF represents.
- Committee members shall maintain strong links between CRAF and its member institutions, the broader scientific community, as well as national network agencies or other regulatory authorities in their home countries.
- Committee members appoint a Chair of the Committee, a Secretary and a Frequency Manager (FM). The ESF nominates a Liaison Officer to CRAF, who will normally attend CRAF's meetings.
- CRAF shall hold regular plenary meetings convened by the CRAF Chair.
- The committee reports to the ESF and the member institutions, as required.

The funding of CRAF is further specified in an appendix to the Charter. As it was recognised that not all member institutes would be able to pay a full share, a reduced share was set for some members, and for some it was even set to zero. Furthermore, a few organisations joined CRAF as "Observers", which allows them to participate in our public meetings, only, and without voting rights. One of the Observer members is the Square Kilometre Array Observatory (SKAO). The role of SKAO and our collaboration is further discussed in Section 4.4. During the period 2016 to 2018 a review of the CRAF Charter was undertaken to adapt it to current needs. For formal reasons the Charter itself was left untouched, but a new *Stakeholder's agreement* was put into place:

- The *Stakeholders Forum (SF)* is established as the platform of directors or equivalent management representatives of all CRAF Member and Observer Institutions, optimising their collective engagement in CRAF. The SF is composed of one named director or equivalent management representative of each CRAF Member and Observer Institution. The SF Chair is appointed for a two-year term, by the representatives of the Member Institutions that pay a full share annual financial contribution to CRAF (known as the *Funders Circle*).
- The SF provides a focused opportunity for the Expert Committee and FM to inform the management of the Member and Observer Institutions of recent and ongoing activities, developments, and issues.
- The SF engages in a collective, strategic discussion on priorities and activities for spectrum management related to radio astronomy observing capabilities.
- The SF oversees the CRAF finances and the employment of the FM. Furthermore, it takes care of institutional memberships.

Furthermore, a *Management Team* (MT) was established, which consists of the CRAF delegates representing the Member Institutions that pay a full share annual financial contribution. The MT normally meets once per week to manage organisational matters, to discuss and decide on strategic issues in the operative CRAF work and prepare CRAF plenary meetings.

Within CRAF, several working groups, the so-called Work Item (WI) teams exist, which take care of all the topics in a particular field of spectrum management, such as satellite applications or mobile networks. This allows CRAF to work more effectively (in smaller groups) and to increase the level of expertise and involvement in these areas. Considering the limited time and resources of each individual CRAF delegate, it is clear that it is not possible for them to get involved in every WI team. Therefore, the WI teams are obliged to produce minutes of their meetings and high-level summaries for the CRAF plenary meetings. An overview on the current WI teams and their topics is given in Table 2.

*Table 2: Current CRAF Work Item (WI) teams.*

| WI team | **Tasks** |
|---|---|
| SEnn | Spectrum engineering topics in CEPT, in particular ECC groups SE7, SE24, SRD/MG |
| SAT | Satellite systems at CEPT and ITU-R, in particular ECC Groups SE40, FM44, and ITU-R SG 4 |
| IMT | IMT-related topics in CEPT and ITU-R, in particular ECC Groups PT1, and ITU-R WP 5D |
| VGOS | VLBI Global Observing System; organise future protection at ITU-R; active at ITU-R WP 7D |
| SWS | Space weather sensors under WRC23 A.I. 9.1a; mainly at ITU-R WP 7C |
| MONIT | Spectrum monitoring and RFI measurements at CRAF observatories |
| PO | Public outreach activities |

Through its members, CRAF represents the spectrum management concerns of the majority of radio astronomical observatories, institutes, and departments in ITU Region 1 (see Figure 1). The three ITU regions designate the extent of common frameworks of frequency allocations.

CRAF's current member institutions and organisations are listed in Appendix 7.1. In Figure 2, a map with the European MIs is shown. The South African Radio Astronomy Observatory (SARAO) is also a member. Furthermore, there are some multinational CRAF members, such as the International VLBI Service for Geodesy and Astrometry (IVS) and the Institut de Radioastronomie Millimétrique (IRAM). The European Space Agency (ESA), the European Southern Observatory (ESO), and the Square Kilometre Array Observatory (SKAO) currently have Observer status in CRAF.

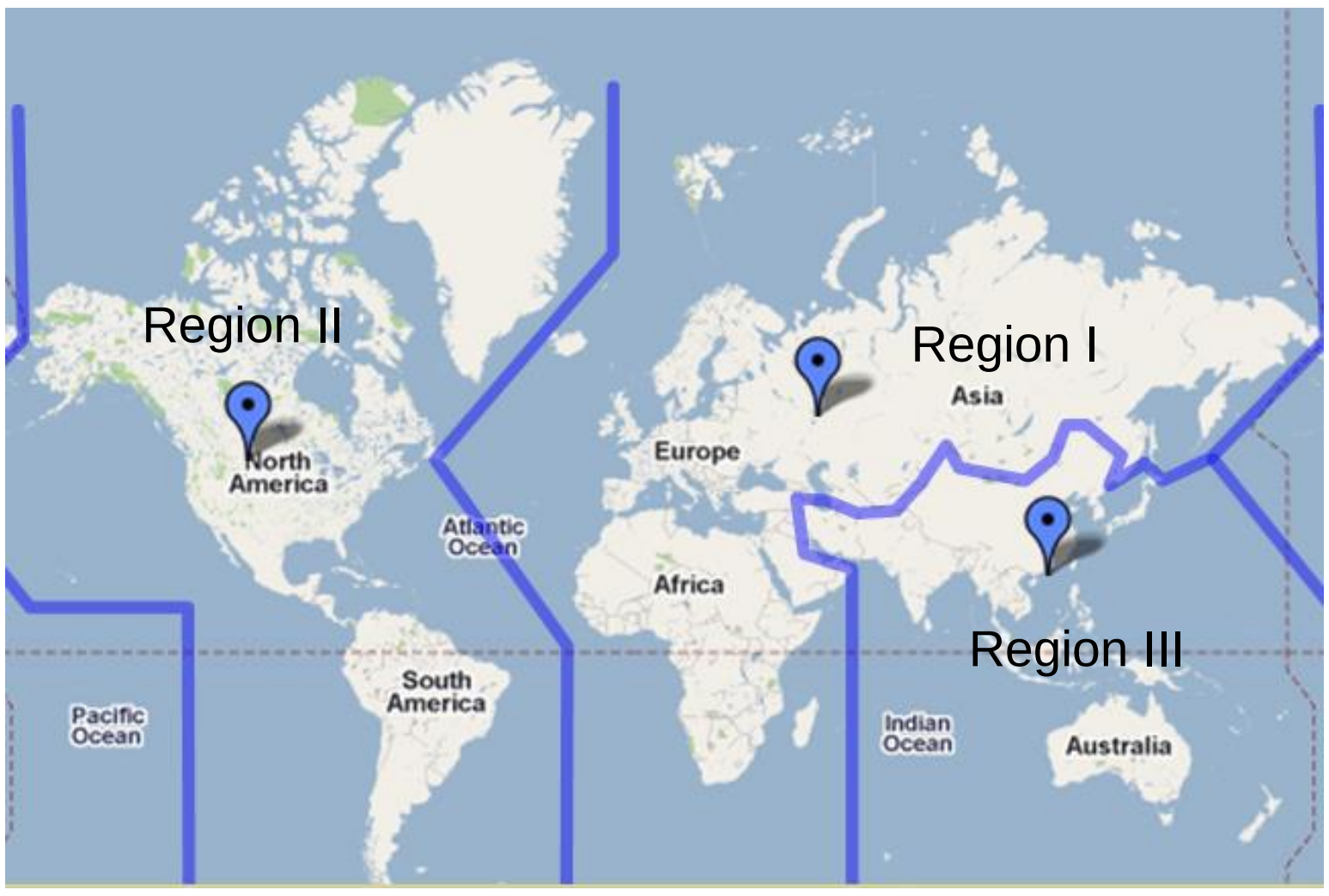

*Figure 1: Administrative Regions of the ITU.*

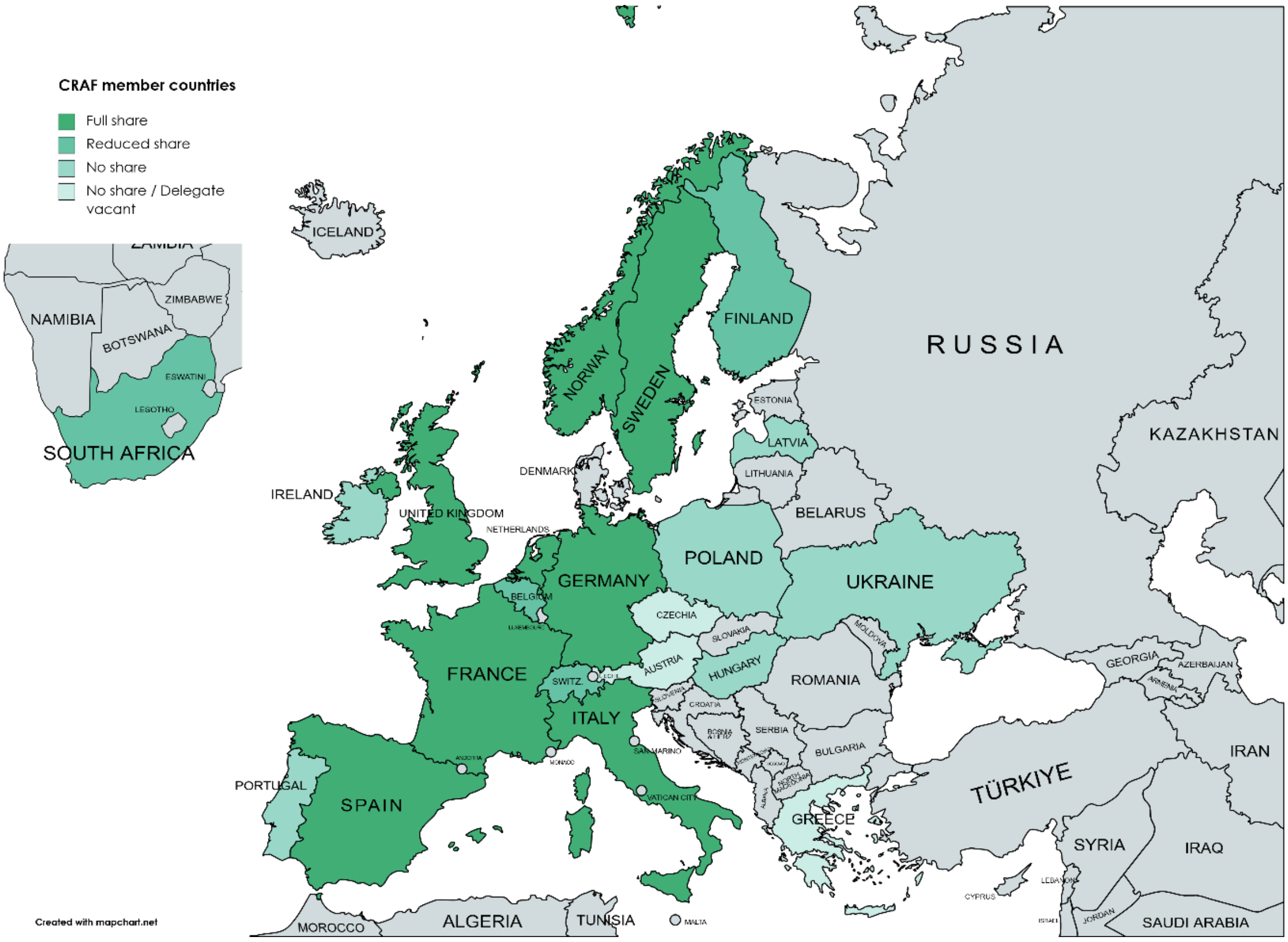

*Figure 2: Countries with at least one CRAF member organisation or institute as of September 2023.*

Before 2019, CRAF's Frequency Managers (FMs) were employed at particular CRAF member institutes, e.g., ASTRON (NL) or the MPIfR (D). Since 2019, CRAF pays an annual contribution to the Joint Institute for VLBI ERIC (JIVE) for spectrum management services.

## 2.3 CRAF mission and strategy

CRAF's mission is to

1. Keep the allocated frequency bands of radio astronomical interest free from interference,
2. Ensure access to and availability of the radio spectrum for scientific needs, and
3. Support the scientific community in their needs for passive use of interference free bands of interest.

For effective work, CRAF committee members (delegates) need to have a strong background in radio astronomy, in particular its aims, scope, and observational methods. Furthermore, knowledge of radio engineering and technology and radio-wave propagation and proficiency with frequency management and its administrative procedures and structures is required.

Delegates are not only expected to participate in CRAF meetings, but also to attend national (preparation) meetings that are organised by spectrum authorities in their home countries and international spectrum management meetings and conferences, e.g., as organised by ITU-R and the CEPT.

Complementary to the work of CRAF, the radio astronomy community invests also a lot of resources into detecting and mitigating radio frequency interference (RFI). As this is obviously closely related to CRAF's work in spectrum management, the delegates follow new developments or are even involved in the work. However, CRAF as a committee is not actively engaged in this topic. It should be mentioned that other spectrum management stakeholders, such as administrations and industry, sometimes propose to relax the RAS protection criteria and make more use of RFI mitigation techniques. The radio astronomy community should be aware of this and be careful how heavily new RFI software tools are advertised. In our experience, even advanced tools are only able to treat a subset of all RFI types and with some software one must be very careful not to mistake real astronomical signals for RFI. Furthermore, most tools focus on flagging RFI, as fully cleaning a dataset ("RFI excision") is very difficult. Finally, no post-processing tool will ever be able to function, when the receiving system was already driven into the non-linear regime or blockage.

## 2.4 CRAF Modus Operandi

The international working groups are usually *input-document driven*, which means that any activities, discussions and outcomes are solely based on text that was provided before a meeting. Thus, for effective work, input documents of other stakeholders and administrations need to be analysed in advance, proper responses have to be worked out and fed back into the groups. This often involves the preparation of spectrum compatibility calculations, to work out the technical conditions under which effective coexistence of two or more services is possible. Radio astronomy, as a passive service, is usually the victim service in such calculations. Based on such studies, contributions for use in regulatory texts (e.g., the ITU-R Radio Regulations, or CEPT/ECC Decisions) have to be prepared by CRAF delegates. Last but not least, the CRAF input documents need to be presented in the meetings and lobbying for RAS positions, in order to maximise the support from national administrations. It is important to note that while CRAF has a seat at the ITU, only the national administrations have a vote and hence the lobbying element is an important one.

The preparation, presentation and advocacy for these input documents to safeguard radio astronomy in Europe is a primary role for CRAF, and (at the ITU and CEPT level) is largely delegated to CRAF by radio observatories, through their CRAF members. The complex interplay between national, regional and international spectrum management stakeholders is sketched in Figure 3.

As the work involves a lot of travelling and is very time consuming, CRAF delegates, who usually can invest only a fraction of their work time, are supported by the CRAF frequency manager. The FM works full-time on spectrum management activities and has an appropriate travel budget. The FM, with help from the CRAF Chair and Secretary, also represents CRAF to the 'outside world' and serves as a point of contact. Furthermore, the FM leads the strategy building efforts in collaboration with the CRAF management team.

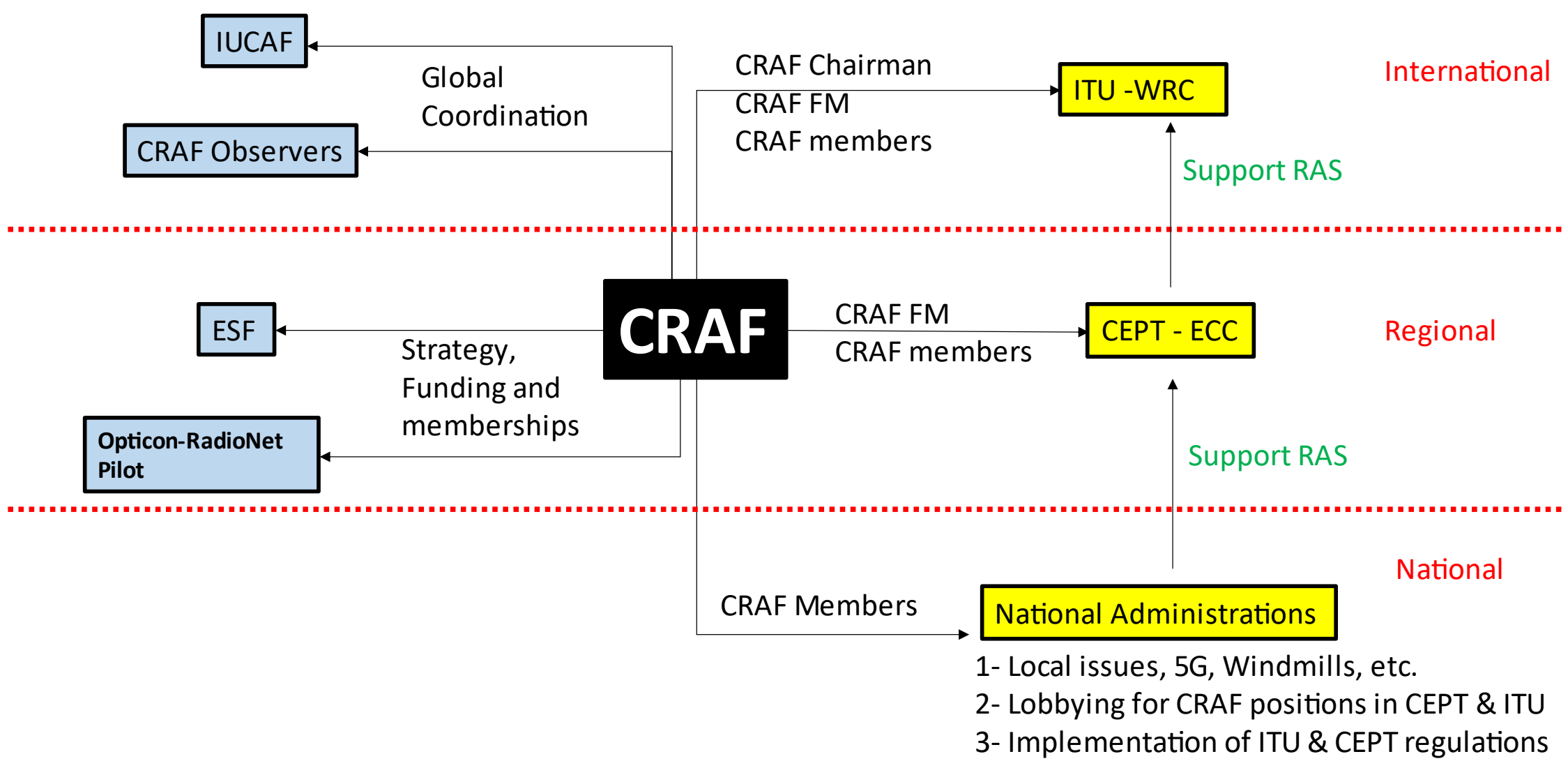

*Figure 3: Structure of radio astronomical frequency management.*

## 2.5 Information on CRAF Budget and on in-kind contributions provided

*This section is omitted in the public version.*

## 2.6 CRAF and its role

### 2.6.1 Relevance of CRAF at current European and global levels

CRAF is the only regional spectrum management organisation for radio astronomy in ITU-R Region 1 (although it should be mentioned that SKAO, being a world-wide organisation, is also actively engaged in spectrum management to protect their observatories in Australia and South Africa). As such CRAF is the link between European radio observatories and spectrum management organisations. This importance has been recognised by European administrations, so that CRAF is one of relatively few Letter-of-Understanding (LoU) partners of the ECC, which has granted it a so-called "observer" status. This means that CRAF's input documents to ECC meetings are never seen as "information-only" documents and have to be considered by the working groups. Likewise,

CRAF is a so-called Sector Member of the ITU-R. Normally, Sector Members pay several thousand Swiss Francs per annum. As a science organisation with a long track-record of engagement in spectrum management, ITU-R granted CRAF a fee-exemption status. As a Sector Member, CRAF can participate in and contribute to ITU-R working groups and the WRCs.

CRAF also represents the European Geo-VLBI community and was very active in an initiative at the ITU-R to improve the level of protection for their VLBI Global Observing System (VGOS). VGOS is the next-generation instrument for geodetic VLBI; see also Section 3.2.7.

Together with IUCAF, CRAF has co-organised several summer schools to teach our (younger) colleagues both about spectrum management issues and solutions, and also about the relevant procedures and techniques involved; see Table 3.

*Table 3: Summer schools organised by IUCAF and CRAF in the past few years.*

| Summer School | Place | Date | Web site |
|---|---|---|---|
| 4th Edition | Santiago, Chile | 7-13 April 2014 | http://www.iucaf.org/sms2014/ |
| 5th Edition | Stellenbosch, South Africa | 2-6 March 2020 | http://www.iucaf.org/sms2020/ |

### 2.6.2 Perception of CRAF activities by European and international actors and stakeholders

CRAF and its delegates are a most important point of contact for all questions of spectrum management, not only for European network agencies and administrations, but also for other stakeholders (e.g., companies and other science-services). Our delegates have excellent relationships with these entities and are known for their high level of expertise.

CRAF is also one of very few Tier-1 supporters of the new IAU Centre "for the Protection of the Dark and Quiet Sky from Satellite Constellation Interference", which is co-led by NOIRLab and SKAO; see Section 3.2.3. With our knowledge of spectrum management and software development (e.g., ability to perform simulations of satellite constellations and their potential impact on radio astronomy), CRAF will be a valuable partner for the IAU and will significantly contribute to the IAU Centre's activities.

Prior to Covid-19 CRAF produced a commercially printed newsletter twice per year, but since the advent of Covid-19, there has only been one such newsletter per year. These newsletters, which take the form of a professionally produced magazine, are distributed worldwide, but mostly within Europe, to over 300 organisations and individuals associated with governments, industry and academia. The newsletters, which are also available for download from the CRAF web site, provide information not only on CRAF's activities, but also relating to new developments at individual observatories and academic research institutes, which can be affected by or have an effect upon radio frequency transmissions, and thus on spectrum management. Examples of these developments include both computer software and hardware, and also hardware associated with radio transmission, but primarily reception.

## 2.7 Previous statutory review

The previous statutory review of CRAF was carried out in 2010/2011, as part of the review of all ESF expert committees, active at that time. The Review Panel concluded at the time, "that most of the recommendations resulting from the previous review have been implemented, although in some cases only partially".

### 2.7.1 Recommendations from the previous review

A recommendation was to improve the communication and outreach activities, in order to "disseminate [CRAF's] findings and publications, but also to accentuate the importance of its mission and activities". The panel considered the quality of CRAF's publications to be good, in particular mentioning the newsletters and the CRAF handbook. However, it noted that the circulation of the newsletter was rather poor – only 10% of senior staff at European radio observatories received the newsletter. It should be noted though, that the printed version of the newsletter is sent to all institutes and that appropriate distribution within the institutes is beyond the control of CRAF. The Review Panel also acknowledged that the "resources to tackle [the outreach] problem were also sparse".

Some other points were raised, such as that CRAF could explore whether other passive spectrum users could be future members. It was also suggested that CRAF might seek a closer relationship with the European Commission. Finally, the Panel greatly welcomed the increased involvement in the Square Kilometre Array (SKA) project, which will be a major new radio astronomy instrument and of utmost importance for European scientists.

### 2.7.2 Progress

Regarding public outreach activities, there is still room for improvement. The newsletter continues to contain high-quality articles, reporting on the work of CRAF on particular spectrum management topics, software and hardware developments at our member institutes, which are related to CRAF's work, or on scientific meetings that CRAF members attended. The electronic version of the newsletter is now also published on our new website. The website itself, which is frequently updated, contains a newsfeed and other useful information for astronomers and other spectrum users. CRAF also has a Twitter account to engage with the community on social media. However, it still seems to be the case, that we reach only a rather small fraction of the radio astronomical community.

CRAF's collaboration with the SKA Observatory (SKAO) has further improved in recent years. SKAO became an Observer member in CRAF and the SKAO representative now works closely with CRAF on a large variety of topics. For example, satellite-related activities at ITU-R and CEPT are very relevant to SKAO and new software tools to perform compatibility calculations have been developed by SKAO/CRAF resulting in the joint submission of numerous documents. Currently, the SKAO representative is leading the CRAF work item team for satellite systems. Furthermore, former CRAF Chairmen were heavily involved in sophisticated studies of electromagnetic interference and compatibility for the SKA site and its construction (e.g., A. Jessner, R.T. Lord, H. van der Marel, R.P. Millenaar, H.C. Reader, F. Schlagenhaufer; "EMI Protection and threshold levels for the SKA", C. Wilson, 2014.)

The final proposal from the previous review was that CRAF shall engage more with the European Commission. This has proved challenging. CRAF members have experienced that high-ranking representatives sometimes tend to speak only to counterparts at similar hierarchical level, which would be institute directors, or even national science organisation managers. Therefore, CRAF invites the radio astronomical community as a whole to get more active in the political activities related to spectrum management issues. It should be noted though that CRAF sometimes submits comments to public consultations of the Radio Spectrum Policy Group (RSPG) of the European Commission, but this group is somewhat less open to active participation than CEPT.

# 3 Review of the 2011-2021 period

## 3.1 Spectrum management activities

CRAF's activities in the past decade have steadily increased, as the number of meetings in which our delegates participated, and the number of input documents to these meetings reveal; see Figure 4 to Figure 6.

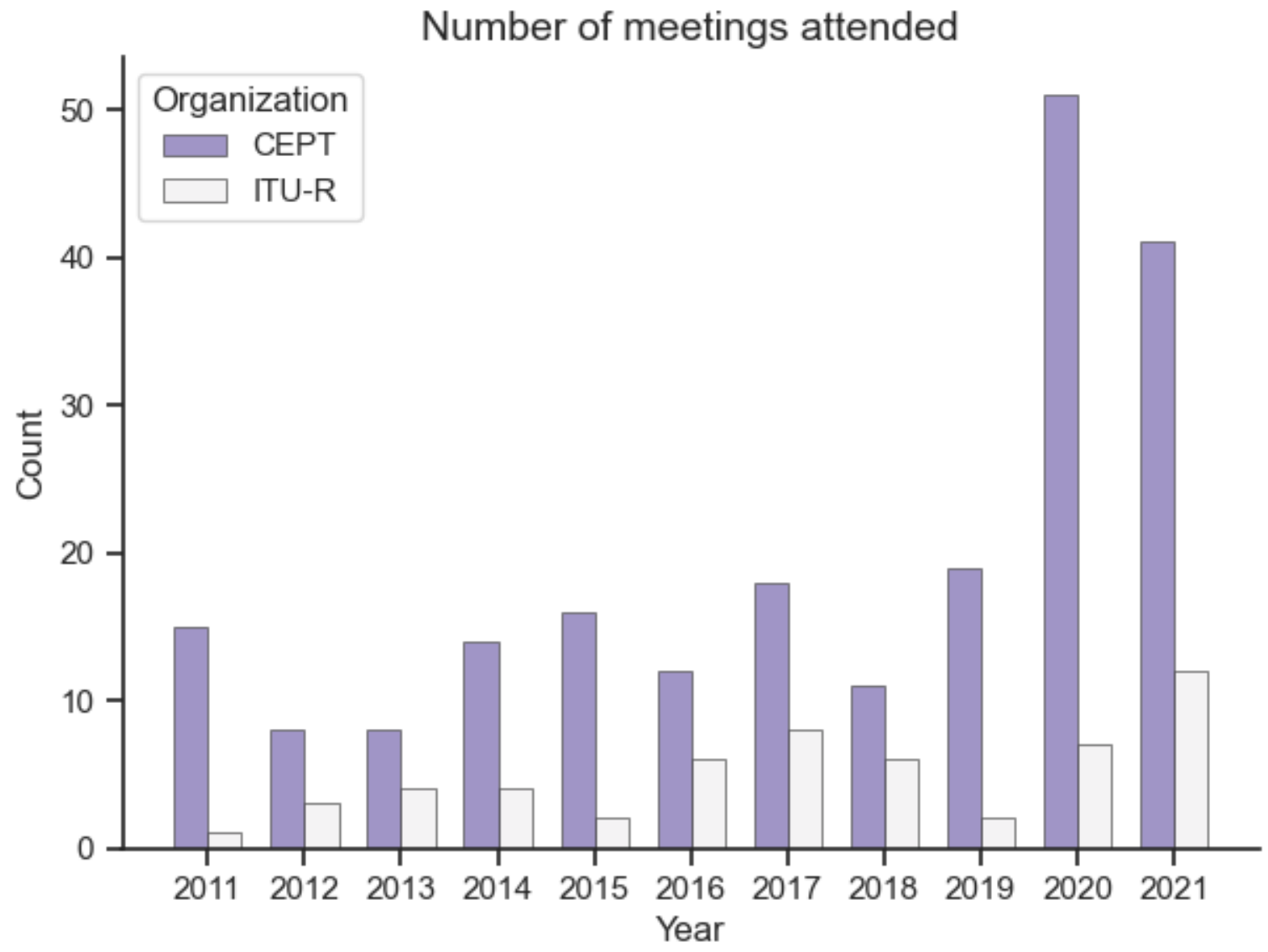

*Figure 4: Number of attended meetings attended per year.*

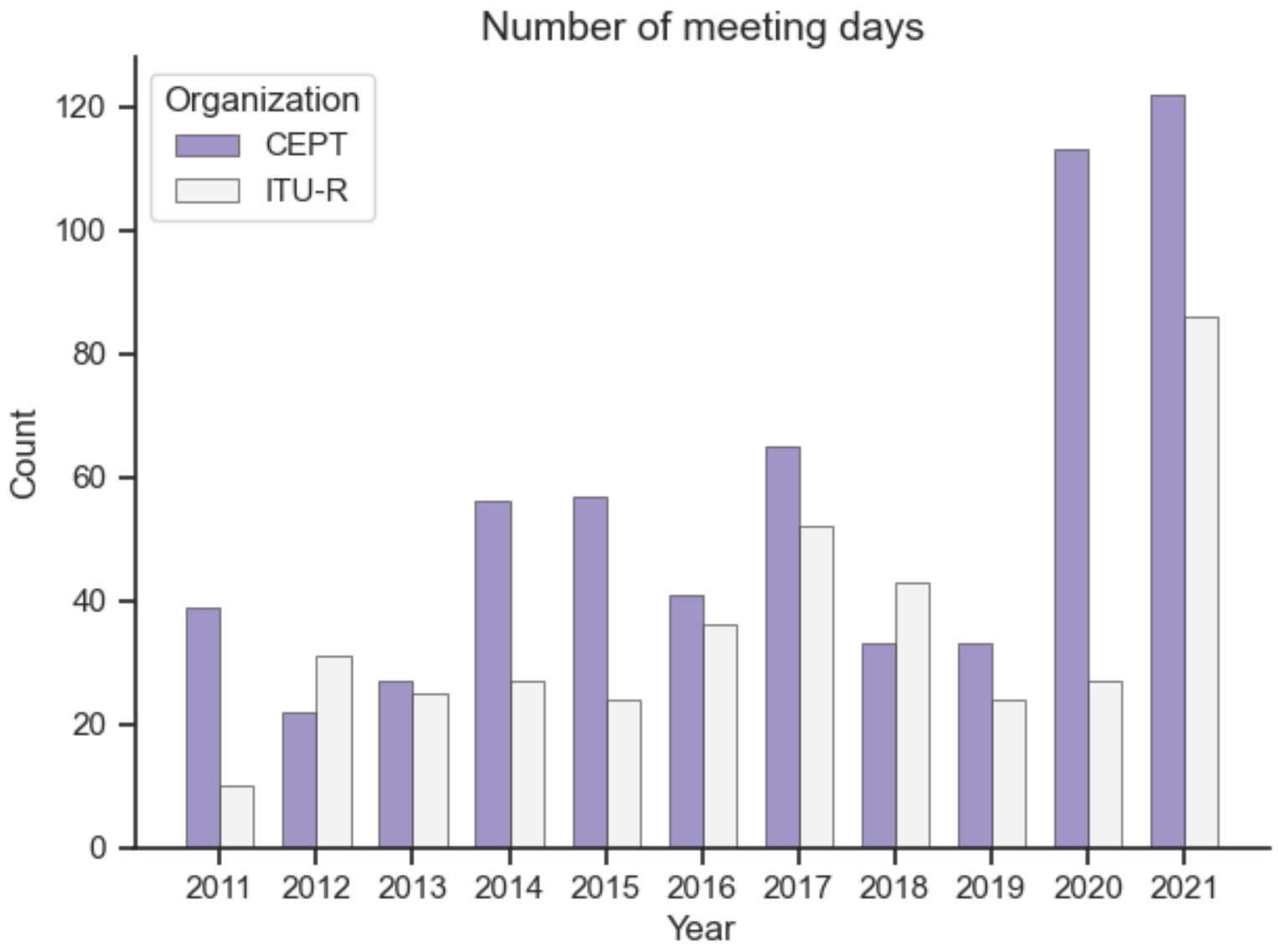

*Figure 5: Number of total meeting days per year.*

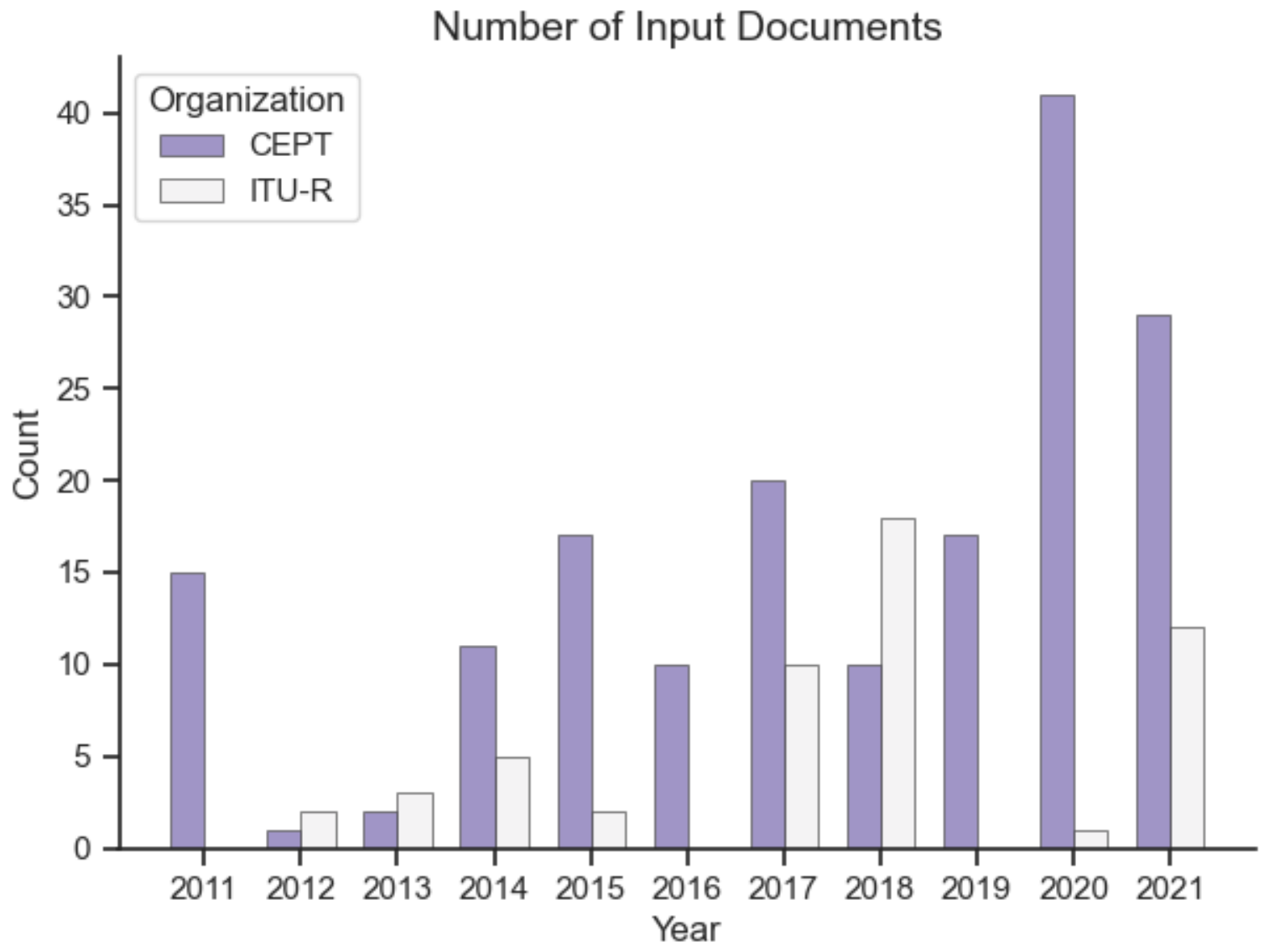

*Figure 6: Number of input documents per year.*

The nearly two-fold increase of input documents in 2020 and 2021 compared with earlier years can be attributed to the introduction of the work-item (WI) teams in 2018/19, which allowed us to activate more of our members and thus more human resources were available to work on the various spectrum management issues. The considerable increase in meetings attended in 2020/21 is also in part owing to the WI teams. However, even more relevant was the Covid-19 situation, as almost all meetings were now virtual and it was therefore possible to easily attend almost all meetings, which were of interest. Previously this had not been possible because of CRAF's limited travel budget and limited time availability. It should be mentioned, that CRAF always tried to send delegates to physical meetings who live nearby – delegates are based all over Europe, which makes this viable – but the effort to brief people sufficiently well was very large. With online meetings, a particular topic can be assigned to one or two delegates, who would then follow all relevant meetings.

From CRAF's perspective the virtual meetings had a very positive impact on our possibilities to take part in spectrum management. However, it is already clear that most administrations are very keen to return to mostly physical meetings as soon as the Covid-19 situation allows. Many stakeholders and representatives of administrations complain about the lack of opportunity for "coffee-break discussions", indicating that compromises on difficult topics are facilitated by face-to-face informal discussions outside the main meeting.

Some interesting conclusions can be drawn from a look at the topics, types of meetings and associated input documents. In Figure 7, the number of meetings, in which CRAF delegates participated, is displayed divided into the various spectrum management groups. These European (CEPT) and international (ITU-R) groups can be technical and/or regulatory orientated. While in earlier times (2011 – 2013) a rather small number of group meetings was attended, this has changed, recently. Even in 2017 – 2019 (before the Covid-19 crisis), more and more ITU-R working party (WP) meetings appear in the graph. The same trend is visible in the topics of submitted input documents; see Figure 8. Furthermore, in every time period a large fraction of our activities was concerned with a rather small number of topics. For example, during all times, in particular in 2014 – 2017 (with almost 50%!), the compatibility issues with the Iridium satellite constellation (under MSS, mobile satellite service) took a lot of CRAF's time and effort (CEPT/ECC FM44 & SE40). The same can be said about IMT (international mobile telecommunication), which deals with cell phone networks, in CEPT/ECC PT1 and ITU-R TG 5/1 & WP 5D. While in the beginning of the review period there were relatively few mobile phone bands, which were also usually well-separated from RAS bands, IMT was allocated many GHz of additional spectrum in the past five years, partly even overlapping RAS bands. This is a huge concern for radio astronomers. Also, the advent of the SpaceX/Starlink and OneWeb satellite mega-constellations, which are subsumed under the FSS (fixed-satellite service) can be seen, starting in 2016, when CEPT/ECC SE40 (which is the technical satellite-study group of ECC) began the work on ECC Report 271. Finally, the work on additional spectrum assignments for short-range radars above 116 GHz (under RDS, radio-determination service) has had a lot of our attention in CEPT/ECC SE24. These topics clearly were of high interest to CRAF and will be described in more detail in the next section.

## 3.2 Most important spectrum management topics

### 3.2.1 Iridium

A long-term issue for CRAF is the well-known case of the Iridium satellite constellation, which leaks power into the RAS band at 1610.6–1613.8 MHz, featuring one of the very important spectral lines of the hydroxyl radical. The Iridium satellites operate within the band 1618–1626 MHz. The leakage power is caused by intermodulation products and results in interference, which significantly exceeds the permitted power flux density thresholds, as defined in Recommendation ITU-R RA.769-2. This incompatibility has persisted for almost three decades and has been discussed on countless occasions at all levels in CEPT and ITU-R.

Originally, Iridium and CRAF signed an agreement in 1998 in which Iridium committed themselves to fully protect the RAS from Jan 1st, 2006. Unfortunately, interference in excess of RA.769 levels could still be detected after that date at a number of European observatories and also at the German satellite monitoring station in Leeheim. CEPT/ECC working groups, in particular SE40, have been involved in the matter and several ECC Reports on this topic have been produced (ECC Reports 171, 226, and 247) since then. CRAF was heavily involved in all of these activities, from contributing to the ECC Reports, participation in software development for data analysis and satellite-network simulations (so-called EPFD simulations) and advising Leeheim staff members on the measurement campaigns.

One key point was the publication of ECC Decision (09)02, which allows CEPT administrations to authorise MSS terminals only if both the uplink (Earth-to-space) and downlink (space-to-Earth) transmissions are compatible with the RAS. For the terminals, this meant that protection zones around CEPT RAS stations were established. Most CEPT countries reported to the CEPT office (ECO) that the Decision was implemented nationally. However, according to recent claims from Iridium, only in Germany was a protection zone actually established (around the Effelsberg 100-m telescope). Furthermore, Iridium still did not meet the downlink requirements, as new measurements demonstrated. As a consequence, the German network agency Bundesnetzagentur threatened to revoke Iridium's licence in Germany.

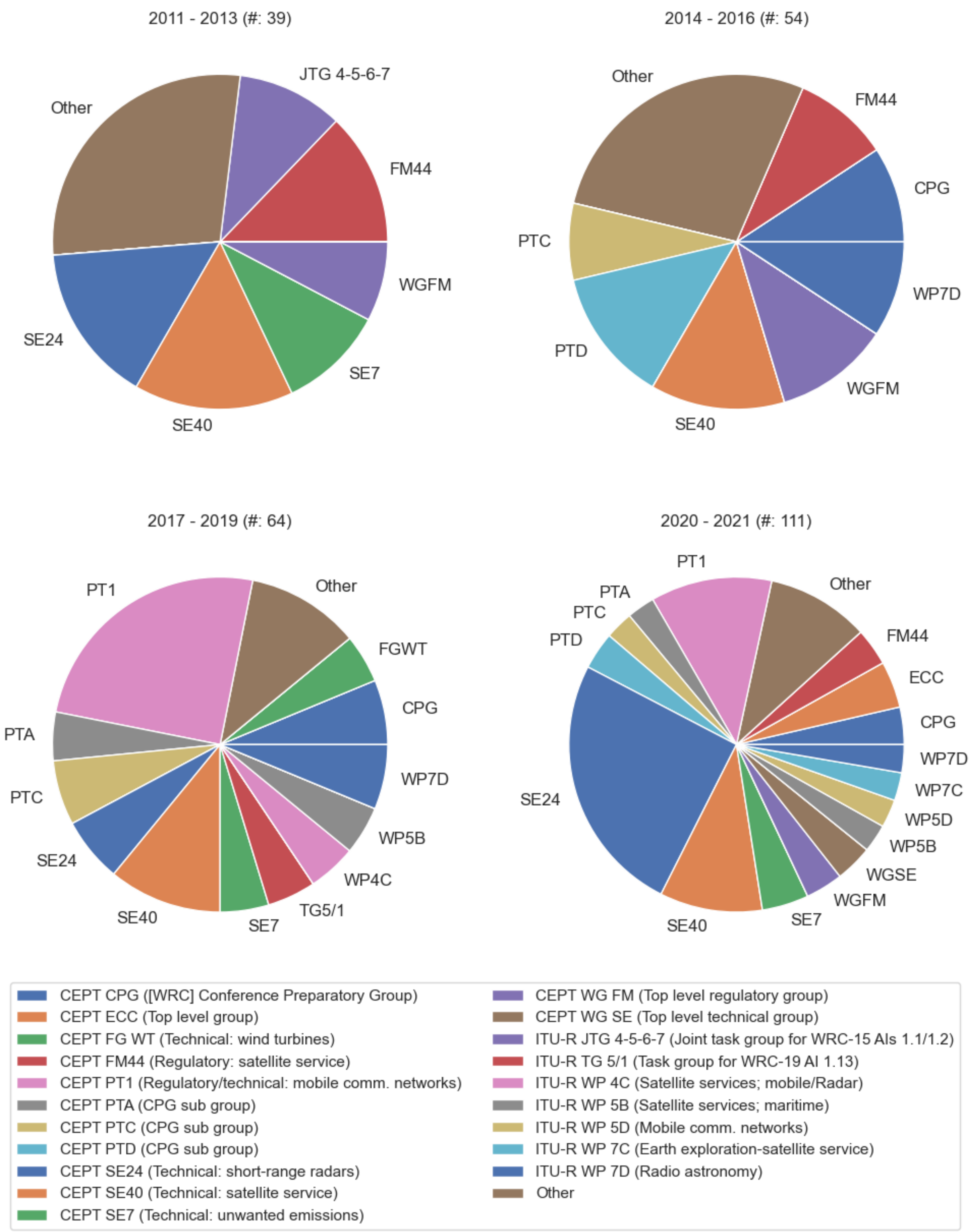

*Figure 7: Spectrum management group meetings attended in various time periods.*

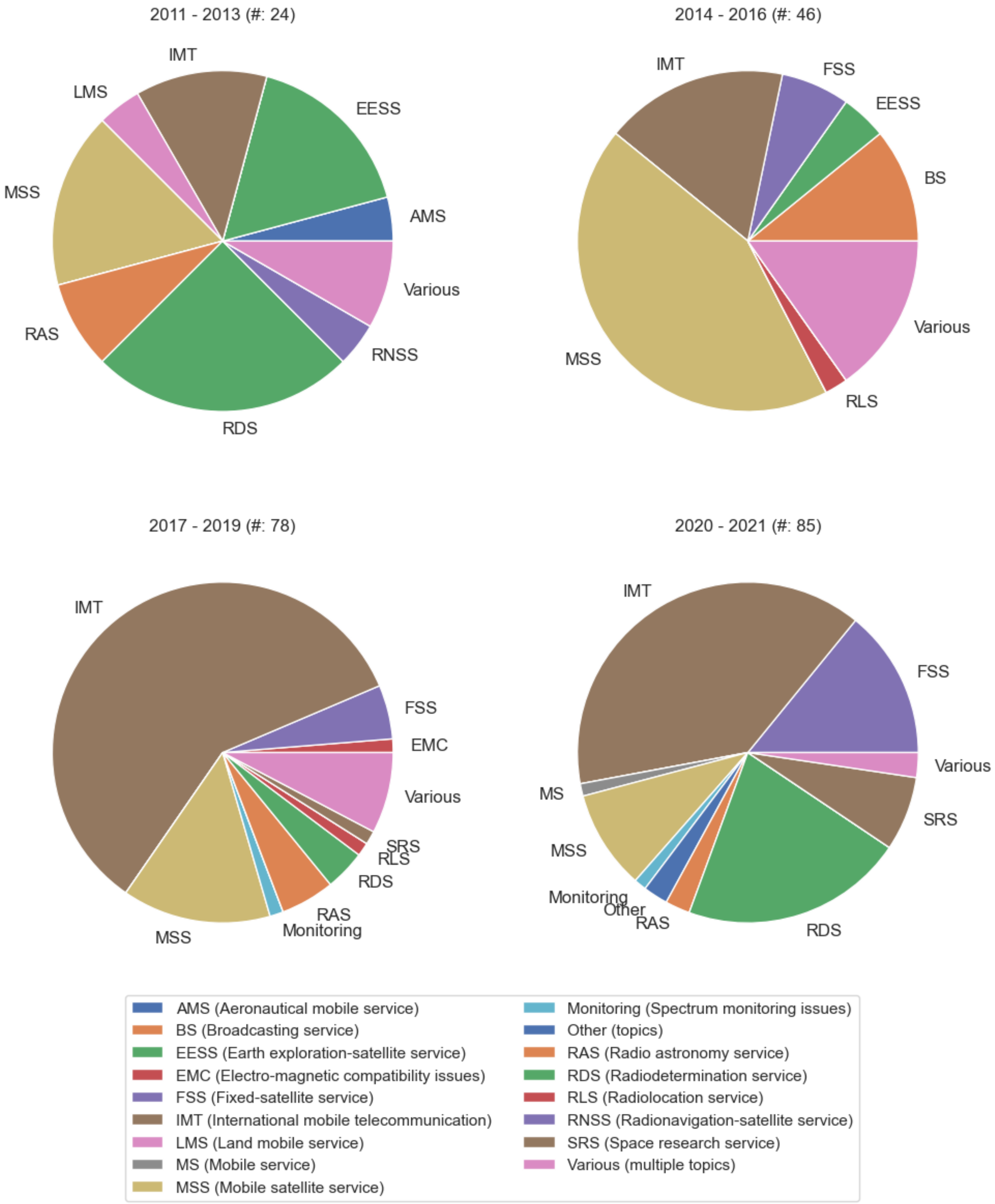

*Figure 8: Topics of input documents submitted to spectrum management working groups in various time periods.*

In the meantime, Iridium has worked on a second generation of satellites, Iridium NEXT. The company promised that the new hardware would be compatible with RAS from the beginning by the introduction of a "RAS protection" (RASP) mode. Simulations provided by Iridium demonstrated, that the resulting data loss would finally be below 2% (the acceptable level of data loss, according to Rec. ITU-R RA.1531-2). CEPT administrations prepared themselves for action, should Iridium NEXT be still incompatible with the RAS.

First Iridium NEXT satellites were launched in 2017. CEPT/ECC group SE40 tasked Leeheim to carry out monitoring observations. It turned out that Iridium once more broke their promises, although one has to admit that the level of interference was indeed reduced. Since then, CRAF has made additional measurements, filed radio frequency interference (RFI) reports with network agencies and contributed to another Draft ECC Report, while Iridium NEXT still continues to operate. Iridium claims that they will further improve the firmware and operations software and that they are confident to finally meet the required interference levels soon.

During all these years of significant efforts, Iridium never stopped operating their satellites, which in CRAF's view is a clear violation of the Radio Regulations.[3]

### 3.2.2 International Mobile Telecommunication, IMT (cell phone networks)

In the following the most important CRAF activities involving IMT, will be summarised, focusing on the past five years (beginning with the WRC-19 study cycle), which is dominated by the introduction of the 5th generation of cell phone networks (5G).

#### *3.2.2.1 WRC-19 Agenda Item 1.13: Introduction of mm-wave 5G bands (IMT2020)*

In the study cycle between 2015 and 2019, to prepare WRC-19 Agenda Item 1.13, ITU-R Study Group 3 assigned the work to a dedicated task group TG 5/1. This is because AI 1.13 was concerned with a huge portion of the spectrum, within which potential new IMT bands in the range between 24.25 and 86 GHz were to be identified. CRAF provided compatibility and sharing studies for all bands under study adjacent to or overlapping with RAS bands, i.e., for 23.6–24, 31.3–31.8, 42.5–43.5, and 81–86 GHz, which all became part of the Chairman's Report of TG 5/1. As IMT/5G technology features new active antenna systems, which allow for beam-forming (i.e., antenna pattern beams, which can be electronically steered), the required compatibility calculations were the most complex and sophisticated calculations that CRAF had done to date. The software developed for this purpose is published as open source software (see Section 3.3.2.3) and the methods have also been described in a scientific paper (Winkel, B. & Jessner, A., Advances in Radio Science Vol 16, 177, 2018).

The Conference Preparatory Meeting, CPM-19, did acknowledge the RAS concerns but did not want to establish restrictions on IMT, as the studies showed that in almost all cases, RAS compatibility could be achieved by national-only regulation. At least, CRAF convinced administrations to task ITU-R to work on new deliverable (Report or Recommendation), which would help interested administrations to coordinate the 5G deployment in the vicinity of RAS stations. Currently this is drafted by WP 7D to which CRAF has contributed and continues to contribute.

[3] In 2023, a new report was published by the European Communications Committee, ECC Report 349, which contains the latest measurements and simulations. The RAS thresholds are still exceeded, though Iridium could steadily reduce the level of excess over the last years. Whether the remaining excess can be further reduced, is currently under investigation.

At the WRC-19 the bands 24.25–27.5, 37–43.5, 45.5–48.2[4] and 66–71 GHz were finally identified for IMT/5G. Both, the 24 and 43 GHz bands are very unfortunate choices from a radio astronomer's perspective, as they are heavily used at CEPT RAS observatories.

As the European Union wants to harmonise the use of these new IMT bands, CEPT/ECC PT1 was tasked to work on two new ECC Decisions for the 24 and 43 GHz bands. The former has been published already (ECC Decision (18)06, while the latter is still in drafting phase[5]). CRAF participated in many correspondence-group and PT1 meetings to make sure that proper RAS protection mechanisms are established in both documents. One of the challenges here is that the technical parameters and deployment scenarios, which were assumed for the TG 5/1 compatibility studies, are not part of the ITU-R regulation. However, some of these, such as the prediction that almost all base stations would be installed below the roof-tops (which enormously increases clutter losses) or that deployment would mostly be done in urban and sub-urban regions (which are usually well separated from RAS stations) are now disputed by pro-IMT stakeholders.

#### *3.2.2.2 Pioneer 5G band in Europe (3400–3800 MHz)*

Independent of the efforts for new allocations to the IMT above 24 GHz, the first band to feature 5G technology in CEPT was at 3.6 GHz. The European Commission wanted to introduce a harmonised band to ensure a streamlined deployment over the full continent and assigned the frequencies between 3400–3800 MHz even before ITU-R allocated the band to IMT in the Radio Regulations. The ECC then initiated the work as part of its "CEPT roadmap for 5G". ECC/PT1 performed compatibility studies, which led to ECC Report 281. CRAF contributed single-entry and aggregation studies for the IMT/5G vs. RAS scenario. At first, there was some reservation to include these in the Report, as the RAS bands at 3.3 GHz are only protected by means of RR Footnote 5.149, which urges administrations to protect RAS operations, and not by a proper allocation in this frequency range. Fortunately, after successful lobbying, the top-level group ECC decided to accept the studies. The required coordination distances are rather small, compared with other IMT vs. RAS cases, as the spurious emission levels below 3.4 GHz, with which IMT must comply, are much stricter than usual. This is to protect military radars, which also operate in the band below 3.4 GHz.

#### *3.2.2.3 Use of mobile-fixed communication network user equipment on-board aircraft*

This topic was initiated in 2018 in CEPT/ECC PT1 (as work item PT1_18). It concerns the possible use of existing mobile-fixed communication network (MFCN) bands (mainly) below 5 GHz by aerial vehicles (AV). The topic was originally asking for the use of "command & control" (C&C) and wireless payload communication of unmanned aerial vehicles (UAV), aka drones, in existing IMT networks. Later, it was extended to cover all kinds of aircraft. PT1 first analysed which bands are most favourable for aerial MFCN use and restricted the further studies to the bands 700, 800, 900, 1800, 2100, 2600, and 3600 MHz.

CRAF contributed compatibility studies in which the potential impact of MFCN UE on-board AVs is analysed for both the single-interferer case and for aggregation scenarios. In particular the 2$^{nd}$ harmonics of LTE700 and LTE800, which fall into the 1400–1427 MHz and 1660–1666 MHz RAS bands are a serious threat to radio observatories operating at L-band, as the potential harmonic emission can have up to –30 dBm/MHz spectral power output. Thus, large separation distances would be required. This topic was debated at length at the ECC/PT1 meetings in Billund (PT1 #62) and Riga (PT1 #63). Proponents of the new applications fought hard to avoid the inclusion of the CRAF studies in the ECC Draft Report. However, at PT1 #63 (Manchester) a compromise was achieved. In the published ECC Report 309 it is now recommended that administrations who want to protect their RAS stations could do this via some kind of "no-transmit" zones around the radio telescopes.

[4] This band (or parts of this band) is only identified for IMT in Region 2 and several other countries, including CRAF countries Hungary, Latvia, South Africa, and Sweden.
[5] In late 2022, ECC Decision (22)06 was published, which contains several clauses for the protection of RAS in the 43-GHz band.

Currently, ECC/PT1 is working on a Draft new ECC Decision (ECC/PT1_30)[6], which is based on ECC Report 309[7]. The introduction of "no-transmit" zones is a complex issue, as it overlaps with other aeronautical regulations. Some administrations proposed "no-transmit" zones, but so far no conclusion has been reached. One very important aspect, to which CRAF needs to pay close attention, is that aerial equipment must use a special SIM card. Otherwise, cell phone base stations could not make a distinction between terrestrial and aerial devices and "no-transmit" zones could not be policed by the network operators.

#### *3.2.2.4 WRC-23 Agenda Item 1.2: Potential new allocation of 5G in mid-band (6650–6675 MHz)*

WRC-23 AI 1.2 is concerned with a possible new allocation of spectrum to IMT in the frequency bands 3300–3400 MHz, 3600–3800 MHz, 6425–7025 MHz, 7025–7125 MHz and 10.0–10.5 GHz. The 6 GHz band is being considered for ITU-R Region 1 (Africa, Europe) only, and the 3.3 and 10 GHz bands are only foreseen for Region 2 (Americas). The 6 GHz band is extremely important for radio astronomy for the study of methanol in the interstellar medium, but unfortunately the RAS has no specific allocation in this frequency range. It is only "protected" via a footnote in the Radio Regulations (No. 5.149), which urges administrations to protect the RAS in the specified bands. As the ITU-R resolution that defines AI 1.2 emphasises that only primary services should be studied versus the candidate IMT applications, administrations are of the view that the 6 GHz band cannot be studied with respect to RAS protection. In the beginning, only Germany has supported CRAF at the relevant ITU-R WP 5D meetings, later joined by several other European administrations.

CRAF studies show that separation distances of up to 500 km are required, which would have a huge impact on the deployment of 5G in this band, especially in Europe (if RAS is to be protected).[8]

#### *3.2.2.5 WRC-23 Agenda Item 1.4: IMT base stations on aircraft and stratospheric platforms*

Another IMT-related topic is to study the potential use of high-altitude platforms (between 20 and 50 km altitude) for use as mobile communication base stations (AI 1.4). As these installations would be in direct line-of-sight out to hundreds of kilometres, the potential threat to the RAS is very high. Even if not directly in the main beam of a telescope, the number of visible platforms could be high enough that contributions through the side lobes of RAS antennas would exceed the permitted threshold levels significantly. However, only existing IMT bands are under investigation, which are mostly well separated in frequency from RAS bands, with the exception of the 2.7 GHz passive band, which is adjacent to an IMT band. That said, second-harmonic emission, if not sufficiently suppressed by filters, could appear in some of the L-band allocations.

---

[6] This was published in November 2022 as ECC Decision (22)07.

[7] In late 2022, another document, ECC Report 348, was published, which addresses the use of beamforming arrays (active antenna systems) by the IMT base stations at 1.8, 2 and 2.6 GHz to serve aerial terminals.

[8] As of September 2023, it is still not finally decided whether CEPT will fully support an IMT identification in the upper 6 GHz band. Some are of the view, that it should be reserved for WiFi or for a mixed scenario. A first analysis by CRAF indicated that WiFi likely has significantly lower impact on European RAS stations, though they may be harder to coordinate with (as WiFi equipment is usually not individual licensed, while IMT base stations are in many countries).

In collaboration with CRAF and SKAO, IUCAF has submitted an input document to WP 5D with a compatibility calculation for adjacent band operation of HIBS (aerial IMT base station) in the 2.6-GHz band and the passive RAS band at 2.69–2.7 GHz. Owing to relatively strong out-of-band emission of the former and the potential main-beam coupling, the resulting separation distances are very large. In order to ensure compatibility, an additional ~50 dB attenuation would be necessary. This could perhaps be achieved with better spectral filtering, but technical solutions have not yet been discussed. Another study, which was submitted by CRAF and SKAO, involved potential second harmonic emission of HIBS into the 1610–1613 MHz RAS band. Several administrations were previously of the opinion that such a study would be out of the scope of the agenda item. The physical reality, however, predicts a violation of the RAS thresholds by up to 30 dB, which is why the study was nevertheless submitted to WP 5D. After long discussions, a few administrations continued to heavily oppose. Germany and South Africa spoke for inclusion of the study in the AI-1.4 report.[9]

### 3.2.3 Satellite mega-constellations

In 2016, CEPT/WG FM approved a new work item for the study and harmonisation of the earth stations of NGSO satellite systems operating in the 10.7–12.75 GHz (space-to-Earth) and 14–14.5 GHz (Earth-to-space) bands in FSS allocations (for fixed and moving platforms). One such NGSO satellite system is OneWeb, which originally planned to deploy a constellation of 720 Low Earth Orbit ("LEO") satellites in 18 orbital planes in near-polar circular orbits at an altitude of 1,200 km. The other system was SpaceX/Starlink, featuring thousands of planned satellites. Starlink is already widely known, even outside of the radio astronomical community, as the first satellites, launched in Spring 2019, are already visible as bright "trains of light" in photographs of the night sky. This event triggered a huge movement in the professional and amateur astronomical communities and beyond, which led to a new IAU Centre "for the Protection of the Dark and Quiet Sky from Satellite Constellation Interference".

Independent of these activities, and well before the optical astronomical community became involved, CRAF was engaged in the work of CEPT/ECC SE40 to produce ECC Report 271, which has become a prime resource of information for spectrum managers all around the world, who are interested in mega-constellations. Based on the Report, CEPT/ECC FM44 drafted ECC Decisions (17)04 and (18)05 on the harmonised use, free circulation and exemption from individual licensing of fixed earth stations as well as those in motion (ESIM) operating with NGSO FSS satellite systems in the frequency band 14.0–14.5 GHz, providing that these earth stations comply with the protection criteria of the incumbent service. Although RAS has a secondary allocation in the band 14.47 – 14.5 GHz, it was agreed to urge the providers to protect the RAS in this band and include the protection criteria in the ECC Decision.

Later, OneWeb increased the desired number of its satellites to more than 1200. SE40 should consider whether the earlier studies require revision. Their competitor, Starlink, has already put more than one thousand satellites into orbit and currently has plans to increase the constellation to several tens of thousands of satellites. SpaceX/Starlink first started bilateral discussions with CRAF, and then agreed to participate in the SE40 studies.

[9] In the final Report of the Conference Preparatory Meeting, which took place in Spring 2023, one (of several) methods was included as a possible solution to the Agenda Item, which would address the 2$^{nd}$ harmonics issue. It is up to the WRC-23 to decide, whether it will be implemented in the final outcome of the Agenda Item. 1.4.

While both systems comply with the RAS protection criteria (on paper), there was still the threat that the aggregated power of the carrier signals may create blocking or non-linear behaviour in the RAS receivers. This was studied in detail by CRAF and led to additional protection requirements in some CEPT countries (to date: Germany, Spain, and Italy), where Starlink was forced to avoid direct illumination of the areas around the Effelsberg, Yebes, Sardinia, Medicina and Noto radio telescopes.

In collaboration with the SKAO, CRAF has also developed an advanced suite of software for satellite simulations and compatibility calculations, so-called EPFD (equivalent power flux density) studies, as described in Rec. ITU-R M.1583-1, which is available under open-source licence and was contributed to SE40.

Despite all these successes, one has to accept that a huge portion of the spectrum, the downlink bands used by the mega-constellations, will be completely lost – at least for single-dish telescopes. For interferometers, observations may still be possible but will take a significant hit in effective sensitivity. It should also be noted that OneWeb and Starlink are just the first two of dozens of planned projects, such as Amazon/Kuiper or Kepler, which will use different frequencies. The mega-constellations pose a huge threat even to the SKA and its pre-cursors, which, despite being located in radio quiet zones, cannot avoid satellite interference.

Finally, it should be mentioned that CRAF is a contributor (highest tier level) to the new IAU Centre, which is a joint organisation of NSF's NOIRLab and the SKAO. Their proposal received 67 support letters from astronomical observatories, research institutes and individual astronomers world-wide. CRAF expressed its intention to contribute to the centre with its expertise on compatibility calculations for radio astronomy and the possibility of coordinating observing campaigns with European observatories. The idea of the Centre arose from several "Dark and Quiet Skies" conferences, which took place in 2020 and 2021 and in which CRAF members participated.

### 3.2.4 Short-range radar and violation of passive bands

This topic is one of many, with which our work-item team SEnn has been involved. WI-SEnn is responsible for monitoring and contributing to the discussion in several technical spectrum engineering and associated frequency management groups in CEPT/ECC. In particular SE24 and SRD/MG (Short Range Devices) and SE7 (Compatibility and sharing issues of mobile systems) kept us very busy during the past few years. Most importantly and relating to work item SE24_71 (UWB radio determination applications in the frequency range 116–260 GHz), the request from ECC to include the passive RR 5.340 bands under certain conditions in these studies has absorbed a lot of the team's resources. RR footnote No. 5.340 forbids any emission in a number of RAS bands and is the strongest level of protection that RAS can have. Needless to say, any initiative to weaken the status of these bands is considered as extremely dangerous by the radio astronomy community. Even if administrations and stakeholders claim that this shall be only for very exceptional cases, future applications will be more inclined to ask for similar exceptions. In fact, this has already been happening during the work of SE24 (UWB security radars). Consequently, CRAF has invested a lot of effort into fighting this. However, owing to strong support from some CEPT countries, the Draft ECC Report 334 includes compatibility studies of radio-determination systems for industry automation in shielded environments (RDI-S) overlapping with some passive RAS bands. Work has now started on a Draft new ECC Decision[10], which shall regulate the use of these systems. Administrations have claimed previously, that within the territory of each ITU-R member state, countries have the sovereignty to use the spectrum as they wish (providing that neighbour countries are not affected). However, the Rules of Procedures of the ITU-R Radio Regulations on RR N 5.340 state in Article 2.2 that "The [ITU-R] Board considers that, in view of this prohibition [(No. 5.340)], a notification concerning any other use than those authorised in the band or on the frequencies concerned cannot be accepted even with a reference to No. 4.4; furthermore the administration submitting such a notice is urged to abstain from such usage." This is only logical, as the footnote ought to also protect the Earth-sensing science satellites. If countries would violate 5.340 in bulk, this would have a disastrous effect on this extremely important application.

For completeness, it must be mentioned that it is not the first time that in CEPT a violation of No. 5.340 has been put into place. In 2007, the European Commission published EC Decision 2007/131 (later revised in EC Decision 2019/785), which permit ultra-wide-band (UWB) devices at any frequency. Of course, CRAF opposed this heavily at the time, but also without success.

There are also activities at the ITU-R, which seek to study possible sharing scenarios between passive bands and active services above 71 GHz. The ITU-R Resolution 731 was established after WRC-2000 after a portion of the spectrum above 100 GHz was identified for RAS usage to allow future applications to have access to this part of the spectrum. (In the year 2000, not many potential applications were known, which could operate at such high frequencies.)

### 3.2.5 Space weather

Space weather refers to physical processes in the space environment, mainly resulting from solar activity, which ultimately impact Earth and human activities. Examples of these are satellites, space missions, terrestrial electrical networks and systems, telecommunications and navigation systems. Interest in this domain, which is growing, is related to the development of human activities. However, space weather is currently not recognised as a radiocommunication service at the ITU-R and thus radio observations with space weather sensors are not protected by the international Radio Regulations.

Recognition of space weather is being considered under WRC23 AI 9.1.A. The characteristics of active (emitting and receiving) and passive (only receiving) space weather sensors (SWS) have been summarised during the previous ITU-R WRC cycle (2015-2019) in Report ITU-R RS.2456-0. In the current cycle (2019-2023) and possibly the next one as well (2023-2027), the recognition of SWS at ITU-R is under consideration. This should include:

- Identification of SWS categories that need to be protected by appropriate regulation,
- Determination of the appropriate radio-communication service under which SWS will be operating,
- Determination of highest priority frequency bands used by SWS, and
- Define the appropriate protection criteria for SWS.

Among the active or passive, terrestrial or space-based SWS, there are several radio-telescopes. These RAS sites are common for both RAS and SWS observations and use several overlapping frequency bands. A new CRAF work item team for covering the space weather topic was initiated in April 2021 and several input documents for the WP 7C meeting in September 2021, were prepared, in which this AI has been intensively discussed. CRAF's position evolved throughout the progress of discussions at ITU-R and CEPT. The current view of CRAF is that the RAS sites simultaneously operating radio astronomical and space weather observations could gain extra protection in the extended frequency bands used by SWS. Given that the stricter RAS protection criteria would not be required for SWS applications, CRAF's preferred position is to consider SWS under the MetAids service to avoid confusion with RAS protection thresholds for the overlapping frequencies. Several considerations are being addressed under this WRC agenda item, which CRAF is closely following through its newly created team.

[10] In November 2022, the Decision (22)03 was published, deciding that "usage within the bands covered by RR No. 5.340 shall be on an exceptional basis". Furthermore, exclusion zones around some RAS stations were established.

### 3.2.6 Wind Turbines

Wind farms can have an impact on RAS observations, especially at frequencies below a few GHz. They feature high-power electrical devices and a multitude of electronic equipment for controlling the turbines. These are usually placed in the nacelles at significant altitudes of up to 170 m in more modern models. It is well known that electrical and electronic equipment can radiate power, even without dedicated antennas, simply by means of oscillations in electrical/electronic circuits. Nevertheless, wind turbines represent a somewhat unusual case in spectrum management, as they are not subject to the Radio Regulations and they are not associated with a radio communication service. Therefore, it took some considerable effort to establish some coordination at a local level for some particularly vulnerable CEPT RAS stations, such as WSRT (NL), SRT (I) and Effelsberg (D). In these cases, the local network agencies were involved, although they were not really the responsible authority for the coordination.

Despite this difficult regulatory situation, ECC initiated a forum group on wind turbines (FG WT), which later published ECC Report 321. This report contains compatibility calculations and measurements for wind turbines in relation to the RAS. CRAF contributed a section on the former, while the measurement sections were provided by administrations. The case of wind farm compatibility studies was also published in a scientific journal (Winkel, B. & Jessner, A., Journal of Astronomical Instrumentation, Vol. 8, Issue 1, 2019).

In the Netherlands, ASTRON has sought a solution for a wind park (and nine solar parks) in the near vicinity of the LOFAR Core by making compliance with EMC standards mandatory. This obligation has been included in several permits. In addition, covenants were agreed on with all wind and solar parks, in which they also committed to adhere to the EMC standards. This has led to the development of an almost radio-silent wind turbine for the wind park near the LOFAR Core and the development of several radio-silent solar parks. Tests in the lab and in the field have confirmed this. The wind farm has also committed to go into full shutdown mode several days a year. Based on these experiences all other municipalities in the North of the Netherlands were asked to take the EMC standards into account when issuing permits for wind parks, solar parks as well as small individual wind turbines (15 m).

### 3.2.7 The VLBI Global Observing System (VGOS)

The United Nations (UN) General Assembly's Resolution 69/266 recognised the importance of a Global Geodetic Reference Frame for sustainability and development, and that the use of radio astronomical Very Long Baseline Interferometry (VLBI) systems was probably the only means to achieve the required 1-mm reference-frame accuracy at a global level. The International VLBI Service for Geodesy and Astrometry (IVS) represents a multinational cooperation, which has the mission to provide the essential data for the regular determination of Earth rotation parameters, the International Celestial Reference Frame (ICRF) and the Global Geodetic Reference Frame (GGRF) with high accuracy. As such, VGOS will play a key-role in a large number of future applications, such as satellite missions, aircraft navigation, Earth and climate studies, and many more.

VGOS sites are distributed around the Earth over all continents and usually feature 10-m class antennas with state-of-the-art receivers, which enable observations in several wide bands up to 14 GHz. Thus, for the protection of the VGOS system, a world-wide harmonised approach is required, which is a challenge, as the protection of RAS sites is usually under the responsibility of national authorities. To improve this situation, IVS, in collaboration with CRAF, has started a process at ITU-R to address this. The first step was to prepare a "Question" document under the umbrella of ITU-R Study Group 7 (SG 7). A Question in ITU-R terms is a document that describes a technical, operational, or procedural problem and which is usually addressed via a new ITU-R Recommendation, Handbook, or Report. The technical characteristics of VGOS are not well known outside the astrometric and geodetic communities, but need to be understood by spectrum agencies, who may have plans to use the VGOS observing bands for other applications. Thus, two points were addressed in the Draft Question, submitted by CRAF:

1. What are the technical and operational characteristics of geodetic VLBI?
2. How does geodetic VLBI use the radio spectrum to achieve the accuracy needed to fulfil its mission?

The Draft[11] was submitted to WP 7D for further discussions, before it can be handed over to the SG 7 plenary. Furthermore, a correspondence group (led by a CRAF member) was initiated. In the meantime, our Work Item team WI-VGOS has started working on a Draft Report, which would address the questions. The ultimate goal is to work towards world-wide protection of the VGOS stations by modifying the Radio Regulations at a future WRC.

## 3.3 Main achievements and highlights

In this Section, the main achievements of CRAF in the past years will be highlighted. As most of the very specific topics have been explained in detail in the previous section already, in the following the focus will be on bigger picture.

### 3.3.1 (ITU-R achievements)

As a sector member at ITU-R, CRAF actively participated in previous cycles of the WRC through technical compatibility studies and regulatory work. CRAF, together with other RAS spectrum management organizations, played a key role in securing the appropriate RAS protection from several radio-communication services during the previous WRC cycles. In the following, some of the topics are briefly summarised, to which CRAF significantly contributed.

#### *3.3.1.1 WRC-15*

The IMT allocations in several RAS bands (608–614 MHz, 1330–1400 MHz, 1427–1518 MHz, 2700–2900 MHz, 4800–5000 MHz) under footnote 5.149 has been studied under Agenda Item 1.1. CRAF contributed to the studies which were then converted to a published report, ITU-R RA.2332-0. The 608 MHz band was untouched, in the end, but unfortunately some L- and C-band frequencies were identified for IMT. Therefore, the Report ITU-R RA.2332-0 is very important, as it contains protection requirements.

[11] Report ITU-R RA.2507-0 was published in October 2022.

The WRC also decided that broadband public protection and disaster relief applications can use larger parts below 1 GHz, including adjacent bands of the 408 MHz RAS band. However, the adopted Resolution 646 mentions the RAS and the requirement to protect it.

CRAF also contributed to Recommendation ITU-R RS.2066-0 (incorporated by reference in the RR), which is about the protection of RAS at 10.6–10.7 GHz from EESS (active) in X-band under Agenda Item 1.12.

##### *3.3.1.2 WRC-19*

CRAF provided several compatibility studies under Agenda Item 1.13 (see Section 3.2.2.1 for details) that led to inviting administrations to consider RAS protection in the relevant WRC-19 resolutions. Also, CRAF was very active before and during the conference to ensure sufficient protection for RAS is provided following the recognition of Iridium as a new GMDSS provider under Agenda Item 1.8. The protection measures from HAPS (Agenda item 1.14) was also achieved through compatibility studies and CEPT coordination during the conference. CRAF contributed to several other agenda items at WRC-19. The outcome of WRC-19 was largely in line with CRAF positions and is detailed in the CRAF report for WRC-19.

##### *3.3.1.3 WP 7D*

At WP 7D, CRAF led the efforts towards the recognition of VGOS through proposing new ITU-R questions and initiating a new ITU-R report on VGOS characteristics. CRAF continues to lead the correspondence group on VGOS at WP 7D with the objective of proposing an agenda item to future WRCs to recognize VGOS. CRAF also contributed to several published ITU-R reports at WP 7D such as Report ITU-R RA.2332-0 and a new report on IMT vs. RAS at 43 GHz.

### 3.3.2 Regional CEPT Level

##### *3.3.2.1 RAS protection from 2nd harmonics*

One issue, which can be very important for the RAS is that of 2nd harmonics. Harmonic emissions are a physical consequence of electronic circuits. Only a perfectly linear system (an ideal device) would not produce any harmonic emissions. Normally, harmonics are not considered an issue for compatibility studies because of their relatively low intensity level compared with the carrier signals. Nevertheless, the high sensitivity of radio telescopes and the fact that transmissions from air- and spaceborne systems can enter the RAS receiver via the main beam requires attention. This has also been recognised by ITU-R throughout its history. For example, Recommendation ITU-R SM.329-12 states that second harmonics, as well as intermodulation products, represent a very important aspect when it comes to spurious emissions. Even the WRC has acknowledged this previously, e.g., in RR No. 5.402[12]. Finally, real-world measurements in CEPT demonstrated that IMT user equipment can indeed reveal relatively strong 2nd harmonic features (Report ITU-R SM.2421-0; their Section 5.3). It is also important to mention that in more classical equipment, high-quality physical filters were sometimes used to suppress this kind of leakage radiation. That assumes that cost considerations by the sales departments of a company were not valued more important, as filters are not necessarily very cheap. Furthermore, a big problem with new active antenna systems

[12] RR **No. 5.402**: "The use of the band 2 483.5–2 500 MHz by the mobile-satellite and the radiodetermination-satellite services is subject to the coordination under No. 9.11A. Administrations are urged to take all practicable steps to prevent harmful interference to the radio astronomy service from emissions in the 2 483.5–2 500 MHz band, especially those caused by second-harmonic radiation that would fall into the 4 990–5 000 MHz band allocated to the radio astronomy service worldwide."

(AAS) is that all antenna elements – of which AAS have many – would need to be equipped with filters, which is a huge cost factor. Therefore, it can be expected that for AAS, spurious emissions will become an even bigger problem than they are today.

As discussed in Section 3.2.2.3, one of the topics, which took a lot of CRAF's effort, was the introduction of cell-phone network user equipment on-board aerial vehicles. A major achievement of CRAF was to successfully push for suitable protection of RAS sites from potential 2nd harmonics emission from such equipment, despite very strong opposition (ECC Report 309).

Other examples are that of private LTE networks at 2.5 GHz, the 2nd harmonics of which would fall into the 4990–5000 MHz RAS band (see ECC Report 325) or WRC-23 AI 1.4 (IMT at 800 MHz as flying base stations, so-called HIBS; see Section 3.2.2.5).

#### *3.3.2.2 ECC Report 271 and additional protection of several European RAS stations from Starlink*

Another recent major achievement of CRAF concerns its work on the topic of satellite mega-constellations. This has already been described in detail in Section 3.2.3, in which it was explained that additional protection from the Starlink satellite downlink transmissions could be organised to the benefit of radio astronomy in some European countries to date, namely Germany, Italy and Spain. In particular, Starlink was tasked to avoid direct illumination of the RAS sites in these countries with their (steerable) antenna beams. The background for this was not the protection of the RAS band from out-of-band emission originating from the satellites, but to avoid bringing the RAS receivers into the non-linear regime owing to the strong Starlink carrier signals, which could cause intermodulation products in the RAS bands. This is an absolute novelty in spectrum management, as one can usually only claim protection from signals in frequency bands that one 'owns', but in this case, RAS receivers are being protected from signals at frequencies outside the RAS bands. This also shows a typical problem of spectrum management, namely that regulation is too much based on relatively formal (legal) arguments, and often neglects the physical reality. RAS receivers cannot be constructed to cope with these very intense incident levels of power, unless one is willing to significantly trade-in sensitivity, which would easily require a factor of 5 or more of observing time to compensate.

It should also be noted that ECC Report 271 itself can be considered a success. Many other stakeholders, world-wide, apparently use it as a reference to assess the potential impact of satellite mega-constellations on radio astronomy. CRAF was a major contributor to this ECC Report.

#### *3.3.2.3 ECC Decisions and Reports*

As a regional spectrum management organisation, CRAF's main activities are focused on the CEPT. That does not mean that CRAF ignores ITU-R, as a lot of the work at the international level has direct consequences for RAS stations in CEPT. However, many new applications and new allocations today are harmonised within Europe (in particular in the European Union). Even if an ITU-R regulation is not in favour of the RAS, CRAF still has the chance to work towards European solutions. In Annexes 7.3 and 7.4, lists of all ECC Decisions and Reports (since 2011), have been produced, which are relevant to the RAS. This applies to 27 of 96 ECC Decisions and to 40 of 180 ECC Reports (from 2011 to today). In most of these Reports, RAS compatibility studies are included, although it should be mentioned that a small fraction of them were not provided by CRAF (but even in these cases, where other stakeholders provided the studies, it has usually been as a direct consequence of CRAF activities in CEPT).

A few highlights regarding ECC deliverables are mentioned in the following. In the wake of CRAF's successful work relating to the OneWeb and Starlink mega-constellations (ECC Report 271; see Sections 3.2.3 and 3.3.2.2), based on the software developed in collaboration with SKAO, it was possible to provide very sophisticated compatibility studies for satellite personal communication systems (S-PCS) at 150 and 400 MHz. These provide satellite-based IoT services. Several medium-sized constellations were in the pipeline, with up to one hundred and more satellites. Based on the ECC Report (not published yet)[13], the older ECC Decision (99)06, which was amended, now includes protection requirements for RAS. Another noteworthy topic is that of automotive car radars at 24 GHz, which were moved to higher frequencies (77 GHz), also to avoid interference to the RAS. The ECC Report 325 was already mentioned (see Section 3.3.2.1). It concluded that no-transmit zones around RAS stations are necessary for RAS protection to avoid interference from cell phone equipment on aircraft and unmanned aerial vehicles (e.g., drones). Currently, CEPT is also working on a new ECC Decision to implement these rules.

### 3.3.3 National level

Our activities are not exclusively aimed at the international or European level. We can and will only be successful if we maintain strong links to national and local administrations, who usually have a strong interest to protect their investments into RAS observatories. CRAF delegates participate in national meetings (also to prepare CEPT and ITU-R work) and maintain good relationships. We also engage with other stakeholders, such as cell phone network operators, to coordinate their networks locally.

At a number of CEPT stations, wind farms were or are planned. This required extra efforts, as the wind farms also represent a potential threat owing to electromagnetic interference. As this is not subject to ITU-R Radio Regulations, an alternative regulatory approach had to be found – which is even very different in the various member countries. In Germany, the counties in the vicinity of the Effelsberg station were very supportive and took concerns seriously. The German CRAF delegate, who is based at Effelsberg, is now consulted whenever a new wind farm is under consideration in the region. This success story will hopefully also influence projects in other countries, e.g., in Sweden where an offshore wind farm is planned not far from Onsala observatory. In the Netherlands, ASTRON has even managed to convince a wind farm company to develop radio silent turbines, to make them compatible with an operation near the LOFAR Core station; see Section 3.2.6.

### 3.3.4 Software development in CRAF: pycraf

Whereas some people tend to think spectrum management is mostly a question of lobbying and economic value, the 'bread and butter' of successful spectrum sharing is to work out the technical conditions under which a peaceful coexistence of two or more services is possible. In most cases, a new service or application, which seeks access to the spectrum, has to demonstrate that existing services at the same or adjacent frequencies are not affected in a way that limits their operations.

ITU-R has published a large number of recommendations, which contain procedures and models to be used in compatibility calculations. While these are mostly written in a way, which allows straight forward implementation of the algorithms, some models – in particular for radio path propagation – can be extensive and require long software programs.

[13] ECC Report 322 was published in January 2022.

CRAF has undertaken considerable work to produce Python software to ease and streamline the creation of compatibility calculations. This software tool is named pycraf and is not a stand-alone software, but a library for the Python programming language (a so-called package). While this demands some familiarity with the Python language, it makes the software very flexible and versatile, easy to extend, and allows us to share our studies under permissive open-source licences with other parties. In the following, a small overview will be given of the features included in pycraf, together with some examples.

Probably the most-used feature of pycraf is its implementation of the Rec. ITU-R P.452-17 path propagation model. The method includes line-of-sight (free-space) loss including correction terms for multipath and focussing effects, diffraction (at terrain features), tropospheric scatter, and anomalous propagation (ducting, reflection from elevated atmospheric layers). One should emphasise that the model is not fully derived from physics, but is based to a large extent on empirical modelling, which best describes the results from a huge number of measurement campaigns. As the P.452 model requires topographical information along the path of propagation, pycraf provides easy access to terrain-height data measured by the space shuttle radar mission (SRTM; Farr, T. G. et al., 2007, Rev. Geophys., 45, RG2004). In Figure 9 an example path loss map is shown for an area around the 100-m telescope at Effelsberg (Germany). The colour in each pixel of the map describes how much path loss/attenuation is predicted by P.452 if the receiver is in the centre of the map and the transmitter is located at the respective pixel. Such attenuation maps are a key ingredient for all sorts of compatibility studies.

Another useful pycraf feature is its implementation of the atmospheric absorption model provided in Rec. ITU-R P.676-16. Whereas this is not as sophisticated as some of the models used in scientific applications, it provides predictions with sufficient accuracy. The P.676 algorithm performs a raytracing of a ray of (radio) light through a layered atmosphere and integrates the overall attenuation along the ray.

In collaboration with SKAO, the pycraf team has worked towards integrating functionality to perform simulations of entire satellite constellations and their aggregated power received at a radio telescope. This becomes more and more important as mega constellations such as SpaceX/Starlink or OneWeb, consisting of thousands of small satellites, are being launched into low-Earth orbits, with the aim of providing broadband Internet all over the world. As satellites can cross the main beams of radio telescopes, they represent a high potential for harmful interference, and astronomers are very worried about the situation.

The pycraf package is hosted on GitHub under open-source license (GPL v3). There is also a lot of documentation, with a user manual and several tutorial notebooks (for the Jupyter web frontend) available.

In the 33rd issue (Autumn 2020) of the CRAF Newsletter, a more detailed article about the pycraf software was published.

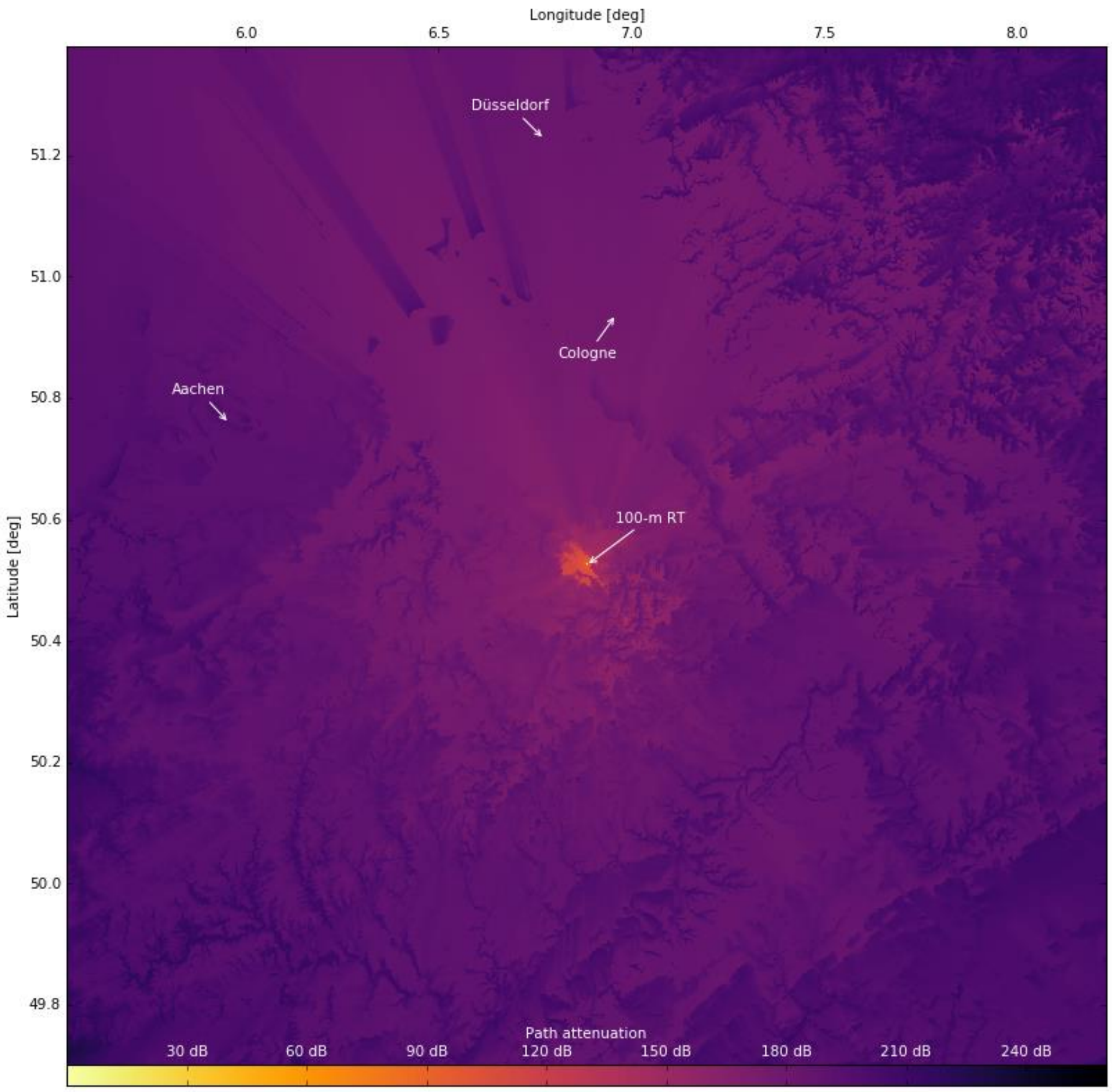

*Figure 9: Path attenuation map for the region around the 100-m radio telescope at Effelsberg (Germany) for a frequency of 3.5 GHz. Transmitter height was assumed to be 40 m. Terrain heights play an important role representing obstacles for the diffraction loss calculation in the model (the telescope is situated in a valley in the Eifel mountains).*

## 3.4 Main problems and challenges encountered and means to address them

### 3.4.1 Satellite mega-constellations

Recent developments in large low earth orbit (LEO) satellite constellations, with plans for several thousand of satellites, have changed the forecast on how astronomical observations will be conducted all across the electromagnetic spectrum in the future. From an observer's point of view, we are seeing a shift from most of the satellites being located at the geo-stationary orbit, with less than 1500 satellites all around the globe, to a massive mesh of LEO satellites. Current constellations under deployment and planned ones show a forecast where more than 2000 satellites could be visible (above the horizon) at any latitude and instance of time.

Radio astronomy has a long history of dealing with satellite signals, with some successes like the case of GLONASS where the operator agreed to move the frequency of transmission away from the OH band at 1612 MHz and fitted filters in a new generation of satellites, and others with more tragic results such as the Iridium case where the same OH band was first affected in 1999 (and still is to this day). These two examples of LEO satellite constellations causing harmful interference into radio astronomy bands involved satellite constellations with less than 100 satellites.

These new large satellite constellations planning on using X, Ku, Ka, V and E bands (from 8 GHz up to 80 GHz) have not only the potential to affect protected radio astronomy bands, but also the sheer number of satellites could make opportunistic observations in wider frequency bands impossible, for example for telescopes sited in radio quiet zones (RQZ) where the use of large parts of the radio spectrum is regulated with an aim of benefiting radio astronomy observations. A prime example of this is the SKA-Mid telescope under construction in South Africa.

Furthermore, the projected large number of spacecrafts in LEO can pose a potential interference to radio astronomy due to "second-order" emissions. Unintended electromagnetic radiation (or radio noise generated by the electronics within the satellites) and reflections of terrestrial RFI at the satellites can potentially affect long integration observations in low frequency instruments such as LOFAR, SKA-LOW and others.

As discussed in Sections 3.2.3 and 3.3.2.2, CRAF has engaged in this topic since many years and SKAO and CRAF developed software to perform satellite system compatibility studies. At the international level, several CRAF delegates have participated in the Dark and Quiet Skies for Science and Society Workshops I and II providing input to the RAS working group that was included in the final reports. Also, in these two workshops CRAF members and observers provided presentations on the impact of satellite constellations on radio astronomy and the possibility of impact from unintended electromagnetic radiation from satellite electronic systems (not related to intended satellite downlinks). CRAF is also involved in groups related to satellite mega-constellations such as ORP-JA1.5, ORP-JA2.5, EAS, or the Royal Astronomical Society, and it is a contributing member of the recently created IAU Centre for the Protection of the Dark and Quiet Sky from Satellite Constellation Interference.

Several CRAF observatories have also started to conduct observations of satellite constellations to understand better their downlink signatures and possibly their unintended emissions.

### 3.4.2 Iridium

In Section 3.2.1 we explained the history of the "Iridium problem" in detail. Despite almost three decades of effort, we have still not achieved an acceptable solution to this. On the contrary, many CEPT administrations and stakeholders (including CRAF) are getting exhausted from the work on this topic. Even some of the strongest supporters of our cause are not willing to put endless efforts into this. This is not surprising, given that even letters to the FCC and the ITU-R Bureau could not convince Iridium to stop the interference into the RAS band at 1610 MHz.

In addition, the severe flooding that occurred in the region around the Effelsberg station in Germany (in July 2021) created some negative media coverage. Owing to the 30-km protection zone around Effelsberg, disaster relief forces could not use their Iridium phones. While this was a well-known fact for every organisation involved in the incident (military, police, etc.), forces on site were not aware of the protection zone. Four days after the initial flooding, the station staff was contacted by the military. Of course, the German network agency was immediately contacted by Effelsberg and within a day, the protection zone was temporarily suspended. Nevertheless, some (poorly informed) newspapers covered this with the headline that Effelsberg impeded the disaster relief forces.

### 3.4.3 IMT2020

In the past two decades, cell phone networks have become ubiquitous. Before the new millennium started, cell phones were a rather niche application, while today on a world-wide basis more than five billion unique mobile subscriptions exist. This enormous increase has also led to a huge demand for additional spectrum. In Figure 10, the spectrum allocations to the IMT (MFCN, mobile-fixed communication networks) and the RAS is displayed. For RAS, a distinction has been made between primary and secondary allocations. For information, also the bands subject to RR Footnote 5.149 is included, which urges administrations to protect RAS operations in a number of frequency intervals. However, the actual protection of secondary and 5.149-bands depends heavily on the national administrations. In come CEPT countries, there is virtually no protection, while in a few countries, national authorities take the RAS protection more serious.

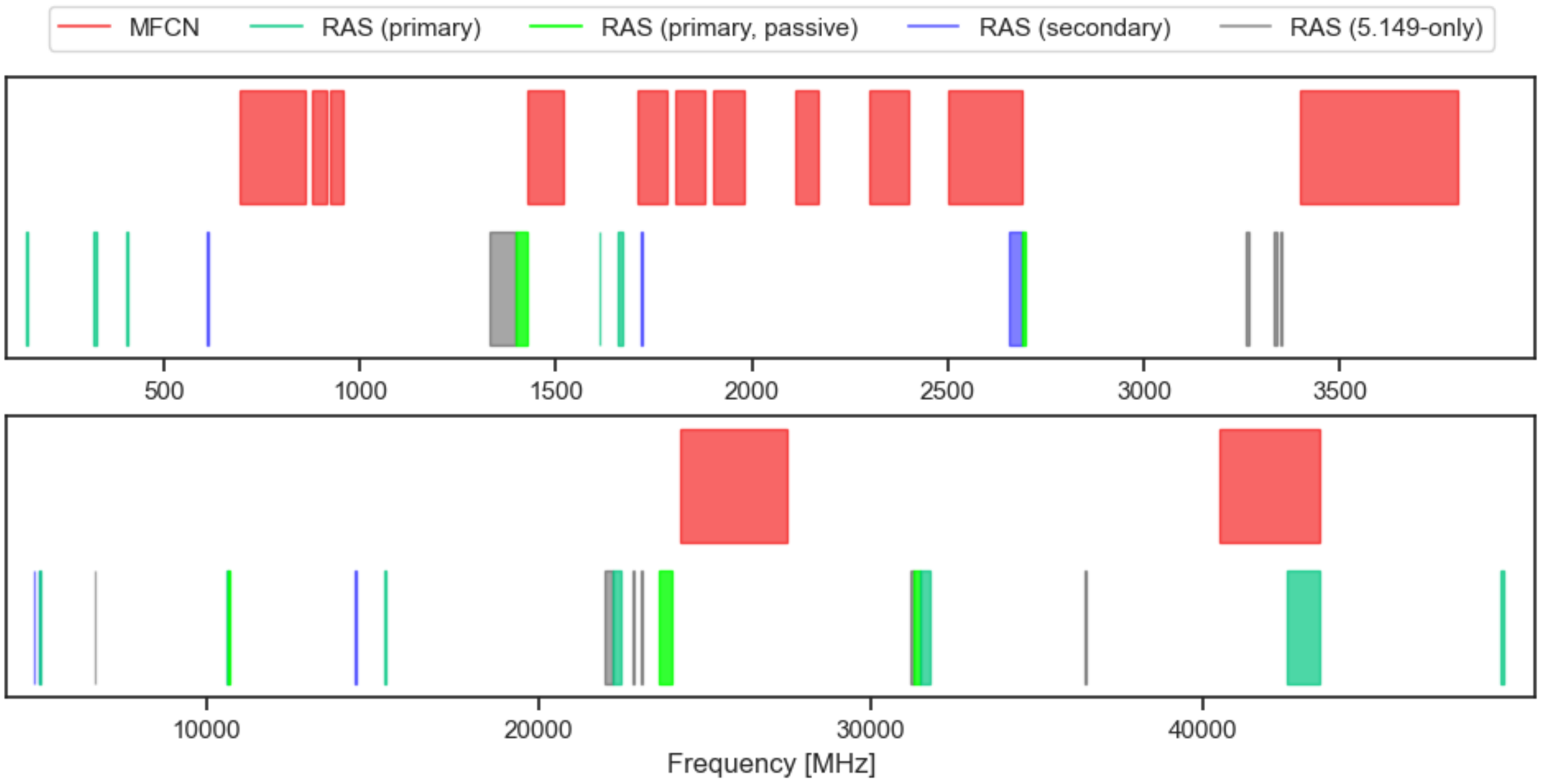

*Figure 10: Spectrum allocated to IMT (MFCN) and the RAS below 50 GHz in CEPT countries.*

Figure 10 also demonstrates that especially below 4 GHz a huge portion (one third) of the spectrum is assigned to IMT/MFCN networks, whereas RAS has only 5%. If one only counts primary bands, as little as 1.5% is allocated to the RAS. Owing to some larger blocks above 20 GHz, the fraction is much higher if one counts all bands below 100 GHz, with approximately 19%[14]. IMT is not yet operating at the mm-wave frequencies, but WRC-19 has already allocated several bands totalling 17 GHz of spectrum indicating that the pressure on radio astronomy at these high frequencies will also increase. In the current WRC study cycle, there are additional agenda items (AI 1.2 and 1.5; see also Section 3.2.2.4), which explore the possibility of even more IMT bands below 1 GHz and in the so-called mid band (between approximately 3 and 11 GHz).

IMT networks are one of the biggest if not the biggest threat to ground-based radio astronomy. As they feature a huge number of high-power base stations, and an even larger number of mobile user devices (which are usually licence-exempt), even in remote locations, use of spectrum for IMT most normally means that opportunistic radio astronomy observations at these frequencies are impossible. Furthermore, new applications are in the planning stage, which involve aerial IMT transmitters – either flying base stations (so-called HIBS and HAPS; compare Section 3.2.2.5) or the permission to use terminals on aerial vehicles (see Section 3.2.2.3). The latter could either be cell phones on normal commercial aircraft or for the remote control of equipment on unmanned aircraft or drones. These aerial transmitters are an even bigger problem, as the likelihood to have one of the devices within or close to the main beam of a RAS telescope is relatively high. In such cases, the large antenna gain would boost even normally relatively unimportant spurious emission features (e.g., harmonics or intermodulation products) to well above the RA.769 thresholds and lead to incompatibilities, even if the IMT and RAS frequency bands are well separated.

[14] In many of these frequencies, RAS has not an exclusive allocation, but needs to coordinate with other so-called “co-primary” services.

The high transmitter power, in particular of IMT base stations, can also easily drive RAS receivers into the non-linear regime or even completely block them, if they are operated close to our observatories. This is becoming more and more a problem, as radio astronomers started to build very large bandwidth receivers, which inevitably overlap with one or more cell-phone bands. We know of numerous cases in CEPT RAS stations, where newly built receivers were affected by this. In several cases, re-designs were necessary, either reducing the total bandwidth of the receivers or placing stop-band filters in front of the cooled low-noise amplifiers, which had severe consequences on the sensitivity of the receivers.

### 3.4.4 Passive band (RR No. 5.340) use

The technological developments of active radiocommunication services, particularly the ultra-wide band (UWB) applications, may demand large contiguous bandwidths for accurate measurements. In many cases, the bandwidth required might overlap with the passive bands protected by footnote 5.340 that strictly prohibits all emissions. The pressure is recently increasing to allow conditional emissions in those bands that will not cause interference to the passive services. It is challenging now to convince administrations on the long-term risks of sharing those bands given the high commercial value offered by industry on the other hand.

### 3.4.5 Lack of support for secondary allocations and RR No. 5.149 bands

The footnote RR No. 5.149 states that "in making assignments to stations of other services to which the bands: […] are allocated, administrations are urged to take all practicable steps to protect the radio astronomy service from harmful interference. Emissions from spaceborne or airborne stations can be particularly serious sources of interference to the radio astronomy service (see Nos. 4.5 and 4.6 and Article 29)". Some of the bands listed in this, however, do not have a primary or even secondary allocation in the radio regulations. This is mostly the case for some spectral-line frequencies that were detected relatively late in the history of radio astronomy. The RAS has been unable to convince the ITU-R countries to allocate a new band for many years.

Over many years, when the pressure on the spectrum was less heavy than today, most administrations were willing, however, to take the wording in No. 5.149 serious and protected RAS operations. In particular, the very important "Methanol band" at 6.65 GHz was respected at many WRCs in the past. But in the current study cycle, where IMT is asking for new spectrum in the mid-band, CRAF was not even allowed to submit compatibility studies anymore (see Section 3.2.2.4). Another case is the 22–22.21 GHz interval, which is under study in AI 1.10. It is still unclear if RAS will be protected. The band 1718.8–1722.2 MHz band was previously lost to IMT in CEPT, already.

One difficulty is that the footnote No. 5.149 aims at the "assignment" process (in contrast to "allocation"), which refers to the national licensing. The problem with many of the WRC-23 agenda items is, however, that the active services in question would need multi-lateral coordination to properly protect the RAS. If one country is not interested in this, the country operating the RAS station has absolutely no means to ask for coordination/protection, as Article 4.4 of the RR clearly states that any uses under 5.149 are to be considered as not in line with the RR.

Even when it comes to secondary allocations, where the regulatory status is at least well-defined, CRAF lost an important band, 2655–2690 MHz, in Europe in the past years, again to IMT.

#### 3.4.6 Administrations ignoring existing rules and regulations

A significant effort is spent on RAS compatibility studies for the ECC technical reports, as well as on incorporating the results into ECC decisions. It is mandatory for administrations implementing the ECC decisions to fully implement the RAS protection measures at the national levels. It was noted that some administrations tend to ignore the RAS protection while implementing the ECC decisions. An example is the ECC decision for Iridium that urges administrations to implement exclusion zones around RAS telescopes. Only few administrations are currently applying this. CRAF members will have to follow up at the national level to ensure the decisions are fully implemented.

Another example is the introduction of 5G at 26 GHz in Germany. Although the German CRAF delegate was heavily involved in ITU-R and CEPT preparatory groups, and ECC Decision (18)06 as well as EC Decisions 2019/784 and 2020/590 conclude that coordination with RAS is necessary, the German network agency (Bundesnetzagentur) initially did not establish sufficient protection of the Effelsberg station during licensing. The German delegate had to engage multiple times and with a lot of effort until the involved department was willing to properly coordinate with the Effelsberg observatory.

### 3.5 Synergy and interactions with other (scientific) bodies at European and global levels

The following three organisations have observer status in CRAF: SKA Observatory (SKAO), European Southern Observatory (ESO) and European Space Agency (ESA). CRAF has a very close collaboration with the SKAO, in particular on issues related to satellite systems and IMT (see Sections 3.2.2.4, 3.2.2.5 and 3.2.3). Both ESA and ESO participate in the general meetings of CRAF. ESA, is an important observer as their Earth exploration satellites share a number of frequency bands with radio astronomy. With ESO observer we have a strong collaboration on common issues in the millimetre and sub-millimetre wavelength ranges for European mm-wave instruments and ALMA, e.g. CloudSat a NASA Earth observation satellite operating at 94 GHz.

Under the umbrella of International VLBI Service (IVS), many observatories in Europe also perform geodetic VLBI observations, often with the same telescope as the RAS. It is therefore natural that IVS has a member in CRAF representing their network. Geodetic VLBI data products play a key role for many applications, from satellite networks, to navigation and Earth science. It can be seen as a part of the critical infrastructure and as such should be a duty of national spectrum authorities to protect Geodetic VLBI to a larger extent than today. The VGOS work item team, led by the IVS representative, deals with strategic goals to protect geodetic observatories against RFI. For this, we started an initiative at ITU-R WP 7D, e.g. by: (1) writing a reference documents at ITU-R on "Geodetic VLBI" for administrations and (2) raise awareness of the need for protection by the registration of geodetic VLBI sites at the ITU-R through the national administrations. Naturally, this work also helps RAS.

CRAF also collaborates with the following external global spectrum management organisations: Committee on Radio Frequencies (CORF) of the National Academies of Sciences, Engineering, and Medicine of USA, The Scientific Committee on Frequency Allocations for Radio Astronomy and Space Science, IUCAF (IUCAF), the Radio Astronomy Frequency Committee in the Asia-Pacific region (RAFCAP), and the United States National Science Foundation (NSF). This is done at ITU-R level both in preparation for World Radio Conferences (WRCs) and for non-WRC agenda items.

# 4 CRAF's views on the evolution of the programmatic landscape and availability of new instruments and how this evolution will impact its activities

Radio astronomy has always been a very innovative science and embraced technological developments. The volume of measurement data can be huge. Pulsar data acquired in single dish mode easily amount to many TBytes per observing session The total volume of recorded data for the Event Horizon Telescope observations in April 2017, which resulted in the image of the black hole in M87, was about 3.5 PB and the data rates of the upcoming SKAO are gigantic, in the order of TBytes per second. Nowadays, VLBI data can in many cases be transferred via optical fibre cables from the radio telescopes to the central processing facility. In the European VLBI Network this is even possible in real time. Nevertheless, the traditional method of shipping the recorded data, stored on hard-disks, to the correlator is also still being used owing to lack of high-speed networks at remote telescope locations. The high data rates allow astronomers to capture spectra with enormous time and frequency resolution. This is enabled by state-of-the-art equipment, making use of field programmable gate array chips (FPGA) and graphic card units (GPU). Both are extremely flexible and their functionality can easily be (re-)programmed. This also opens up completely new ways to flag or mitigate radio frequency interference already during data recording. Also, analogue receiver technology is still improving, allowing low receiver noise figures even for uncooled receivers. Usually, most RAS receivers are cryogenically cooled, such that the radiometer noise of the instruments is below 5 K in some bands at GHz frequencies. Uncooled low-noise amplifiers, on the other hand, are quite interesting for so-called phased-array feeds – large boxes featuring dozens of antenna elements the outputs of which can be combined electronically. With proper phase delays applied, artificial antenna beams can be formed and be freely steered. Based on this, true multi-pixel "cameras" are finally a real possibility for decimetre-wave single dish telescopes, such as the Green Bank-, Parkes- or the Effelsberg telescopes. In addition, the available bandwidths of receivers grow steadily, with up to many GHz of simultaneously recorded spectrum.

## 4.1 Broadband receivers and susceptibility to RFI

All these new techniques and advances enable new fantastic science projects, allow for improvements in RFI detection and mitigation, but some also come with draw-backs regarding RFI robustness. Large bandwidths also mean that there is a high potential to have some very intense transmitters in the band, e.g., cell phone towers, communication satellite downlinks, or Earth-sensing satellite-borne Radars. This can especially impact the low-noise amplifiers (LNAs), but also the receiving system as a whole (including analogue to digital converters and potentially optical fibre links for signal transportation). In less severe cases, the receiving system may "only" produce intermodulation products, however it could also be saturated or even destroyed.

For this purpose, receiver engineers are engaged in the design of microwave filters to be located in front of the LNAs. In addition to attenuating the undesired signals, such filters are required to have extremely low-loss performance to limit any added noise to the receiver. As a result, and because of their excellent electrical conductivity, planar microwave components based on high-temperature superconductors (HTS) have been intensively studied and installed in several radio astronomical receivers. More details on this technology are available in the article "Low-loss Superconducting Filters for the Sardinia Radio Telescope" in the CRAF Newsletter No. 32 (Autumn 2019). A challenge remains, though. The HTS operate at fixed frequencies, therefore without the flexibility that would be required given the rapidly evolving spectrum situation.

New technology can also enable more robust receiver operation. Phased array feeds, for example, could be utilised to put antenna-pattern Nulls into the direction of strong emitters. For ground-based transmitters, this should be straight-forward, as soon as the electronic beam weights can be updated on timescales in the order of seconds. However, for satellite interference, much quicker updates would be necessary, which is currently not yet feasible to our knowledge.

CRAF will continue to inform the scientific community about these challenges and concerns related to broadband observations, in particular about the activities in spectrum management that have the potential to cause saturation effects.

## 4.2 A radio observatory on the Moon?

Given the huge pressure on the radio spectrum on Earth, the idea to run a radio astronomy station on the far side of the Moon, shielded by the Moon itself from Earth-based emissions is now being seriously discussed. As private companies initiated huge space programmes, with re-usable and more cost-effective rockets, and NASA is planning to operate a base on the Moon, the idea of a Moon observatory seems more realistic. CRAF would like to point out, however, that this would (still) be extremely expensive and cannot replace all the functions of terrestrial observatories (e.g., VLBI and geodesy by definition). First activities seem to be aimed at low frequency observations, for which the Moon undoubtedly is a much better place than any location on Earth. Also, the more the Moon is being settled, the more lunar RFI will be present, which would impact the observatory as well. Nevertheless, CRAF actively engages with all involved parties; also at the ITU-R a report concerned with a radio quiet zone on the far side of the Moon is already in the queue.

## 4.3 Non-disclosure agreements

With more and more satellite-borne threats to radio astronomy, even the largest-scale RAS projects, such as the SKAO, ALMA or the ngVLA, which are situated in very remote places, will become more and more affected. The Radio Quiet Zones that have been established around these multi-billion-dollar projects will not help against interference from satellites. As a consequence, the operators of these instruments are trying to find constructive ways to collaborate with the satellite operators to find other solutions. CRAF fully supports this approach. However, what is perceived to be of great concern, is that more and more non-disclosure agreements are being signed by these stakeholders. While this may work for the few large observatories, this represents a huge threat to the many smaller observatories. The RAS community should try to seek solutions for everybody.

## 4.4 The relationship between CRAF and SKAO

A special and very close relationship exists between CRAF and SKAO. SKAO is an Observer member of the Committee. It is important to point out, however, that CRAF is not immediately responsible for SKAOs spectrum management, as they have their own staff dedicated to this. There was a long debate whether SKAO should become a full (paying) CRAF member organisation, but SKAO decided against this and instead became only a CRAF Observer. The main reason for this was because of political implications. SKAO is stretched over two ITU-R regions, and another regional organisation (RAFCAP) is responsible for the Australian part of the SKAO. That said, CRAF and SKAO have closely collaborated and work very effectively together on a large number of topics, such as satellite systems (in particular mega-constellations) or airborne interferers and appropriate regulation. The SKAO member is in fact one of CRAF's work item team leaders (WI-SAT). We frequently discuss upcoming issues and topics, collaborate on input documents to CEPT and ITU-R meetings and inform each other on strategies. In fact, the collaboration will even be intensified by collaboration on the new IAU Centre "for the Protection of the Dark and Quiet Sky from Satellite Constellation Interference" (see Section 3.2.3).

# 5 SWOT Analysis – Looking forward

*This section is omitted in the public version.*

# 6 Acknowledgements

*In 2021, the European Science Foundation set up a review panel of four experts,*

- ***Dr. Pierre Cox (Review Panel Chair),***
  *Former Director of IRAM & ALMA,*
  *Institut d'Astrophysique de Paris, CNRS, France,*
- ***Markus Dreis,***
  *ITU-R WP 7C Chair,*
  *EUMETSAT, Germany,*
- ***Prof. Hermann Opgenoorth,***
  *COSPAR Vice-Chair Panel on Space Weather,*
  *Umea University, Sweden,*
- ***Prof. Liese van Zee,***
  *Former Chair of Committee on Radio Frequencies, US National Academies,*
  *Indiana University Bloomington, USA,*

*with the mandate of reviewing the activities of CRAF over the period 2016-2021. The panel was specifically asked to comment on the activities and achievements of CRAF and to provide recommendations for the next period, taking into account the CRAF mandate as well as the European and global research landscape.*
*This self-evaluation report was provided to the review panel members in March 2022, and the panel members also participated in a CRAF plenary meeting in May 2022. In addition, interviews with several CRAF representatives were conducted. The review panel presented their findings in a report to the Committee in August 2022.*
*CRAF and its Stakeholders Forum would like to sincerely thank the Review Panel for their excellent work, and the detailed and constructive feedback, which required a large amount of time and effort.*
*CRAF also wish to thank the European Science Foundation (in particular Nicolas Walter and Eliana Legoux) for the most professional management and organisation of the whole process and for creating an inspiring and conducive atmosphere.*

# 7 Appendices

## 7.1 List of CRAF member institutions (as of September 2023)

In the following the CRAF member institutions are listed

- Austrian Academy of Sciences, **Austria**; Schmiedlstr. 6, 8042 Graz
- STCE – Royal Observatory of Belgium, **Belgium**; Ave Circulaire 3, 1180 Bruxelles
- Astronomical Institute of the Czech Academy of Sciences, **Czech Republic**; Fričova 298, 25165 Ondrejov
- Academy of Finland, **Finland**; Hakaniemenranta 6, P.O.B. 131, 00531 Helsinki
- CNRS-INSU (Nançay Radioastronomy Facility & Paris Observatory), **France**; Route de Souesmes, 18330 Nançay
- Max-Planck-Institut für Radioastronomie, **Germany**; Auf dem Hügel 69, 53121 Bonn
- Aristotele University Thessaloniki, Department of Physics, **Greece;** 54124 Thessaloniki
- Hungarian Academy of Science, **Hungary**; Széchenyi István tér 9, 1051 Budapest
- Institut de Radio Astronomie Millimétrique (I.R.A.M.), **International Research Institute**; 300 Rue de la Piscine, Domaine Universitaire, 38406 Saint Martin d'Hères, France
- School of Physics, Trinity College, University of Dublin, **Ireland**; College Green, 2 Dublin
- Istituto Nazionale di Astrofisica (INAF), **Italy**; Via P. Gobetti 101, 40129 Bologna
- International VLBI Service for Geodesy and Astrometry (IVS), **International collaboration**; Coordinating Center, NASA Goddard Space Flight Center, Greenbelt, MD 20771, USA
- Ventspils International Radio Astronomy Centre (VIRAC), Ventspils University of Applied Sciences, **Latvia**; Inzenieru iela 101, 3601 Ventspils
- Joint Institute for VLBI ERIC (JIVE), The Netherlands, Oude Hoogeveensedijk 4, 7991PD Dwingeloo
- Stichting ASTRON, Netherlands Institute for Radio Astronomy, **The Netherlands**; Oude Hoogeveensedijk 4, 7991PD Dwingeloo
- Kartverket Geodetisk Observatorie, **Norway;** Hamnerabben 1, 9173 Ny-Ålesund, Svalbard
- Astronomical Observatory, Jagiellonian University, **Poland**; ul. Orla 171, 30-244 Kraków
- Instituto de Telecomunicações (Polo de Aveiro) – GRITPortugal, Campus Universitário de San, **Portugal**; 193 AVEIRO, 3810 Aveiro
- South African Radio Astronomy Observatory (SARAO), **South Africa**; SKA South Africa, 3rd. Floor, The Park, Park Road, Pinelands, 7405 Cape Town
- Instituto Geográfico Nacional (IGN), Subdirección General de Astronomia, **Spain**; calle General Ibáñez de Ibero 3, 28003 Madrid
- Onsala Space Observatory, Chalmers, **Sweden**; 43992 Onsala
- Institute of Applied Physics, University of Bern, **Switzerland**; Sidlerstrasse 5, 3012 Bern
- Center for Space and Habitability (CSH), University of Bern, **Switzerland**; Gesellschaftsstrasse 6, 3012 Bern
- Department of Astronomy and Space Sciences, Erciyes University, **Turkey**; Kutadgu Bilig Sk. Bina7, 38030 Melikgazi/Kayseri
- Institute of Radio Astronomy of National Academy of Sciences of Ukraine, **Ukraine**; 4, Mystetstv St, 61002 Kharkiv
- Science and Technology Facilities Council (STFC), represented by Manchester University & Cambridge University, **UK**; Polaris House, North Star Avenue, Swindon SN2 1SZ

## 7.2 List of LOFAR stations

*Table 4: List of LOFAR stations.*

| Country | Name | N Latitude | E Longitude |
|---|---|---|---|
| **The Netherlands** ***LOFAR Core*** | CS001 | 52° 54′ 41″ | 6° 52′ 3.5″ |
| | CS002 | 52° 54′ 54″ | 6° 52′ 11.4″ |
| | CS003 | 52° 54′ 57″ | 6° 52′ 6.8″ |
| | CS004 | 52° 54′ 52″ | 6° 52′ 6.5″ |
| | CS005 | 52° 54′ 51″ | 6° 52′ 13.3″ |
| | CS006 | 52° 54′ 53″ | 6° 52′ 17.8″ |
| | CS007 | 52° 54′ 56″ | 6° 52′ 15.7″ |
| | CS011 | 52° 54′ 50″ | 6° 52′ 24.8″ |
| | CS013 | 52° 55′ 5″ | 6° 52′ 0.6″ |
| | CS017 | 52° 54′ 57″ | 6° 52′ 37.6″ |
| | CS021 | 52° 55′ 4″ | 6° 51′ 46.8″ |
| | CS024 | 52° 54′ 29″ | 6° 52′ 28.2″ |
| | CS026 | 52° 54′ 58″ | 6° 52′ 57.6″ |
| | CS028 | 52° 55′ 33″ | 6° 52′ 30.3″ |
| | CS030 | 52° 55′ 20″ | 6° 51′ 41.2″ |
| | CS031 | 52° 55′ 4″ | 6° 51′ 32.5″ |
| | CS032 | 52° 54′ 45″ | 6° 51′ 35.6″ |
| | CS101 | 52° 55′ 21″ | 6° 52′ 52.3″ |
| | CS103 | 52° 54′ 57″ | 6° 53′ 48.8″ |
| | CS201 | 52° 54′ 45″ | 6° 52′ 57.9″ |
| | CS301 | 52° 54′ 21″ | 6° 52′ 4.4″ |
| | CS302 | 52° 54′ 6″ | 6° 50′ 57.3″ |
| | CS401 | 52° 54′ 51″ | 6° 51′ 21.2″ |
| | CS501 | 52° 55′ 35″ | 6° 51′ 57.2″ |
| **The Netherlands** ***Remote stations*** | RS106 | 52° 52′ 27″ | 6° 59′ 6.3″ |
| | RS205 | 52° 51′ 27″ | 6° 53′ 48.3″ |
| | RS208 | 52° 40′ 10″ | 6° 55′ 8.3″ |
| | RS210 | 52° 19′ 51″ | 6° 52′ 24.9″ |
| | RS305 | 52° 53′ 58″ | 6° 46′ 26.2″ |
| | RS306 | 52° 53′ 25″ | 6° 44′ 35.8″ |
| | RS307 | 52° 48′ 12″ | 6° 40′ 51.4″ |
| | RS310 | 52° 45′ 52″ | 6° 8′ 19.1″ |
| | RS406 | 53° 1′ 4″ | 6° 45′ 1.5″ |
| | RS407 | 53° 5′ 33″ | 6° 47′ 3.4″ |
| | RS409 | 52° 58′ 50″ | 6° 21′ 28.9″ |
| | RS503 | 52° 56′ 42″ | 6° 51′ 3.1″ |
| | RS508 | 53° 14′ 24″ | 6° 57′ 11.4″ |
| | RS509 | 53° 24′ 32″ | 6° 47′ 7.0″ |
| **Finland** | FI609 | 69° 4′ 15″ | 20° 45′ 43.5″ |
| **France** | FR606 | 47° 22′ 33″ | 2° 11′ 36.1″ |
| **Germany** | DE601 | 50° 31′ 24″ | 6° 53′ 2.2″ |
| | DE602 | 48° 30′ 4″ | 11° 17′ 17.3″ |
| | DE603 | 50° 58′ 46″ | 11° 42′ 40.6″ |
| | DE604 | 52° 26′ 19″ | 13° 0′ 58.8″ |
| | DE605 | 50° 53′ 50″ | 6° 25′ 24.4″ |
| | DE609 | 53° 41′ 54″ | 9° 58′ 11.4″ |
| **Ireland** | IE613 | 53° 5′ 43″ | –7° 55′ 19.9″ |
| **Poland** | PL610 | 52° 16′ 34″ | 17° 4′ 24.1″ |
| | PL611 | 49° 57′ 52″ | 20° 29′ 24.3″ |
| | PL612 | 53° 35′ 38″ | 20° 35′ 20.0″ |
| **Sweden** | SE607 | 57° 23′ 56″ | 11° 55′ 46.8″ |
| **United Kingdom** | UK608 | 51° 8′ 38″ | –1° 26′ 0.6″ |

## 7.3 ECC Decisions relevant to RAS

| Year (last Rev.) | Decision | Topic | RAS Band | Title | Comments |
|---|---|---|---|---|---|
| **2009** | (04)09 | MSS | 1660–1670 MHz | on the designation of the bands 1518–1525 MHz and 1670–1675 MHz for systems in the Mobile-Satellite Service | Background<br>Considering e) |
| **2009** | (09)04 | MSS; License exemption | 1610.6–1613.8 MHz | on exemption from individual licensing and the free circulation and use of transmit-only mobile satellite terminals operating in the Mobile-Satellite Service allocations in the 1613.8–1626.5 MHz band | Considering j) |
| **2010** | (99)15 | Multimedia- and Fixed-Wireless Systems (WMS) | 42.5–43.5 GHz | on the designation of the harmonised frequency band 40.5 to 43.5 GHz for the introduction of Multimedia Wireless Systems (MWS) and Point-to-Point (P-P) Fixed Wireless Systems | Considering b), e) |
| **2010** | (10)01 | FS, MS, EEES | 10.6–10.68 GHz | on sharing conditions in the 10.6–10.68 GHz band between the fixed service, mobile service and Earth exploration satellite service (passive) | Considering a) |
| **2012** | (09)02 | MSS | 1610.6–1613.8 MHz | The harmonisation of the bands 1610–1626.5 MHz and 2483.5–2500 MHz for use by systems in the Mobile-Satellite Service | Considering d), f), l), m), o)<br>Decides 2, 4<br>Annex |
| **2015** | (04)03 | Automotive SRD | 77–81 GHz | The frequency band 77–81 GHz to be designated for the use of Automotive Short Range Radars | Considering g) |
| **2015** | (12)01 | MSS; mobile terminals | 1660–1670 MHz | Exemption from individual licensing and free circulation and use of terrestrial and satellite mobile terminals operating under the control of networks | Considering w) |
| **2016** | (16)01 | UWB;<br>Helicopter Radar | 76–77 GHz | The harmonised frequency band 76–77 GHz, technical characteristics, exemption from individual licensing and free carriage and use of obstacle detection radars for rotorcraft use | Considering i), j), n)<br>Decides 8, 9, 12<br>Annex 2 |
| **2017** | (11)01 | Protection of EESS (somewhat relevant for RAS) | 1400–1427 MHz | The Protection of the Earth Exploration-Satellite Service (passive) in the 1400–1427 MHz Band | Considering a)<br>Decides 1, 2 |
| **2018** | (17)06 | IMT LTE (MFCN); supplemental downlink | 1400–1427 MHz | The harmonised use of the frequency bands 1427–1452 MHz and 1492–1518 MHz for Mobile/Fixed Communications Networks Supplemental Downlink (MFCN SDL) | Background<br>Considering f) |
| **2018** | (18)04 | GSO FSS; ESIMs | 10.6–10.7 GHz<br>14.47–14.5 GHz | The harmonised use, exemption from individual licensing and free circulation and use of land based Earth Stations In-Motion (ESIM) operating with GSO FSS satellite systems in the frequency bands 10.7–12.75 GHz and 14.0–14.5 GHz | Introduction<br>Background<br>Considering h)<br>Decides 3, 5a), 5b), 6a), 6b), 7c) |
| **2019** | (05)10 | FSS ESIMs | 14.47–14.5 GHz<br>(Earth-to-space) | The free circulation and use of Earth Stations on board Vessels operating in fixed satellite service networks in the frequency bands 14–14.5 GHz | Considering e) |
| **2019** | (05)11 | MMS, Aircraft Earth Stations | 14.47–14.5 GHz<br>(Earth-to-space)<br>10.6–10.7 GHz<br>(space-to-Earth) | The free circulation and use of Aircraft Earth Stations (AES) in the frequency bands 14.0–14.5 GHz (Earth-to-space), 10.7–11.7 GHz (space-to-Earth) and 12.5–12.75 GHz (space-to-Earth) | Background<br>Considering e)<br>Decides 3d)<br>Annex 2 |
| **2019** | (06)04 | UWB; license exemption | Several below 10.6 GHz<br>(incl. 5.340 bands!) | The harmonised use, exemption from individual licensing and free circulation of devices using Ultra-Wideband (UWB) technology in bands below 10.6 GHz | Considering o) |
| **2019** | (07)01 | UWB; license exemption | Several below 10.6 GHz<br>(incl. 5.340 bands!) | The harmonised use, exemption from individual licensing and free circulation of Material Sensing Devices using Ultra-Wideband (UWB) technology | Considering n)<br>Table 1/2, Note 2* |
| **2019** | (16)02 | Broadband Public Protection and Disaster Relief (BB-PPDR) | 406.1–410 MHz | Harmonised technical conditions and frequency bands for the implementation of Broadband Public Protection and Disaster Relief (BB-PPDR) systems | Considering r), nn), qq) |
| **2019** | (19)02 | Land mobile systems | 73.0–74.6 MHz<br>150.05–153.00 MHz<br>406.1–410 MHz | Land mobile systems in the frequency ranges 68–87.5 MHz, 146–174 MHz, 406.1–410 MHz, 410–430 MHz, 440–450 MHz and 450–470 MHz | Considering a), b), v), y)<br>Annex 5 |
| **2019** | (05)05 | IMT 5G | 2690–2700 MHz | Harmonised utilization of spectrum for Mobile/Fixed Communications Networks (MFCN) operating within the band 2500–2690 MHz | Considering r), s), u)<br>Table 1 |
| **2019** | (11)02 | UWB/SRD | 6650–6675.2 MHz<br>23.6–24.0 GHz<br>58.2–59 GHz<br>65–65 GHz<br>76–85 GHz | Industrial Level Probing Radars (LPR) operating in frequency bands 6–8.5 GHz, 24.05–26.5 GHz, 57–64 GHz and 75–85 GHz | Considering g), h)<br>Annex 3 |
| **2020** | (18)06 | IMT 5G (MFCN) | 23.6–24.0 GHz | Harmonised technical conditions for Mobile/Fixed Communications Networks (MFCN) in the band 24.25–27.5 GHz | Background |
| **2020** | (20)01 | WAS/RLAN (WiFi) | 6650–6675.2 MHz | On the harmonised use of the frequency band 5945–6425 MHz for Wireless Access Systems including Radio Local Area Networks (WAS/RLAN) | Considering c) |
| **2021** | (04)10 | Automotive SRD | 23.6–24.0 GHz | The frequency bands to be designated for the temporary introduction of Automotive Short Range Radars (SRR) | Background<br>Considering l), m), t), ee)<br>Decides 3, 4<br>Annex 3 |
| **2021** | (99)06 | S-PCS<1GHz | 150.05–153.00 MHz<br>406.1–410 MHz | On the harmonised introduction of satellite personal communication systems operating in the bands below 1 GHz (S-PCS<1GHz) | Annex 2 (Systems: Hiberband, Argos Kineis, Myriota)<br>Annex 3 |
| **2021** | (17)04 | NGSO FSS; License exemption of fixed Earth stations | 10.6–10.7 GHz (space-to-Earth)<br>14.47–14.5 GHz (Earth-to-space) | The harmonised use and exemption from individual licensing of fixed earth stations operating with NGSO FSS satellite systems in the frequency bands 10.7–12.75 GHz and 14.0–14.5 GHz | Considering g), h), l)<br>Decides c)<br>Table 1/2 |
| **2021** | (18)05 | NGSO FSS; License exemption of fixed ESIMs | 10.6–10.7 GHz (space-to-Earth)<br>14.47–14.5 GHz (Earth-to-space) | The harmonised use, exemption from individual licensing and free circulation and use of Earth Stations In-Motion (ESIM) operating with NGSO FSS satellite systems in the frequency bands 10.7–12.75 GHz and 14.0–14.5 GHz | Considering g), h), l)<br>Decides c)<br>Table 2/3 |
| **2022** | (21)01 | FSS | 48.94–49.04 GHz (Earth-to-space)<br>51.4–54.25 GHz (Earth-to-space) | The use of the bands 47.2–50.2 GHz and 50.4–52.4 GHz by the fixed-satellite service (Earth-to-space) | Considering f), j)<br>Decides 3, 4<br>Annex 2 |
| **2022** | (21)02 | HD-GBSAR (Radar) | 76–77 GHz | The harmonised frequency band 76–77 GHz, technical characteristics, exemption from individual licensing and free circulation and use of High Definition Ground Based Synthetic Aperture Radar (HD-GBSAR) | Considering f), g), h), o)<br>Decides 7, 8<br>Annex 1 |
| **2022** | (22)03 | Radars | Several in 116–260 GHz | On technical characteristics, exemption from individual licensing and free circulation and use of specific radiodetermination applications in the frequency range 116–260 GHz | Considering d), e), f), g), k), l), m), o)<br>Decides 3, 5, 8, |
| | | | | | 9, 13<br>Annex 1, 3 |
| **2022** | (22)07 | Aerial UE, MFCN | 1400–1427 MHz<br>(2nd harm. of 703–733 MHz)<br>1660–1670 MHz<br>(2nd harm. of 832–862 MHz)<br>2690–2700 MHz | On harmonised technical conditions for the usage of aerial UE for communications based on LTE and 5G NR in the bands 703–733 MHz, 832–862 MHz, 880–915 MHz, 1710–1785 MHz, 1920–1980 MHz, 2500–2570 MHz and 2570–2620 MHz harmonised for MFCN | Considering a), c), n), o)<br>Decides 3, 4<br>Annex 1 (esp. A1.2) |
| **2020** | (22)06 | IMT 5G (MFCN) | 42.5–43.5 GHz | Harmonised technical conditions for Mobile/Fixed Communications Networks (MFCN) in the band 40.5–43.5 GHz | Considering i), p), t), u), v), aa), bb), cc), ee), gg)<br>Decides 6, 7, 8<br>Annex 3 (esp. A3.2) |
| **2022** | (18)04 | GSO FSS; License exemption of fixed ESIMs | 10.6–10.7 GHz<br>(space-to-Earth)<br>14.47–14.5 GHz<br>(Earth-to-space) | On the harmonised use, exemption from individual licensing and free circulation and use of land based Earth Stations In-Motion (ESIM) operating with GSO FSS satellite systems in the frequency bands 10.7–12.75 GHz and 14.0–14.5 GHz | Considering h)<br>Decides 1c), 5, 6<br>Annex 2 |
| | | | | | |

## 7.4 ECC Reports relevant to RAS (since 2011)

| Year (last Rev.) | Report | Topic | RAS Band | Title |
|---|---|---|---|---|
| **2011** | 156 | HAPS Gateways | 6650.0–6675.2 MHz | Conditions for possible co-existence between HAPS gateway links and other services/systems in the 5850–7075 MHz band |
| **2011** | 157 | Fixed Radars | 2690–2700 MHz<br>4990–5000 MHz | The impact of spurious emissions of radars at 2.8, 5.6 and 9.0 GHz on other radiocommunication services/systems |
| **2011** | 158 | Automotive UWB/SRD | 23.6–24 GHz | The impact of SRR 26 GHz applications using Ultra-Wide-Band (UWB) Technology on Radio Services |
| **2011** | 159 | Cognitive radio; white space devices (WSDs) | 608 MHz | Technical and operational requirements for the possible operation of cognitive radio systems in the 'white spaces' of the frequency band 470–790 MHz |
| **2011** | 164 | Automotive UWB/SRD | 23.6–24 GHz | Compatibility between Wide Band Low Activity Mode (WLAM) automotive radars in the frequency range 24.25 GHz to 24.5 GHz, and other radiocommunication systems/services |
| **2011** | 165 | MSS; terrestrial terminals; CGC (MSS Complementary Ground Component) | 1610.6–1613.8 MHz | Compatibility study between MSS complementary ground component operating in the bands 1610.0–1626.5 MHz and 2483.5–2500.0 MHz and other systems in the same bands or in adjacent bands |
| **2011** | 166 | Meteorological radars | 23.6–24 GHz | Coexistence between zenith-pointing meteorological radars at 24 GHz and 35 GHz and systems in other radio services |
| **2011** | 168 | Pseudolites (Pseudo satellites, PLs) | 1610.6–1613.8 MHz | Regulatory framework for indoor GNSS pseudolites |
| **2011** | 170 | Location Tracking for Emergency (LAES) | 6650.0–6675.2 MHz | Specific UWB applications in the bands 3.4–4.8 GHz and 6–8.5 GHz Location Tracking Applications for Emergency Services (LAES), location tracking applications type 2 (LT2) and location tracking and sensor applications for automotive and transportation environments (LTA) |
| **2011** | 171 | MSS; downlink; Iridium; interference; monitoring | 1610.6–1613.8 MHz | Impact of unwanted emissions of Iridium satellites on radioastronomy operations in the band 1610.6–1613.8 MHz |
| **2012** | 172 | Broadband Wireless; Geodetic VLBI | 2200–2290 MHz | Broadband Wireless Systems Usage in 2300-2400 MHz |
| **2012** | 175 | UWB; aeronautical; in-flight entertainment | 6650–6675.2 MHz | Co-existence study considering UWB applications inside aircraft and existing radio services in the frequency bands from 3.1 GHz to 4.8 GHz and from 6.0 GHz to 8.5 GHz |
| **2013** | 183 | Pseudolites (Pseudo satellites, PLs) | 1610.6–1613.8 MHz | Regulatory Framework for Outdoor GNSS Pseudolites |
| **2013** | 186 | Cognitive radio; white space devices (WSDs) | 608 MHz | Technical and operational requirements for the operation of white space devices under geo-location approach |
| **2013** | 187 | Mobile communication on aircraft (MCA) | 2655–2690 MHz<br>2690–2700 MHz | Compatibility study between mobile communication services on board aircraft (MCA) and ground-based systems |
| **2015** | 218 | Broadband Public Protection and Disaster Relief (BB-PPDR); LTE/4G | 406.1–410 MHz | Harmonised conditions and spectrum bands for the implementation of future European Broadband Public Protection and Disaster Relief (BB-PPDR) systems |
| **2014** | 222 | UWB; Helicopter Radar | 76–77 GHz | The impact of Surveillance Radar equipment operating in the 76 to 79 GHz range for helicopter application on radio systems |
| **2015** | 226 | MSS; downlink; Iridium; interference; monitoring; epfd software | 1610.6–1613.8 MHz | Unwanted emissions of IRIDIUM satellites in the band 1610.6–1613.8 MHz, monitoring campaign 2013 |
| **2015** | 240 | Broadband Public Protection and Disaster Relief (BB-PPDR); LTE400 | 406.1–410 MHz | Compatibility studies regarding Broadband PPDR and other radio applications in 410–430 and 450–470 MHz and adjacent bands |
| **2016** | 243 | Video Programme Making and Special Events (PMSE) | 2690–2700 MHz | Wireless video links in the frequency bands 2700–2900 MHz and 2900–3400 MHz |
| **2016** | 245 | Audio Programme Making and Special Events (PMSE) | 1330–1400 MHz<br>1400–1427 MHz | Compatibility studies between a PMSE and other systems / services in the band 1350–1400 MHz |
| **2020** | 247 | MSS; downlink; Iridium; epfd software | 1610.6–1613.8 MHz | Description of the software tool for processing of measurements data of IRIDIUM satellites at the Leeheim station |
| **2022** | 249 | Measurements: unwanted emissions, 2$^{nd}$ harmonics | n/a | Unwanted emissions of common radio systems: measurements and use in sharing/compatibility studies |
| **2021** | 271 | NGSO FSS; Mega-constellations (Iridium/OneWeb) | 10.6–10.7 GHz (space-to-Earth) 14.47–14.5 GHz (Earth-to-space) | Compatibility and sharing studies related to NGSO satellite systems operating in the FSS bands 10.7–12.75 GHz (space-to-Earth) and 14–14.5 GHz (Earth-to-space) |
| **2018** | 278 | UWB; vehicular access | 6650.0–6675.2 MHz | Specific UWB applications in the bands 3.4-4.8 GHz and 6.0-8.5 GHz: Location tracking and sensor applications (LTA) for vehicular access systems |
| **2018** | 279 | NGSO FSS; ESIMs; Mega-constellations (OneWeb) | 10.6–10.7 GHz (space-to-Earth) 14.47–14.5 GHz (Earth-to-space) | The Use of Earth Stations In-Motion (ESIM) operating to NGSO Satellite Systems in the 10.7–12.75 GHz and 14–14.5 GHz Band |
| **2018** | 281 | IMT/5G | 3332–3339 MHz 3345.8–3352.5 MHz | Analysis of the suitability of the regulatory technical conditions for 5G MFCN operation in the 3400–3800 MHz band |
| **2018** | 285 | Video Programme Making and Special Events (PMSE) | 2690–2700 MHz | Best practices for Video Programme Making and Special Event (PMSE) in the 2700–2900 MHz band |
| **2019** | 292 | Professional (Private)/Public Access Mobile Radio (PMR/PAMR) | 408 MHz | Current Use, Future Opportunities and Guidance to Administrations for the 400 MHz PMR/PAMR frequencies |
| **2019** | 302 | WAS/RLAN (WiFi) | 6650.0–6675.2 MHz | Sharing and compatibility studies related to Wireless Access Systems including Radio Local Area Networks (WAS/RLAN) in the frequency band 5925–6425 MHz |
| **2020** | 308 | IMT Upgrade 4G to 5G | 2690–2700 MHz | Analysis of the suitability and update of the regulatory technical conditions for 5G MFCN and AAS operation in the 2500–2690 MHz band |
| **2020** | 309 | Aerial use of IMT user equipment | 1400–1427 MHz (2$^{nd}$ harm. of 703–733 MHz) 1660–1670 MHz (2$^{nd}$ harm. of 832–862 MHz) 2690–2700 MHz 3332–3339 MHz 3345.8–3352.5 MHz | Analysis of the usage of aerial UE for communication in current MFCN harmonised bands |
| **2020** | 315 | HD-GBSAR (Radar) | 76–77 GHz | Feasibility of spectrum sharing between High-Definition Ground Based Synthetic Aperture Radar (HD-GBSAR) application using 1 GHz bandwidth within 74–81 GHz and existing services and applications |
| **2020** | 317 | IMT 5G base stations, License exemption | 23.6–24 GHz | Additional work on 26 GHz to address spectrum use under authorisation regimes other than individual rights of use: Technical toolkit to assist administrations |
| **2020** | 321 | Wind farms; measurements, compatibility studies | 30–240 MHz<br>610 MHz, 1420 MHz | Radio frequency test methods, tools and test results for wind turbines in relation to the Radio Astronomy Service |
| **2021** | 325 | Private LTE networks | 4950–5000 MHz<br>(2nd harm. of 2483.5–2500 MHz) | Compatibility and technical feasibility of coexistence studies for the potential introduction of new terrestrial applications operating in the 2483.5–2500 MHz frequency band with existing services / applications in the same band and adjacent bands |
| **2021** | 327 | UWB | 6650.0–6675.2 MHz | Technical studies for the update of the Ultra Wide Band (UWB) regulatory framework in the band 6.0 GHz to 8.5 GHz |
| **2022** | 322 | Satellite-IoT | Several below 1 GHz | Compatibility analysis (inter-service and intra service) for S-PCS below 1 GHz |
| **2022** | 334 | UWB/SRD; also: use of passive bands | Several in 116–260 GHz | UWB radiodetermination applications in the frequency range 116–260 GHz |
| **2022** | 348 | Aerial use of IMT user equipment with AAS base stations | 2690–2700 MHz | Usage of aerial UE in 1.8 GHz, 2 GHz and 2.6 GHz frequency bands with MFCN AAS base stations |
| **2023** | 349 | New Iridium measurements | 1610.6–1613.8 MHz | Unwanted emissions of Iridium Next satellites in the band 1610.6–1613.8 MHz, monitoring campaign of November 2020 to May 2021 |
| **2023** | 350 | Car radars | 77–81 GHz | Radiodetermination equipment for ground based vehicular applications in 77–81 GHz |
| **2023** | 351 | Car radars | Several in 116–148.5 GHz | UWB radiodetermination applications within the frequency range 116 GHz to 148.5 GHz for vehicular use |
| | | | | |

## 7.5 List of input documents contributed to ITU-R and CEPT working groups

*Table 5: Input documents to CEPT and ITU-R work groups. It is noted that a fraction of the documents are joint contributions with other spectrum management stakeholders (in particular other scientific organisations) and (European) administrations. In some cases, CRAF contributed (significantly) to documents submitted by administrations. These cases are not listed in the table. The document naming scheme was taken from CEPT/ECC and ITU-R document servers, which is not always very illustrative. Please note that most of the documents cannot be freely accessed but require an ITU TIES or CEPT website account.*

| Date | Document | Org. | Group | Topic |
|---|---|---|---|---|
| **Aug/2023** | SE24(23)034 | CEPT | SE24 | Security scanners between 3.6–10.6 GHz (SE24_76) |
| **Aug/2023** | FM44(23)Info01 | CEPT | FM44 | Measurements of electromagnetic leakage from SpaceX/Starlink satellites in VHF band |
| **Aug/2023** | PTA(23)INFO022 | CEPT | PTA | Measurements of electromagnetic leakage from SpaceX/Starlink satellites in VHF band |
| **Jul/2023** | SE40(23)INFO07, SE40(23)INFO07A01 | CEPT | SE40 | Measurements of electromagnetic leakage from SpaceX/Starlink satellites in VHF band |
| **Jun/2023** | SE(23)057 | CEPT | WGSE | Update of ECC Report 271 (Starlink) |
| **May/2023** | SE40(23)014 | CEPT | SE40 | Aggregation from (multiple) SatCons (SE40_45) |
| **May/2023** | SE21(23)028 | CEPT | SE21 | Report on Methodology for receiver resilience |
| **May/2023** | FM44(23)016 | CEPT | FM44 | Non-terrestrial networks (direct-to-device via satellites) |
| **May/2023** | PTA(23)074 | CEPT | PTA | New proposed agenda item – Protection of RAS from large satellite constellations |
| **Apr/2023** | SE21(23)016, SE21(23)016A01 | CEPT | SE21 | Report on Methodology for receiver resilience |
| **Mar/2023** | R19-CPM23.2-C-0085 | ITU-R | CPM | Proposed modifications of the draft CPM Report to WRC-23 – Agenda item 1.13 |
| **Mar/2023** | SE24-SWG3(22)010, SE24-SWG3(22)010A1 | CEPT | SE24 | Car radars in 116–260 GHz (SE24_75) |
| **Feb/2023** | SE21(23)005, SE21(23)005A01 | CEPT | SE21 | Recommendation on receiver resilience |
| **Jan/2023** | SE40(23)007 | CEPT | SE40 | Aggregation from (multiple) SatCons (SE40_45) |
| **Jan/2023** | PTD(22)INFO002 | CEPT | PTD | Comparison between flat-terrain and real terrain path propagation |
| **Jan/2023** | PTC(23)019 | CEPT | PTC | WRC23 AI 1.10 CEPT–RAS sites in 22 GHz and 15 GHz |
| **Jan/2023** | PTA(23)INFO003 | CEPT | PTA | New proposed agenda item - Protection of RAS from space services |
| **Jan/2023** | PTA(23)024 | CEPT | PTA | Frequency prioritisation for future RX only space weather allocations |
| **Jan/2023** | PTA(23)023 | CEPT | PTA | Proposed amendments draft CEPT Brief WRC-23 Agenda Item 9.1 topic a |
| **Jan/2023** | PTA(23)022 | CEPT | PTA | Proposed updates to draft ECP WRC23 AI 1.13 |
| **Jan/2023** | PT1(23)026 | CEPT | PT1 | CRAF, SKAO – AI 1.2-related studies IMT vs RAS |
| **Oct/2022** | SE40(22)019, SE40(22)019A1 | CEPT | SE40 | Aggregation from (multiple) SatCons (SE40_45) |

| Date | Document | Org. | Group | Topic |
|---|---|---|---|---|
| **Sep/2022** | R19-WP7D-C-0199 | ITU-R | WP7D | Contribution to PDN Report ITU-R RA.[RAS-IMT-COMPAT-43-GHz] (Sharing and compatibility studies of MFCN 5G/AAS vs. RAS at 43 GHz) |
| **Sep/2022** | R19-WP7D-C-0198 | ITU-R | WP7D | Contribution to PDN Report ITU-R RA.[VGOS] |
| **Sep/2022** | R19-WP7D-C-0197 | ITU-R | WP7D | Contribution to PDN Report ITU-R RA.[EHT] |
| **Sep/2022** | R19-WP7D-C-0196 | ITU-R | WP7D | Contribution to PDN Report ITU-R RA.[CMB] |
| **Sep/2022** | R19-WP7D-C-0195 | ITU-R | WP7D | Contribution to PDN Report ITU-R RA.[67-116GHZ] |
| **Sep/2022** | R19-WP7D-C-0194 | ITU-R | WP7D | Draft rev. of Rec. RA.314 (preferred frequency bands) |
| **Sep/2022** | R19-WP7C-C-0445 | ITU-R | WP7C | Rev. of Rep. RS.2489 (water vapor radiometers) |
| **Sep/2022** | R19-WP7C-C-0438 | ITU-R | WP7C | Rev. of Rep. RS.2456 (space weather sensors) |
| **Sep/2022** | R19-WP7B-C-0227 | ITU-R | WP7B | Proposal for draft CPM text (AI 1.13) |
| **Sep/2022** | SE(22)103, SE(22)103A01, SE(22)103A02 | CEPT | WGSE | Car radars in 77–81 GHz (SE24_73) |
| **Sep/2022** | SE24-SWG3(22)038, SE24-SWG3(22)038A1, SE24-SWG3(22)038A2 | CEPT | SE24 | Car radars in 77–81 GHz (SE24_73) |
| **Sep/2022** | PTA(22)012 | CEPT | PTA | Proposed updates to Draft CEPT Brief on WRC-23 agenda item 9.1 topic a |
| **Aug/2022** | R19-TG6.1-C-0108 | ITU-R | TG6/1 | Sharing and compatibility studies for WRC-23 AI 1.5 (Potential IMT identification in 470-694 MHz, protection of RAS in 606-614 MHz) |
| **Jul/2022** | SE40(22)014 | CEPT | SE40 | Aggregation from (multiple) SatCons (SE40_45) |
| **Jul/2022** | SE40(22)013 | CEPT | SE40 | New Iridium measurements (SE40_12) |
| **Jun/2022** | SE45(22)015 | CEPT | SE45 | WiFi in upper 6 GHz band |
| **Jun/2022** | SE40_45(22)003 | CEPT | SE40 | Aggregation from (multiple) SatCons (SE40_45) |
| **Jun/2022** | SE24_WI73#10_01, SE24_WI73#10-01A1, SE24_WI73#10-01A2 | CEPT | SE24 | Car radars in 77–81 GHz (SE24_73) |
| **Jun/2022** | WI75#7-03 | CEPT | SE24 | Car radars in 116–260 GHz (SE24_75) |
| **Jun/2022** | SE24_WI73#08-02 | CEPT | SE24 | Car radars in 77–81 GHz (SE24_73) |
| **May/2022** | PT1(22)127 | CEPT | PT1 | CRAF – Positions for section 6 in CEPT brief on AI1.2 and 1.4 |
| **May/2022** | PT1(22)126 | CEPT | PT1 | CRAF–SKAO Proposal on Draft CEPT Brief on WRC-23 agenda item 1.4 |
| **Apr/2022** | R19-WP7D-C-0161 | ITU-R | WP7D | Contribution to PDN Report ITU-R RA.[RAS Harmonics] |
| **Apr/2022** | R19-WP7D-C-0150 | ITU-R | WP7D | Contribution to PDN Report ITU-R RA.[RAS-IMT-COMPAT-43-GHz] (Sharing and compatibility studies of MFCN 5G/AAS vs. RAS at 43 GHz) |
| **Apr/2022** | R19-WP7D-C-0149 | ITU-R | WP7D | Contribution to WD PDN Report ITU-R RA.[RAS below 300 MHz] |
| **Apr/2022** | R19-WP7D-C-0148 | ITU-R | WP7D | Contribution to WD PDN Report ITU-R RA.[mm bolometers] |

| Date | Document | Org. | Group | Topic |
|---|---|---|---|---|
| **Apr/2022** | R19-WP7C-C-0353 | ITU-R | WP7C | Contribution do WD PDN Report ITU-R RS.[GROUND SENSOR] |
| **Apr/2022** | R19-WP7C-C-0351 | ITU-R | WP7C | Proposal for draft CPM text (AI 9.1 a) |
| **Apr/2022** | R19-WP7C-C-0338 | ITU-R | WP7C | Rev. of Rep. RS.2456 (space weather sensors) |
| **Apr/2022** | R19-WP7B-C-0186 | ITU-R | WP7B | Contribution do WD PDN Report ITU-R SA.[15 GHz SRS SHARING] (WRC-23 AI 1.13) |
| **Apr/2022** | SRDMG(23)035 | CEPT | SRD/MG | UWB/SRD applications, 116–260 GHz, 5.340 use |
| **Apr/2022** | SRDMG(22)058 | CEPT | SRD/MG | UWB/SRD applications, 116–260 GHz |
| **Apr/2022** | SE40(22)006 | CEPT | SE40 | Aggregation from (multiple) SatCons (SE40_45) |
| **Apr/2022** | PT1(22)110 | CEPT | PT1 | CRAF – 40 GHz CEPT Report LRTC |
| **Mar/2022** | SE40_45(22)001 | CEPT | SE40 | Aggregation from (multiple) SatCons (SE40_45) |
| **Mar/2022** | PTA(22)012 | CEPT | PTA | Proposed updates to Draft CEPT Brief on WRC-23 agenda item 9.1 topic a |
| **Mar/2022** | CG-5.340(22)05 | CEPT | CG 5.340 | UWB/SRD applications, 5.340 use |
| **Mar/2022** | UWB(22)12 | CEPT | CG UWB | UWB/SRD applications, 116–260 GHz, 5.340 use |
| **Feb/2022** | UWB(22)006 | CEPT | CG UWB | UWB/SRD applications, 116–260 GHz, 5.340 use |
| **Jan/2022** | SRDMG(22)021 | CEPT | SRD/MG | UWB/SRD applications, 116–260 GHz, 5.340 use |
| **Jan/2022** | SRDMG(22)021 | CEPT | SRD/MG | Comments on sharing of UWB and passive RR 5.340 bands |
| **Jan/2022** | PT1(22)066 | CEPT | PT1 | Contribution to WP 5D on 2nd harmonics (WRC-23 AI 1.4, HIBS at 800 MHz) |
| **Nov/2021** | SE40(21)040 | CEPT | SE40 | Draft new Report on aggregated interference into RAS bands (proposed structure and work plan) |
| **Oct/2021** | R19-TG6.1-C-0056 | ITU-R | TG6/1 | Sharing and compatibility studies for WRC-23 AI 1.5 (Potential IMT identification in 470-694 MHz, protection of RAS in 606-614 MHz) |
| **Sep/2021** | R19-WP7D-C-0107 | ITU-R | WP7D | Considerations regarding a possible designation of receive-only SWS systems to the RAS |
| **Sep/2021** | R19-WP7D-C-0106 | ITU-R | WP7D | Contribution to PDN Report ITU-R RA.[RAS-IMT-COMPAT-43-GHz] (Sharing and compatibility studies of MFCN 5G/AAS vs. RAS at 43 GHz) |
| **Sep/2021** | R19-WP7D-C-0104 | ITU-R | WP7D | Contribution to PDN Report ITU-R RA.[RAS 6-7-GHz] (Sharing and compatibility, MFCN and other systems vs. RAS at 6.6 GHz) |
| **Sep/2021** | R19-WP7C-C-0271 | ITU-R | WP7C | Considerations regarding a possible designation of receive-only SWS systems to the RAS |
| **Sep/2021** | R19-WP7C-C-0267 | ITU-R | WP7C | Views on PDN Report ITU-R RS.[SPEC_REQTS_RX_SPACE _WEATHER] (Spectrum requirements and applicable radio service designations) |
| **Sep/2021** | R19-WP7C-C-0266 | ITU-R | WP7C | Views on PDN Report ITU-R RS.[RXSW_INTERF_CRITERIA] (Interference criteria of receive-only space weather sensors) |
| **Sep/2021** | R19-WP7C-C-0265 | ITU-R | WP7C | Updates to elements regarding WRC-23 AI 9.1 (a) |

| Date | Document | Org. | Group | Topic |
|---|---|---|---|---|
| **Sep/2021** | R19-WP5D-C-0814 | ITU-R | WP5D | Compatibility studies for WRC-23 AI 1.4 (RAS vs. HIBS, second harmonics of HIBS into 1 610.6-1 613.8 MHz RAS band) |
| **Sep/2021** | R19-WP5D-C-0788 | ITU-R | WP5D | Compatibility studies for WRC-23 AI 1.4 (RAS vs. HIBS, OOB of HIBS into 2690-2700 MHz RAS band) |
| **Sep/2021** | PT1(21)229 | CEPT | PT1 | Compatibility studies (WRC-23 AI 1.2, IMT vs RAS at 6650 MHz) |
| **Sep/2021** | PTD(21)028R1 | CEPT | PTD | Compatibility studies (WRC-23 AI 1.5, Potential IMT identification in 470-694 MHz, protection of RAS in 606-614 MHz) |
| **Sep/2021** | PTA(21)039 | CEPT | PTA | Contribution to Draft CEPT Brief on WRC-23 AI 9.1 Topic A (Receive-only space weather sensors and the radio astronomy service) |
| **Jun/2021** | SE40(21)019 | CEPT | SE40 | Aggregate interference from satellite constellations to RAS |
| **Jun/2021** | PTD(21)013 | CEPT | PTD | Contribution to Draft CEPT Brief on WRC-23 AI 1.5 (Potential IMT identification in 470-694 MHz, protection of RAS in 606-614 MHz) |
| **May/2021** | SE24(21)022R1 | CEPT | SE24 | Contribution to Draft new ECC Report 334 (UWB/SRD applications, 116-260 GHz) |
| **May/2021** | FM44(21)020 | CEPT | FM44 | Propose editorial changes to ECC Decision (17)04 to align it with ECC Decision (18)05 (regarding RAS protection) |
| **May/2021** | FM44(21)019 | CEPT | FM44 | Propose text regarding RAS protection for Draft new ECC Decision (21)01 (FSS Q&V band uplinks) |
| **Apr/2021** | R19-WP7D-C-0058 | ITU-R | WP7D | Draft New Question ITU-R XXX/7 (Towards recognition of the geodetic VLBI Global Observing System, VGOS) |
| **Apr/2021** | R19-WP7D-C-0057 | ITU-R | WP7D | Draft Reply LS to TG 6/1 on preparations of WRC-23 AI 1.5 (Information on RAS in the band 470-960 MHz in Region 1) |
| **Apr/2021** | SE(21)076 | CEPT | WGSE | Draft Reply LS from WG SE to WG FM on S-PCS below 1 GHz |
| **Apr/2021** | SE(21)068R1 | CEPT | WGSE | Comments on aggregate interference from satellite systems to the RAS |
| **Apr/2021** | SE(21)068A01 | CEPT | WGSE | Propose new Work Item on satellite system apportionment to WGSE |
| **Apr/2021** | PT1(21)101 | CEPT | PT1 | Preliminary compatibility studies (WRC-23 AI 1.4, HIBS at 700-800 MHz, 2900 MHz) |
| **Apr/2021** | PT1(21)100 | CEPT | PT1 | Preliminary compatibility studies (WRC-23 AI 1.2, IMT vs RAS at 6650 MHz) |
| **Mar/2021** | SRDMG(21)026 | CEPT | SRD/MG | Contribution to Draft new ECC Decision (21)02 (HD-GBSAR within 74-81 GHz) |
| **Mar/2021** | SE40(21)008 | CEPT | SE40 | Revision ECC Report 271; PC (Annex 6) |
| **Mar/2021** | SE40(21)008 | CEPT | SE40 | Revision ECC Report 271; PC (Annex 9) |
| **Mar/2021** | SE40(21)008 | CEPT | SE40 | Revision ECC Report 271; PC (Annex 8) |
| **Mar/2021** | SE40(21)008 | CEPT | SE40 | Revision ECC Report 271; PC (Annex 5) |

| Date | Document | Org. | Group | Topic |
|---|---|---|---|---|
| **Mar/2021** | WI63#8-02 | CEPT | SE24 | Contribution to Draft new ECC Report 327 (Update existing UWB regulatory framework) |
| **Mar/2021** | SE24(21)022_Rev1 | CEPT | SE24 | Contribution to Draft new ECC Report 334 (UWB/SRD applications, 116-260 GHz) |
| **Mar/2021** | PTA(21)INFO003 | CEPT | PTA | Update on CRAF positions on WRC-23 agenda items |
| **Feb/2021** | WI71#15-05 | CEPT | SE24 | Contribution to Draft new ECC Report 334 (UWB/SRD applications, 116-260 GHz) |
| **Feb/2021** | SE24_WI73#03-06 | CEPT | SE24 | Contribution to Draft new ECC Report (SRD automotive applications, 77-81 GHz) |
| **Feb/2021** | PTD(21)INFO003 | CEPT | PTD | Preliminary CRAF positions on WRC-23 Agenda Item 1.5 |
| **Feb/2021** | PTD(21)007 | CEPT | PTD | On RAS spectrum needs within the frequency band 470-960 MHz in Region 1 (related to WRC-23 AI 1.5) |
| **Jan/2021** | WI71#14-07 | CEPT | SE24 | Comments on sharing of passive RR 5.340 bands |
| **Jan/2021** | WI71#14-06 | CEPT | SE24 | Contribution to Draft new ECC Report 334 (UWB/SRD applications, 116-260 GHz) |
| **Jan/2021** | WI63#4-02 | CEPT | SE24 | Contribution to Draft new ECC Report 327 (Update existing UWB regulatory framework) |
| **Dec/2020** | SE7(20)090A1 | CEPT | SE7 | Contribution to Draft new ECC Report 325 (Private LTE networks at 2.5 GHz) |
| **Dec/2020** | SE7(20)090 | CEPT | SE7 | Contribution to Draft new ECC Report 325 (Private LTE networks at 2.5 GHz); Cover page |
| **Dec/2020** | SE24(20)124 | CEPT | SE24 | Contribution to Draft new ECC Report (SRD automotive applications, 77-81 GHz) |
| **Dec/2020** | Temp 2 | CEPT | PT1 | Contribution to Draft new ECC Decision (MFCN/5G at 40 GHz, protection of RAS at 42.5-43.5 GHz) |
| **Dec/2020** | PT1_CG40(20)036 | CEPT | PT1 | Contribution to Draft new ECC Decision (MFCN/5G at 40 GHz, protection of RAS at 42.5-43.5 GHz) |
| **Dec/2020** | PT1_CG40(20)030 | CEPT | PT1 | Contribution to Draft new ECC Decision (MFCN/5G at 40 GHz, protection of RAS at 42.5-43.5 GHz) |
| **Dec/2020** | PT1_CG40(20)018 | CEPT | PT1 | Contribution to Draft new ECC Decision (MFCN/5G at 40 GHz, protection of RAS at 42.5-43.5 GHz); Cover page |
| **Dec/2020** | PT1_CG40(20)018 | CEPT | PT1 | Contribution to Draft new ECC Decision (MFCN/5G at 40 GHz, protection of RAS at 42.5-43.5 GHz) |
| **Dec/2020** | PT1_CG40 GHz (20)038 | CEPT | PT1 | Contribution to Draft new CEPT Report (MFCN/5G at 40 GHz, protection of RAS at 42.5-43.5 GHz); Cover page |
| **Dec/2020** | PT1_CG40 GHz (20)037 | CEPT | PT1 | Contribution to Draft new CEPT Report (MFCN/5G at 40 GHz, protection of RAS at 42.5-43.5 GHz) |
| **Dec/2020** | FM44(20)078 | CEPT | FM44 | Proposal regarding aggregate effects from satellite constellations (S-PCS below 1 GHz) for revision of ERC Decision 99(06) |
| **Nov/2020** | SE7(20)081A1 | CEPT | SE7 | Contribution to Draft new ECC Report 325 (Private LTE networks at 2.5 GHz) |
| **Nov/2020** | SE7(20)081 | CEPT | SE7 | Contribution to Draft new ECC Report 325 (Private LTE networks at 2.5 GHz); Cover page |

| Date | Document | Org. | Group | Topic |
|---|---|---|---|---|
| **Nov/2020** | WI71#13-02 | CEPT | SE24 | Considerations regarding clutter losses for car radar applications and the RAS (Draft new ECC Report 334) |
| **Nov/2020** | PT1_CG40(20)022 | CEPT | PT1 | Contribution to Draft new CEPT Report (MFCN/5G at 40 GHz, protection of RAS at 42.5-43.5 GHz) |
| **Nov/2020** | PT1_CG40(20)018_An1 | CEPT | PT1 | Contribution to Draft new ECC Decision (MFCN/5G at 40 GHz, protection of RAS at 42.5-43.5 GHz) |
| **Nov/2020** | PT1_CG40(20)018 | CEPT | PT1 | Contribution to Draft new ECC Decision (MFCN/5G at 40 GHz, protection of RAS at 42.5-43.5 GHz); Cover page |
| **Nov/2020** | PT1_CG40(20)017_An1 | CEPT | PT1 | Contribution to Draft new CEPT Report (MFCN/5G at 40 GHz, protection of RAS at 42.5-43.5 GHz) |
| **Nov/2020** | PT1_CG40(20)017 | CEPT | PT1 | Contribution to Draft new CEPT Report (MFCN/5G at 40 GHz, protection of RAS at 42.5-43.5 GHz); Cover page |
| **Nov/2020** | PT1_CG40(20)013 | CEPT | PT1 | Contribution to Draft new CEPT Report (MFCN/5G at 40 GHz, protection of RAS at 42.5-43.5 GHz) |
| **Nov/2020** | PT1_CG40(20)006 | CEPT | PT1 | Information on protection of RAS at 42.5-43.5 GHz from MFCN/5G at 40 GHz |
| **Nov/2020** | CPG(20)INFO 21 | CEPT | CPG | Preliminary CRAF positions on WRC-23 agenda items |
| **Oct/2020** | Info22 | CEPT | WGFM | Questions on implementation status of ECC Decision (09)02 in CEPT countries (related to Iridium) |
| **Oct/2020** | WI71#12-04 | CEPT | SE24 | Considerations regarding clutter losses for car radar applications and the RAS (Draft new ECC Report 334) |
| **Sep/2020** | R19-WP7D-C-0022 | ITU-R | WP7D | Proposal for Draft New Question ITU-R XXX/7 (Towards recognition of the geodetic VLBI Global Observing System, VGOS) |
| **Sep/2020** | SE40(20)069 | CEPT | SE40 | Contribution to Draft ECC Report 322 (S-PCS below 1 GHz) |
| **Sep/2020** | SE24(20)084 | CEPT | SE24 | Comments on sharing of passive RR 5.340 bands with RDI-S (Draft new ECC Report 334) |
| **Aug/2020** | PT1(20)153 | CEPT | PT1 | Information on protection of RAS at 42.5-43.5 GHz from MFCN/5G at 40 GHz |
| **Jun/2020** | SE40(20)044 | CEPT | SE40 | Draft LS to FM 44 on aggregate interference from satellite constellations to RAS |
| **Jun/2020** | SE40(20)043 | CEPT | SE40 | Contribution to Draft ECC Report 322 (S-PCS below 1 GHz) |
| **May/2020** | SE40(20)032 | CEPT | SE40 | EPFD Software for RAS calculations involving satellite constellations |
| **May/2020** | WI71#7-06 | CEPT | SE24 | Comments on sharing of passive RR 5.340 bands with RDI-S (Draft new ECC Report 334) |
| **Apr/2020** | SE40(20)026 Annex 1 | CEPT | SE40 | Contribution to Draft ECC Report 322 (S-PCS below 1 GHz) |
| **Apr/2020** | SE40(20)026 | CEPT | SE40 | RAS compatibility studies (S-PCS below 1 GHz) |
| **Apr/2020** | WI71#6-03 | CEPT | SE24 | Information on RAS calculations for Draft new ECC Report 334 (UWB/SRD applications, 116-260 GHz) |
| **Mar/2020** | WI71#5-5 | CEPT | SE24 | Contribution to Draft new ECC Report 334 (UWB/SRD applications, 116-260 GHz) |

| Date | Document | Org. | Group | Topic |
|---|---|---|---|---|
| **Feb/2020** | FM(20)039 | CEPT | WGFM | Proposal to exclude passive RR 5.340 bands from use by radiodetermination applications in 116-260 GHz |
| **Feb/2020** | ECC(20)032 | CEPT | ECC | Supporting letter on German satellite monitoring station Leeheim and its relevance (not only) to RAS |
| **Feb/2020** | ECC(20)031 | CEPT | ECC | Call for action on continued interference by Iridium NEXT into RAS stations at 1610.6-1613.8 MHz |
| **Jan/2020** | SRDMG(29Info 6 | CEPT | SRD/MG | Proposal to exclude passive RR 5.340 bands from use by radiodetermination applications in 116-260 GHz |
| **Jan/2020** | SE24(20)020 | CEPT | SE24 | Contribution to Draft new ECC Report 334 (UWB/SRD applications, 116-260 GHz) |
| **Jan/2020** | PT1(20)023 | CEPT | PT1 | Contribution to Draft new ECC Report 309 (Aerial usage of MFCN user equipment, protection of RAS in a number of bands) |
| **Dec/2019** | SE24(19)WI71#4-6A1 | CEPT | SE24 | Contribution to Draft new ECC Report 334 (UWB/SRD applications, 116-260 GHz) |
| **Dec/2019** | SE24(19)WI71#4-6 | CEPT | SE24 | Contribution to Draft new ECC Report 334 (UWB/SRD applications, 116-260 GHz) |
| **Nov/2019** | WI70#3-2 | CEPT | SE24 | Contribution to Draft new ECC Report 315 (HD-GBSAR within 74-81 GHz) |
| **Sep/2019** | PT1(19)230 | CEPT | PT1 | Contribution to Draft new ECC Report 309 (Aerial usage of MFCN user equipment, protection of RAS in a number of bands) |
| **Aug/2019** | CG_UAS_10_A2 | CEPT | PT1 | Contribution to Draft new ECC Report 309 (Aerial usage of MFCN user equipment, single-entry studies) |
| **Aug/2019** | CG_UAS_10_A1 | CEPT | PT1 | Contribution to Draft new ECC Report 309 (Aerial usage of MFCN user equipment, aggregation studies) |
| **Aug/2019** | CG_UAS_10 | CEPT | PT1 | Contribution to Draft new ECC Report 309 (Aerial usage of MFCN user equipment, protection of RAS in a number of bands) |
| **Aug/2019** | CPG(19)138 | CEPT | CPG | Contribution to Draft CEPT Brief on WRC-19 AI 1.14 (HAPS; RAS protection in RR No. 5.149-only band at 6650-6675 MHz) |
| **Jul/2019** | CG_UAS_5_Annex1 | CEPT | PT1 | Contribution to Draft new ECC Report 309 (Aerial usage of MFCN user equipment, aggregation studies) |
| **Jul/2019** | CG_UAS_5 | CEPT | PT1 | Contribution to Draft new ECC Report 309 (Aerial usage of MFCN user equipment, protection of RAS in a number of bands) |
| **Jul/2019** | ECC(19)076 Rev1 (1) | CEPT | ECC | MCP on continued contributions to SatMOU to fund Leeheim measurements |
| **Jun/2019** | PT1(19)145 | CEPT | PT1 | Contribution to Draft new ECC Report 309 (Aerial usage of MFCN user equipment, single-entry studies) |
| **Mar/2019** | WTFG(19)003 | CEPT | FG WT | Contribution to Draft new ECC Report 321 (Methods, tools and test results for wind turbines in relation to RAS) |
| **Feb/2019** | PT1(19)109 | CEPT | PT1 | Contribution to Draft new ECC Report 308 (Technical conditions for MFCN 5G/AAS operation, RAS protection at 2690 MHz) |

| Date | Document | Org. | Group | Topic |
|---|---|---|---|---|
| **Jan/2019** | PT1(19)071 | CEPT | PT1 | Contribution to Draft new ECC Report 308 (Technical conditions for MFCN 5G/AAS operation, RAS protection at 2690 MHz) |
| **Jan/2019** | PT1(19)053 | CEPT | PT1 | Contribution to European Common Proposal for WRC-19 AI 1.13 (MFCN 5G at 43 GHz, protection of RAS) |
| **Jan/2019** | 1 | CEPT | PT1 | Contribution to Draft new ECC Report 308 (Technical conditions for MFCN 5G/AAS operation, RAS protection at 2690 MHz) |
| **Dec/2018** | 1 | CEPT | PT1 | Contribution to Draft new ECC Report 308 (Technical conditions for MFCN 5G/AAS operation, RAS protection at 2690 MHz) |
| **Nov/2018** | CPG(18)INFO67 | CEPT | CPG | Updated CRAF positions on WRC-19 agenda items |
| **Aug/2018** | R15-TG5.1-C-0439 | ITU-R | TG5/1 | Contribution to Draft CPM text for WRC-19 AI 1.13 (MFCN 5G between 24 and 86 GHz, protection of RAS) |
| **Aug/2018** | PTA(18)082 | CEPT | PTA | Contribution to Draft CEPT Brief on WRC-19 AI 1.14 (HAPS; RAS protection in RR No. 5.149-only band at 6650-6675 MHz) |
| **Jun/2018** | R15-WP7D-C-0169 | ITU-R | WP7D | Views on necessary elements of HAPS - RAS compatibility studies (WRC-19 AI 1.14, RAS at 21, 24 and 31 GHz) |
| **Jun/2018** | R15-WP5C-C-0535 | ITU-R | WP5C | Views on necessary elements of HAPS - RAS compatibility studies (WRC-19 AI 1.14, RAS at 21, 24 and 31 GHz) |
| **May/2018** | R15-WP7D-C-0161 | ITU-R | WP7D | Comments on PDN Report ITU-R M.[RAS-COMPAT] (Compatibility Iridium NEXT vs. RAS at 1610.6-1613.8 MHz) |
| **May/2018** | R15-WP7D-C-0160 | ITU-R | WP7D | Proposal for additional studies in PDN Report ITU-R S.[50/40 GHZ ADJACENT BAND STUDIES] (WRC-19 AI 1.6, non-GSO FSS) |
| **May/2018** | R15-WP7D-C-0159 | ITU-R | WP7D | Comments on Draft CPM text for WRC-19 AI 1.6 (Modernization of GMDSS, introduction of new systems, Iridium NEXT) |
| **May/2018** | R15-WP7D-C-0158 | ITU-R | WP7D | Proposal for Draft Reply LS to WP 5C (WRC-19 AI 1.14, HAPS; RAS protection in RR No. 5.149-only band at 6650-6675 MHz) |
| **May/2018** | R15-WP7D-C-0154 | ITU-R | WP7D | Proposal for revision of Recs. ITU-R RA.314 and ITU-R RA.1860 (Preferred RAS frequency bands) |
| **May/2018** | R15-WP5C-C-0476 | ITU-R | WP5C | Contribution to PDN Report ITU-R F.[HAPS-31GHZ] (related to WRC-19 AI 1.14, RAS at 28 and 31 GHz) |
| **May/2018** | R15-WP5C-C-0475 | ITU-R | WP5C | Contribution to PDN Report ITU-R F.[HAPS-21GHZ] (related to WRC-19 AI 1.14, RAS at 22 GHz) |
| **Apr/2018** | R15-TG5.1-C-0362 | ITU-R | TG5/1 | Contribution to Draft CPM text for WRC-19 AI 1.13 (MFCN 5G between 24 and 86 GHz, protection of RAS) |

| Date | Document | Org. | Group | Topic |
|---|---|---|---|---|
| **Apr/2018** | R15-TG5.1-C-0296 | ITU-R | TG5/1 | Contribution to TG5/1 Chairman's Report on WRC-19 AI 1.13 (Sharing and compatibility studies MFCN 5G/AAS vs. RAS at 86 GHz) |
| **Apr/2018** | R15-TG5.1-C-0292 | ITU-R | TG5/1 | Contribution to TG5/1 Chairman's Report on WRC-19 AI 1.13 (Compatibility studies MFCN 5G/AAS vs. RAS at 31 GHz) |
| **Apr/2018** | R15-TG5.1-C-0291 | ITU-R | TG5/1 | Contribution to TG5/1 Chairman's Report on WRC-19 AI 1.13 (Sharing and compatibility studies MFCN 5G/AAS vs. RAS at 43 GHz) |
| **Apr/2018** | Observations | CEPT | SE40 | Comments on Leeheim observations and Iridium analyses |
| **Apr/2018** | PT1(18)097 | CEPT | PT1 | Contribution to Draft CEPT Brief on WRC-19 AI 1.13 (MFCN 5G between 24 and 86 GHz, protection of RAS) |
| **Apr/2018** | PT1(18)032 | CEPT | PT1 | Compatibility studies (WRC-19 AI 1.13, MFCN 5G/AAS vs. RAS at 86 GHz) |
| **Apr/2018** | PT1(18)031 | CEPT | PT1 | Update on compatibility studies (WRC-19 AI 1.13, MFCN 5G/AAS vs. RAS at 43 GHz) |
| **Apr/2018** | PT1(18)030 | CEPT | PT1 | Update on compatibility studies (WRC-19 AI 1.13, MFCN 5G/AAS vs. RAS at 31 GHz) |
| **Feb/2018** | R15-WP4C-C-0334 | ITU-R | WP4C | Contribution to Draft CPM text for WRC-19 AI 1.8 (Modernization of GMDSS, introduction of new systems, Iridium NEXT) |
| **Feb/2018** | PTA(18)046 | CEPT | PTA | Contribution to Draft CEPT Brief on WRC-19 AI 1.15 (Potential new identification of LMS and FS in 275-450 GHz; RAS protection) |
| **Feb/2018** | PTA(18)045 | CEPT | PTA | Contribution to Draft CEPT Brief on WRC-19 AI 1.14 (HAPS; RAS protection at 31.3-31.8 GHz) |
| **Feb/2018** | PTA(18)044 | CEPT | PTA | Compatibility studies (WRC-19 AI 1.7, Short-duration non-GSO satellites vs. RAS in VHF/UHF band) |
| **Jan/2018** | R15-TG5.1-C-0240 | ITU-R | TG5/1 | Contribution to TG5/1 Chairman's Report on WRC-19 AI 1.13 (Sharing and compatibility studies MFCN 5G/AAS vs. RAS at 43 GHz) |
| **Jan/2018** | R15-TG5.1-C-0239 | ITU-R | TG5/1 | Contribution to TG5/1 Chairman's Report on WRC-19 AI 1.13 (Compatibility studies MFCN 5G/AAS vs. RAS at 31 GHz) |
| **Jan/2018** | R15-TG5.1-C-0238 | ITU-R | TG5/1 | Contribution to TG5/1 Chairman's Report on WRC-19 AI 1.13 (Compatibility studies MFCN 5G/AAS vs. RAS at 24 GHz) |
| **Jan/2018** | FM44(18)Info3 | CEPT | FM44 | On Iridium NEXT out-of-band emission measurements |
| **Dec/2017** | SE40(17)067 | CEPT | SE40 | Analysis of the first Leeheim measurement of Iridium NEXT |
| **Dec/2017** | PT1(17)217 | CEPT | PT1 | Update on compatibility studies (WRC-19 AI 1.13, MFCN 5G/AAS vs. RAS at 43 GHz) |
| **Dec/2017** | PT1(17)216 | CEPT | PT1 | Contribution to Draft new ECC Report 281 (Technical conditions for MFCN 5G/AAS operation, RAS protection at 3350 MHz) |

| Date | Document | Org. | Group | Topic |
|---|---|---|---|---|
| **Dec/2017** | CPG(18)016 | CEPT | CPG | Elements to consider for CEPT position on WRC-19 AI 1.8 (Modernization of GMDSS, introduction of new systems, Iridium NEXT) |
| **Oct/2017** | R15-WP7D-C-0116 | ITU-R | WP7D | Proposal for modifications to Rec. ITU-R RS.2066-0 (List of radio telescopes capable of operating in the band 10.6-10.7 GHz) |
| **Oct/2017** | R15-WP7D-C-0115 | ITU-R | WP7D | Contribution to PDN Report ITU-R A.2259 (Characteristics of radio quiet zones, IRAM 30 m radio telescope in Spain) |
| **Oct/2017** | R15-WP7D-C-0114 | ITU-R | WP7D | Proposal for Draft Reply LS to TG 5/1 (Apportionment of interference between services) |
| **Oct/2017** | R15-WP7D-C-0113 | ITU-R | WP7D | Contribution to PDN Report ITU-R RA.[COEXISTENCE] (Coexistence between RAS and radiolocation the band 76-81 GHz) |
| **Oct/2017** | R15-WP7D-C-0111 | ITU-R | WP7D | Proposal for Draft LS to WP 4C (Compatibility Iridium NEXT vs. RAS at 1610.6-1613.8 MHz) |
| **Oct/2017** | R15-WP7C-C-0188 | ITU-R | WP7C | Contribution to PDN Report ITU-R RS.[SPACE_WEATHER_SENSORS] (Technical and operational characteristics of SWS) |
| **Sep/2017** | R15-TG5.1-C-0164 | ITU-R | TG5/1 | Contribution to TG5/1 Chairman's Report on WRC-19 AI 1.13 (Sharing and compatibility studies MFCN 5G/AAS vs. RAS at 43 GHz) |
| **Sep/2017** | R15-TG5.1-C-0163 | ITU-R | TG5/1 | Contribution to TG5/1 Chairman's Report on WRC-19 AI 1.13 (Compatibility studies MFCN 5G/AAS vs. RAS at 31 GHz) |
| **Sep/2017** | R15-TG5.1-C-0162 | ITU-R | TG5/1 | Contribution to TG5/1 Chairman's Report on WRC-19 AI 1.13 (Compatibility studies MFCN 5G/AAS vs. RAS at 24 GHz) |
| **Sep/2017** | PTA(17)050 | CEPT | PTA | Compatibility studies (WRC-19 AI 1.7, Short-duration non-GSO satellites vs. RAS in VHF/UHF band) |
| **Aug/2017** | SE40(17)044 | CEPT | SE40 | Aggregate effect from NGSO FSS systems on RAS |
| **Aug/2017** | PT1(17)159 | CEPT | PT1 | Contribution to Draft CEPT Brief on WRC-19 AI 1.13 (MFCN 5G between 24 and 86 GHz, protection of RAS) |
| **Aug/2017** | PT1(17)158 | CEPT | PT1 | Compatibility studies (WRC-19 AI 1.13, MFCN 5G/AAS vs. RAS at 43 GHz) |
| **Aug/2017** | PT1(17)157 | CEPT | PT1 | Compatibility studies (WRC-19 AI 1.13, MFCN 5G/AAS vs. RAS at 31 GHz) |
| **Aug/2017** | PT1(17)156 | CEPT | PT1 | Update on compatibility studies (WRC-19 AI 1.13, MFCN 5G/AAS vs. RAS at 24 GHz) |
| **Jul/2017** | PTB(17)INFO15rev | CEPT | PTB | Updated CRAF positions on WRC-19 agenda items |
| **Jun/2017** | SE40(17)021 (1) | CEPT | SE40 | Comments on the proposed modification of the EPFD software (Iridium NEXT) |
| **Jun/2017** | CPG(17)INFO30 | CEPT | CPG | Updated CRAF positions on WRC-19 agenda items |
| **May/2017** | R15-TG5.1-C-0074 | ITU-R | TG5/1 | Contribution to TG5/1 Chairman's Report on WRC-19 AI 1.13 (Compatibility studies MFCN 5G/AAS vs. RAS at 24 GHz) |

| Date | Document | Org. | Group | Topic |
|---|---|---|---|---|
| **Apr/2017** | PT1(17)103 | CEPT | PT1 | Compatibility studies (WRC-19 AI 1.13, MFCN 5G/AAS vs. RAS at 24 GHz) |
| **Mar/2017** | SE7(17)015R1 (1) | CEPT | SE7 | Contribution to Draft new ECC Report 218 (Broadband PPDR networks; LTE vs. RAS at 408 MHz) |
| **Mar/2017** | SE40(17)009_Annex | CEPT | SE40 | Comments on Draft new ECC Report 271 (NGSO FSS, OneWeb) |
| **Mar/2017** | SE40(17)009 | CEPT | SE40 | Comments on Draft new ECC Report 271 (NGSO FSS, OneWeb); Cover page |
| **Mar/2017** | PTD(17)INFO 29 | CEPT | PTD | Updated CRAF positions on WRC-19 agenda items |
| **Mar/2017** | CPG(17)INFO20 | CEPT | CPG | Updated CRAF positions on WRC-19 agenda items |
| **Jan/2017** | PT1 (17)050 | CEPT | PT1 | Contribution to Draft new ECC Decision (17)06 (MFCN supplemental downlink in 1427-1452 MHz, protection of RAS at 1400-1427 MHz) |
| **Jan/2017** | PTB(17)INFO09 | CEPT | PTB | Updated CRAF positions on WRC-19 agenda items |
| **Dec/2016** | SE7(16)072 | CEPT | SE7 | Contribution to Draft new ECC Report 218 (Broadband PPDR networks; LTE vs. RAS at 408 MHz) |
| **Dec/2016** | PTD(17)INFO 24 | CEPT | PTD | Preliminary CRAF positions on WRC-19 agenda items |
| **Dec/2016** | CPG(16)INFO14 | CEPT | CPG | Preliminary CRAF positions on WRC-19 agenda items |
| **Nov/2016** | SE40(16)039 | CEPT | SE40 | Comparison of ECC Reports 112, 171 and 226 (Iridium NEXT) |
| **Sep/2016** | SE40(16)022 | CEPT | SE40 | Iridium NEXT - Measurements |
| **Sep/2016** | PT1(16)115 | CEPT | PT1 | Contribution to Draft CEPT Brief on WRC-19 AI 1.13 (MFCN 5G between 24 and 86 GHz, protection of RAS) |
| **Sep/2016** | PTA(16)INFO04 | CEPT | PTA | Preliminary CRAF positions on WRC-19 Agenda Items 1.2, 1.7, 1.14, and 1.15 |
| **Jun/2016** | SE40(16)013 | CEPT | SE40 | Protection of RAS from NGSO FSS at 14 GHz (OneWeb) |
| **May/2016** | FM(16)132 | CEPT | WGFM | Request to carry out compatibility studies concerning RAS and NGSO FSS at 14 GHz |
| **May/2016** | FM44(16)012 | CEPT | FM44 | Protection of RAS in the Ku Band |
| **Jan/2016** | FM(16)057 | CEPT | WGFM | Comments on the NATO proposal for the revision of the ECA table for a number of frequency bands |
| **Sep/2015** | SE7(15)108 Att1 | CEPT | SE7 | List of RAS stations for Draft new ECC Report 245 (PMSE in the band 1350-1400 MHz) |
| **Sep/2015** | SE7(15)108 Annex1 | CEPT | SE7 | Contribution to Draft new ECC Report 245 (PMSE in the band 1350-1400 MHz) |
| **Sep/2015** | SE7(15)108 | CEPT | SE7 | Contribution to Draft new ECC Report 245 (PMSE in the band 1350-1400 MHz); Cover page |
| **Sep/2015** | SE7(15)107 | CEPT | SE7 | Comments to Public Consultation: Draft new ECC Report 218 (Broadband PPDR networks; LTE vs. RAS at 408 MHz) |
| **Sep/2015** | CPG15(15)INFO47 | CEPT | CPG | Updated CRAF positions on WRC-15 agenda items |

| Date | Document | Org. | Group | Topic |
|---|---|---|---|---|
| **Aug/2015** | INFO 003 | CEPT | SE40 | Effelsberg horizon data for Draft new ECC Report 226 (Iridium NEXT) |
| **Aug/2015** | INFO 002 | CEPT | SE40 | Comments on Draft new ECC Report 226 (Iridium NEXT) |
| **Jun/2015** | SE40(15)29 | CEPT | SE40 | Guidance document on satellite monitoring stations measurement routines |
| **Jun/2015** | SE40(15)24 (3) | CEPT | SE40 | Comments on proposed changes to EPFD analysis software regarding Iridium |
| **Jun/2015** | CPG15(15)INFO22R2 | CEPT | CPG | Updated CRAF positions on WRC-15 agenda items |
| **May/2015** | R12-WP7D-C-0155 | ITU-R | WP7D | Proposal for corrections to Rec. ITU-R RS.2066-0 (Protection of the RAS 10.6-10.7 GHz from EESS (active) around 9.6 GHz) |
| **May/2015** | R12-WP7C-C-0341 | ITU-R | WP7C | Editorial correction to Annex 2 of Rec. ITU-R RS.2066-0 (Protection of the RAS 10.6-10.7 GHz from EESS (active) around 9.6 GHz) |
| **Apr/2015** | SE7(15)082 | CEPT | SE7 | Contribution to Draft new ECC Report 218 (Broadband PPDR networks; LTE vs. RAS at 408 MHz) |
| **Apr/2015** | SE40(15)011_Annex | CEPT | SE40 | Guidance document on satellite monitoring stations measurement routines |
| **Apr/2015** | SE40(15)011 | CEPT | SE40 | Guidance document on satellite monitoring stations measurement routines |
| **Apr/2015** | SE40(15)007 (1) | CEPT | SE40 | Specification of measurements on Iridium NEXT satellites |
| **Mar/2015** | SE7(15)007 | CEPT | SE7 | Contribution to Draft new ECC Report 218 (Broadband PPDR networks; LTE vs. RAS at 408 MHz) |
| **Mar/2015** | SE7(15)006 | CEPT | SE7 | Contribution to Draft new ECC Report 245 (PMSE in the band 1350-1400 MHz) |
| **Feb/2015** | FM(15)037 | CEPT | WGFM | Request for a technical/regulatory solution to force Iridium NEXT to comply with ECC Decision (09)02 |
| **Nov/2014** | SE40(14)042 | CEPT | SE40 | Tools for the evaluation of satellite measurements |
| **Sep/2014** | Info 003 | CEPT | WGFM | Letter: CRAF to FCC concerning Iridium NEXT |
| **Sep/2014** | FM(14)166 | CEPT | WGFM | CRAF position document with respect to Iridium NEXT |
| **Sep/2014** | CPG15(14)040 | CEPT | CPG | Updated CRAF positions on WRC-15 agenda items |
| **Aug/2014** | FM44(14)028 | CEPT | FM44 | CRAF Comments to Minutes of FM44#31 |
| **Jun/2014** | FM44(14)info2 | CEPT | FM44 | CRAF/Iridium documentation WGFM#80 |
| **Jun/2014** | FM44(14)023 | CEPT | FM44 | RAS protection in 1660-1670 MHz from MES-AES |
| **May/2014** | FM(14)091 | CEPT | WGFM | Report on several bilateral meeting between CRAF and Iridium |
| **May/2014** | FM(14)088 | CEPT | WGFM | Conclusions from CRAF-Iridium meetings and request for action |
| **Apr/2014** | R12-WP7D-C-0106 | ITU-R | WP7D | Proposal to modify of RR No. 5.511F (Protection of RAS in the 15.4 GHz band from radiolocation service) |
| **Mar/2014** | FM44(14)006 | CEPT | FM44 | RAS protection in 1660-1670 MHz from MES-AES |

| Date | Document | Org. | Group | Topic |
|---|---|---|---|---|
| **Feb/2014** | R12-JTG4567-C-0477 | ITU-R | JTG4567 | Contribution to PDN Report ITU-R RA.[RAS-IMT] |
| **Feb/2014** | R12-JTG4567-C-0476 | ITU-R | JTG4567 | Contribution to Draft CPM text for WRC-15 AI 1.1 (IMT vs. RAS in several bands between 608 MHz and 5 GHz; related to PDN Report ITU-R RA.[RAS-IMT]) |
| **Feb/2014** | R12-JTG4567-C-0475 | ITU-R | JTG4567 | Contribution to PDN Report ITU-R RA.[RAS-IMT] (Compatibility of IMT vs. RAS in several bands between 608 MHz and 5 GHz) |
| **Feb/2014** | R12-JTG4567-C-0474 | ITU-R | JTG4567 | Contribution to PDN Report ITU-R RA.[RAS-IMT] (Sharing of IMT vs. RAS in 1330-1400 MHz) |
| **Jan/2014** | INFO 007 Rev1 | CEPT | WGFM | Report on a bilateral meeting between CRAF and Iridium on Jan 28, 2014 |
| **Oct/2013** | R12-JTG4567-C-0376 | ITU-R | JTG4567 | Contribution to PDN Report ITU-R RA.[RAS-IMT] (Compatibility and sharing of IMT vs. RAS in several bands between 608 MHz and 5 GHz) |
| **Sep/2013** | INFO 009 | CEPT | WGFM | Correspondence between CRAF and FCC and Iridium |
| **Mar/2013** | R12-WP7D-C-0055 | ITU-R | WP7D | Contribution to PDN Report ITU-R RS.[EESS-9GHZ_OOBE] (Protection of the RAS 10.6-10.7 GHz from EESS (active) around 9.6 GHz) |
| **Mar/2013** | R12-WP7C-C-0121 | ITU-R | WP7C | Contribution to PDN Report ITU-R RS.[EESS-9GHZ_OOBE] (Protection of the RAS 10.6-10.7 GHz from EESS (active) around 9.6 GHz) |
| **Jan/2013** | (13)007 rev1 (1) | CEPT | CPG | Preliminary CRAF positions on WRC-15 agenda items |
| **Sep/2012** | R12-WP7D-C-0028 | ITU-R | WP7D | Compatibility studies (Protection of the RAS 10.6-10.7 GHz from EESS (active) around 9.6 GHz) |
| **Sep/2012** | R12-WP7C-C-0050 | ITU-R | WP7C | Compatibility study (Protection of the RAS 10.6-10.7 GHz from EESS (active) around 9.6 GHz) |
| **May/2012** | SE7(12)037 | CEPT | SE7 | Contribution to Draft new ECC Report 187 (Mobile comm. on board aircraft; MCA vs. RAS at 2700 MHz) |
| **Dec/2011** | ECC(11)082R1 (2) | CEPT | ECC | MCP: Letter to FCC from ECC about harmful interference to RAS in 1610.6-1613.8 MHz caused by Iridium |
| **Oct/2011** | 035 (1) | CEPT | CPG | CRAF positions for WRC-12 |
| **Sep/2011** | SE7(11)info015 | CEPT | SE7 | Presentation on VLBI radio astronomy at SE7 meeting |
| **Sep/2011** | SE7(11)043 | CEPT | SE7 | Contribution to Draft new ECC Report 172 (Broadband wireless systems in the band 2300-2400 MHz; VLBI operations) |
| **Aug/2011** | M61_11R0_SE24 | CEPT | SE24 | Contribution to Draft new ECC Report 175 (Airborne UWB in 6-8.5 GHz) |
| **Jun/2011** | M60_15R0_SE24 | CEPT | SE24 | Potential interference from non-automotive radar applications at 76-77 GHz (ETSI TR 102 704) |
| **Jun/2011** | ECC(11)053 | CEPT | ECC | Unwanted emissions limits from SRR operating in the 24.25-26.65 GHz band to protect EESS and RAS in the 23.6-24 GHz band |
| **Apr/2011** | SE40(11)040 | CEPT | SE40 | Pseudolites and RAS |
| **Apr/2011** | M59_09R0_SE24 | CEPT | SE24 | Potential impact from non-automotive radar applications at 76-77 GHz on RAS (ETSI TR 102 704) |

| Date | Document | Org. | Group | Topic |
|---|---|---|---|---|
| **Apr/2011** | M59_08R1_SE24 | CEPT | SE24 | Potential impact from automotive LTA on RAS (6-7 GHz) |
| **Apr/2011** | FM47(11)014 | CEPT | FM47 | Potential impact from automotive LTA on RAS (6-7 GHz) |
| **Mar/2011** | M59_08R0_SE24 | CEPT | SE24 | Potential impact from automotive MFCN/LTE on RAS (6-7 GHz) |
| **Jan/2011** | SE40(11)025 | CEPT | SE40 | Scaling of interference thresholds |
| **Jan/2011** | SE40(11)024 | CEPT | SE40 | ECC Report: Impact of Unwanted Emissions of Iridium Satellites on Radio Astronomy Operations in the Band 1610.6-1613.8 MHz |
| **Jan/2011** | SE40(11)023 | CEPT | SE40 | Summary of measurements performed by Leeheim |